\documentclass[12pt]{article}
\pdfoutput=1
\usepackage{pstricks}
\usepackage{color}
\usepackage{amssymb,amsmath,bm,bbm}
\usepackage{epsf}
\usepackage{epsfig}
\usepackage{afterpage}
\usepackage{longtable}
\usepackage{cite}
\usepackage{latexsym,mathrsfs,dsfont}
\usepackage{graphics}
\usepackage{url}
\usepackage{paralist}
\usepackage{bbold}
\usepackage{appendix}

\newcommand{\spur}[1]{\not\! #1 \,}

\newcommand{\sm}{\textsc{SM}}

\newcommand{\be}{\begin{equation}}
\newcommand{\ee}{\end{equation}}
\newcommand{\bi}{\begin{itemize}}
\newcommand{\ei}{\end{itemize}}
\newcommand{\ba}{\begin{array}}
\newcommand{\ea}{\end{array}}
\newcommand{\bea}{\begin{eqnarray}}
\newcommand{\eea}{\end{eqnarray}}
\newcommand{\bec}{\begin{center}}
\newcommand{\eec}{\end{center}}
\newcommand{\dd}{\displaystyle}
\newcommand{\nn}{\nonumber}

\newcommand{\mup}{{\mu^\prime}}
\newcommand{\nup}{{\nu^\prime}}
\newcommand{\mupi}{\hat{\mu}_\pi^2}
\newcommand{\muG}{\hat{\mu}_G^2}

\newcommand{\rhoD}{\hat{\rho}_D^3}
\newcommand{\rhoLS}{\hat{\rho}_{LS}^3}

\def\@seccntformat#1{\@ifundefined{#1@cntformat}%
   {\csname the#1\endcsname\quad}
   {\csname #1@cntformat\endcsname}
}
\begin{document}

\begin{flushright} {BARI-TH/20-724}\end{flushright}

\medskip

\begin{center}
{\Large  Inclusive semileptonic $\Lambda_b$ decays in the Standard Model and beyond}
\\[1.0 cm]
{\large P.~Colangelo$^{a}$, F.~De~Fazio$^{a}$ and F.~Loparco$^{a,b}$
 \\[0.5 cm]}
{\small
$^a$
Istituto Nazionale di Fisica Nucleare, Sezione di Bari, Via Orabona 4,
I-70126 Bari, Italy\\
$^b$
Universit\'a degli Studi  di Bari, Via Orabona 4,
I-70126 Bari, Italy
}
\end{center}

\vskip0.5cm

\begin{abstract}
\noindent
Inclusive semileptonic decays of beauty baryons are studied using the heavy quark expansion  to  ${\cal O}(1/m_b^3)$,  at leading order in $\alpha_s$. The case of a polarized decaying baryon is examined,  with  reference to $\Lambda_b$. An extension of the  Standard Model effective Hamiltonian  inducing $b \to U \ell {\bar \nu}_ \ell$ transitions ($U=u,\,c$ and $\ell=e,\,\mu,\,\tau$)  is considered, which comprises the full  set of D=6 semileptonic operators with left-handed neutrinos. The effects of the new operators in several observables are described.
\end{abstract}

\thispagestyle{empty}
\newpage

\section{Introduction}

The  observations of anomalies in  $b \to c$ semileptonic exclusive decays of $B$ mesons, with hints toward  possible violation of lepton flavour universality (LFU) \footnote{For a review see \cite{Bifani:2018zmi}}, require new analyses of related processes involving heavy hadrons with a single $b$ quark, to enlarge the set of observables suitable to test the Standard Model (SM) predictions. The inclusive semileptonic  modes are theoretically appealing, since the nonperturbative effects of strong interactions, which necessarily must be taken into account, can be systematically considered by an expansion in the inverse heavy quark mass \cite{Chay:1990da,Bigi:1993fe}. The expansion involves a set of long-distance hadronic matrix elements of operators of increasing dimension, which  can be classified and parametrized. For each  term in the heavy quark expansion  perturbative QCD corrections can also be computed at increasing order in $\alpha_s$, therefore a double expansion in $1/m_Q$ and $\alpha_s$ is obtained.  Improving the control of  QCD effects,  in the inclusive as well as in the exclusive processes,   is the premise to  disentangle the origin of the  observed anomalies. 

The present study is devoted to the  inclusive $b \to c,u$ semileptonic modes of $b$-flavoured baryons, in particular $\Lambda_b\to  X_{c,u} \ell^- {\bar \nu}_\ell$.
The formalism is developed for a generic baryon, therefore  it can also be applied to $\Xi_b$ and $\Omega_b$.   In our study   the heavy quark mass expansion  is considered at  ${\cal O}(1/m_Q^3)$, and the parametrization of the  baryon matrix elements relevant at this order is provided. Moreover, the case of polarized baryon decays is considered at this order,  the unpolarized case being recovered  averaging over  the initial baryon polarizations.  The semileptonic  transitions are analyzed in the Standard Model  and in an extension of the SM effective weak Hamiltonian  comprising  vector, scalar, pseudoscalar, tensor and axial  operators. Such  Hamiltonian densities have been  scrutinized in connection with the flavour anomaly problem, considering  $B$ meson exclusive modes, see, e.g.,  \cite{Biancofiore:2013ki,Colangelo:2018cnj,Bhattacharya:2018kig},  but less information is available  about their impact on inclusive observables \cite{Colangelo:2016ymy,Mannel:2017jfk,Kamali:2018fhr,Kamali:2018bdp}.

Let us  briefly remind the status of the above mentioned flavour anomaly.  A  small excess in  the ratios $\dd R(D^{(*)})=\frac{{\cal B}(B\to D^{(*)} \tau^- {\bar \nu}_\tau)}{{\cal B}(B \to D^{(*)} \ell^- {\bar \nu}_\ell)}$ ($\ell=e,\mu$) with respect to the SM expectations emerges after the  BABAR \cite{Lees:2012xj,Lees:2013uzd}, Belle \cite{Huschle:2015rga,Sato:2016svk,Hirose:2016wfn,Hirose:2017dxl} and LHCb  
\cite{Aaij:2015yra,Aaij:2017deq,Aaij:2017uff} measurements are combined. The  tension with SM is presently estimated at  $3.1 \, \sigma$  level   \cite{Fajfer:2012vx,Bigi:2017jbd,Bernlochner:2017jka, Jaiswal:2020wer}. Several interpretations  attribute the deviation to the effect of non-SM interactions mainly affecting the  third generation.  New  lepton flavour universality violating interactions could produce 
at low energies additional operators in the $b\to c \tau^- {\bar \nu}_\tau$  effective weak Hamiltonian, which can be scrutinized  using  global quantities, namely the decay branching fractions, and also, more efficiently,  using observables as
the 4d   $\bar B \to D^* (D \pi, D \gamma) \ell^- \bar \nu_\ell$ decay distributions for the three lepton species \cite{Alonso:2016gym,Ligeti:2016npd,Alok:2016qyh,Ivanov:2017mrj,Colangelo:2018cnj,Becirevic:2019tpx,Alguero:2020ukk}. This kind of analyses are also  possible for  $B_s$ modes
\cite{Aaij:2020xjy}.

For  $\Lambda_b$  the decay rates and  the angular distributions can be considered, although the latter measurements are experimentally challenging.  Moreover,
the systematic study of New Physics (NP) effects for polarized and unpolarized  $\Lambda_b$  would provide a wealth of  information. 
The prime purpose  of such investigations is to identify  the correlations between different processes induced by the same short-distance transition.  If the B anomalies are  due to new interactions,  correlated pattern of same size effects must be observed in B and $\Lambda_b$ decays:  we aim at describing such patterns, starting from the same extended low-energy Hamiltonian scrutinized  for $B$ and choosing the same benchmark points studied in that case, so that the size of the possible deviations from SM in the meson and baryon case can be compared. We shall provide the formulae for various observables, they can be used in experimental simulations to assess the sensitivity to the NP operators.

We have to say that measurements of the $\Lambda_b$ polarization at LHC give results compatible with zero \cite{Aaij:2013oxa,Aad:2014iba,Sirunyan:2018bfd,Aaij:2020vzk}, which means that the $b$ quarks hadronizing in $\Lambda_b$ are mainly produced by QCD processes. However, a sizable longitudinal $\Lambda_b$ polarization is expected for $b$ quarks produced in $Z$ and top quark decays, as shown by the measurements at LEP \cite{BUSKULIC1996437,Abbiendi:1998uz,2000205}. For this reason,   the effects beyond the Standard Model (BSM) in the polarized case have been scrutinized for the exclusive $\Lambda_b\to \Lambda_c \ell^- {\bar \nu}_\ell$ modes \cite{Bernlochner:2018bfn,Boer:2019zmp,Penalva:2019rgt,Ferrillo:2019owd},  in addition to the case of unpolarized baryon   \cite{Dutta:2015ueb,Shivashankara:2015cta,Li:2016pdv,Datta:2017aue,DiSalvo:2018ngq,Ray:2018hrx}.

The plan of our study is as follows. In Sec.~\ref{hamiltonian} we  introduce the semileptonic $b \to c,u$ effective Hamiltonian, which generalizes the SM one  by the inclusion of  the set of $D=6$   four-fermion semileptonic operators weighted by  complex  coefficients. The heavy quark expansion (HQE) to describe the inclusive process 
$ H_b(p,s) \to X_{c,u}\ell^- {\bar \nu_\ell}$ is discussed in  Sec.~\ref{OPE} considering the terms up to ${\cal O}(1/m_b^3)$. In
  Sec.~\ref{distributions}  we construct  the fully differential  $\Lambda_b \to  X_{c,u} \ell^- {\bar \nu}_\ell$ decay distributions in the case of polarized $\Lambda_b$.
In Sec.~\ref{results} we analyze several observables in SM and with the extended Hamiltonian density, at a  benchmark point in the parameter space of the effective couplings  to investigate the sensitivity  to  the  new operators.   The last Section contains  the conclusions. 
 
The appendices contain the ingredients developed in our analysis.
In  Appendix \ref{appA} we collect the  baryon matrix elements relevant for the OPE at order $\dd 1/m_b^3$, considering the spin of the baryon.
In  Appendix \ref{appB}  we write the expressions of the structure functions for the hadronic tensor in SM and in the case of the extended  Hamiltonian.
Appendix \ref{appC} contains the coefficients appearing in the $1/m_b$ expansion the full semileptonic decay widths, for the SM and for the generalized Hamiltonian.
\section{Effective Weak Hamiltonian}\label{hamiltonian}
We consider the  inclusive semileptonic decay  of a baryon $H_b$ comprising a single $b$ quark
\be
H_b(p,s) \to X_{c,u}(p_X) \ell^-(p_\ell) {\bar \nu_\ell}(p_\nu) \,\,\, , \label{inclusive-decay}
\ee
with $s$  the spin of the decaying baryon. We assume that the process  is induced by the  low-energy effective Hamiltonian density which extends the SM one,
\bea
H_{\rm eff}^{b \to U \ell \nu}= {G_F \over \sqrt{2}} V_{Ub} &\Big[&(1+\epsilon_V^\ell) \left({\bar U} \gamma_\mu (1-\gamma_5) b \right)\left( {\bar \ell} \gamma^\mu (1-\gamma_5) {\nu}_\ell \right)
\nn \\
&+& \epsilon_S^\ell \, ({\bar U} b) \left( {\bar \ell} (1-\gamma_5) { \nu}_\ell \right)
+ \epsilon_P^\ell \, \left({\bar U} \gamma_5 b\right)  \left({\bar \ell} (1-\gamma_5) { \nu}_\ell \right)\label{hamil} \\
&+& \epsilon_T^\ell \, \left({\bar U} \sigma_{\mu \nu} (1-\gamma_5) b\right) \,\left( {\bar \ell} \sigma^{\mu \nu} (1-\gamma_5) { \nu}_\ell \right)  \nn \\
&+&  \epsilon_R^\ell \left({\bar U} \gamma_\mu (1+\gamma_5) b \right)\left( {\bar \ell} \gamma^\mu (1-\gamma_5) {\nu}_\ell \right)   \Big] + h.c.\,\,\, .  \nn
\eea
$H_{\rm eff}$ consists of  D=6 four-fermion operators with complex  lepton-flavour dependent  coefficients  $\epsilon^\ell_{V,S,P,T,R}$.  Only  left-handed neutrinos are considered. $U$ can be either  the $u$ or the $c$ quark,  $V_{Ub}$ is the corresponding Cabibbo-Kobayashi-Maskawa (CKM) matrix element.  $V_{Ub}$ and $\epsilon_V^\ell$ are independent parameters:  the product  $V_{Ub} (1+\epsilon_V^\ell)$ is not  a mere redefinition of the SM $V_{Ub}$, due to the lepton-flavour dependence of  $\epsilon_V^\ell$.  

 A comment is in order,  concerning the operator with  right-handed (RH) quark vector current $O_R=({\bar U} \gamma_\mu (1+\gamma_5) b) ( {\bar \ell} \gamma^\mu (1-\gamma_5) {\nu})$.
This operator is in the set of $D=6$ operators constituting the effective Hamiltonian \eqref{hamil}, and 
 has been previously considered   \cite{Voloshin:1997zi,Dassinger:2007pj,Dassinger:2008as,Crivellin:2009sd,Buras:2010pz,Feger:2010qc,Bernlochner:2014ova}. We shall give analytic formulae that comprise  its contribution,  providing  general expressions for the $\Lambda_b$ fully differential decay distributions and for the integrated distributions. 
 However, in the Standard Model Effective Field Theory  the only D=6 operator with a RH quark current, invariant under the SM gauge group, is nonlinear in the Higgs field: $i\left({\bar U}_R \gamma_\mu  b_R \right) ( \tilde H^\dagger D_\mu H)$, with $D_\mu$ the electroweak (ew) covariant derivative,
$H$ the SU(2) Higgs doublet,  $\tilde H^i=\epsilon^{ij} H^{j*}$, and $\epsilon^{ij}$ the totally antisymmetric tensor
\cite{Buchmuller:1985jz,Cirigliano:2009wk,Aebischer:2020lsx}.  At the ew symmetry breaking scale this  operator modifies the $WUb$ coupling, but   the resulting low energy four-fermion $O_R$ operator in the effective  $b \to c, u$ semileptonic Hamiltonian  does not violate LFU.  Hence,  in the  framework of this effective field theory  it is not involved in $B$ flavour anomalies,  and it has been omitted in several analyses  \cite{Alonso:2015sja,Shi:2019gxi,Feruglio:2018fxo}. Modifications of the $WUb$ vertex are connected to modifications of the quark $Z$ vertices, which are tightly constrained by the electroweak precision observables \cite{Aebischer:2018iyb,Aebischer:2020lsx}. Stringent bounds can also be obtained from different processes \cite{Bernard:2006gy,Alioli:2017ces}. For these reasons  we shall not include  $ O_R$  in our phenomenological analysis, since this would require a dedicated study beyond  the purposes of the present work.

The Hamiltonian  \eqref{hamil} can be written as
\be
H_{\rm eff}^{b \to U \ell \nu}= {G_F \over \sqrt{2}} V_{Ub} \sum_{i=1}^5 C_i^\ell \, J^{(i)}_M\, L^{(i)M} + h.c.\,\,\, ,\label{hnew}\ee
with  $C_1^\ell=(1 +\epsilon_V^\ell)$ and  $C_{2,3,4,5}^\ell=\epsilon_{S,P,T,R}^\ell$.   $J_M^{(i)}$ indicates  the hadronic and $L^{(i)M}$ the leptonic current in each operator,  $M$ generically denotes the set of Lorentz indices contracted between $J$ and $L$. The SM Hamiltonian corresponds to $i=1$ and $\epsilon^\ell_{V,S,P,T,R}=0$. We keep  the mass of the charged lepton $\ell=e,\mu,\tau$ different from zero.

\section{Inclusive decay width}\label{OPE}
The  decay width of the processes  \eqref{inclusive-decay}  is given by

\be
d\Gamma= d\Sigma \, \frac{G_F^2 |V_{Ub}|^2}{4m_H} \sum_{i,j} C_i^* C_j (W^{ij})_{MN} (L^{ij})^{MN} ,
\ee
where $G_F$ is the Fermi constant, $q=p_\ell+p_\nu$  the lepton-pair momentum, and   $d\Sigma$  the phase space element  $d\Sigma=(2\pi) d^4q \, \delta^4(q-p_\ell-p_\nu) [dp_\ell]\,[dp_\nu] $, with   $[dp]=\displaystyle\frac{d^3 p}{(2\pi)^3 2p^0}$.  The leptonic tensor is $(L^{ij})^{MN}= L^{(i)\dagger M} L^{(j)N}$.
The hadronic tensor $(W^{ij})_{MN}$ is obtained from
 the discontinuity of the forward  amplitude
\bea
(T^{ij})_{MN}&=&i\,\int d^4x \, e^{-i\,q \cdot x} \langle H_b(p,s)|T[ J^{(i)\dagger}_M (x) \,J^{(j)}_N (0)] |H_b(p,s) \rangle\,\,\label{Tij-gen}
\eea
 across the cut corresponding to the process \eqref{inclusive-decay}:
\be
(W^{ij})_{MN}=\frac{1}{\pi}{\rm Im}(T^{ij})_{MN} .
\ee 
$T^{ij}$ and $W^{ij}$ can be computed  exploiting an operator product expansion (OPE)  with  expansion parameter the inverse $b$ quark mass  \cite{Chay:1990da,Bigi:1993fe}.
To construct the OPE, the hadron momentum $p=m_H v$, with  four-velocity $v$, is expressed in terms of the heavy quark mass $m_b$ and of a residual momentum $k$: $p=m_b v+k$. The QCD $b$ quark field is written as  $b(x)=e^{-i\,m_b v \cdot x} b_v(x)$, with  $b_v(x)$ still defined in QCD and  satisfying the equation of motion:
\be
b_v(x)=\left(P_+ +\frac{i {\spur D}}{2m_b} \right) b_v(x) \,\,,
\ee
where $P_+$ is the velocity projector   $P_+=\displaystyle\frac{1+ \spur v}{2}$. 
In terms of $b_v(x)$ one has:
\be
(T^{ij})_{MN}=i\,\int d^4x \, e^{i\,(m_b v -q) \cdot x} \langle H_b(v,s)|T[ {\hat J}^{(i)\dagger}_M (x) \, \hat J^{(j)}_N (0)] |H_b(v,s)\rangle\,\,
\ee
with ${\hat J}^{(i)}$ containing the field $b_v$. 
The heavy quark expansion is obtained from 
\be
(T^{ij})_{MN}=\langle H_b(v,s)|{\bar b}_v(0) \Gamma_M^{(i)\dagger} S_U(p_X) \Gamma_N^{(j)}b_v(0) |H_b(v,s)\rangle\,\, ,
\ee
with $p_X=m_bv+k-q$ and $S_U(p_X)$  the $U$ quark propagator.  Replacing $k \to iD$, with $D$  the QCD covariant derivative, the $U$ quark propagator  can be expanded:
\be
S_U(p_X)=S_U^{(0)}-S_U^{(0)}(i {\spur D})S_U^{(0)}+S_U^{(0)}(i {\spur D})S_U^{(0)}(i {\spur D})S_U^{(0)}+ \dots\,\, \label{exp-prop}
\ee
where $S_U^{(0)}=\displaystyle\frac{1}{m_b {\spur v}-{\spur q} -m_U}$.
With the definitions $p_U=m_b v -q$, ${\cal P}=({\spur p}_U+m_U)$ and $\Delta_0=p_U^2-m_U^2$,  the expansion  at order $1/m_b^3$ is
\bea
&&\frac{1}{\pi}{\rm Im}(T^{ij})_{MN}= \nn \\
&& \frac{1}{\pi}{\rm Im}\frac{1}{\Delta_0}\langle H_b(v,s)|{\bar b}_v [\Gamma_M^{(i)\dagger} {\cal P} \Gamma_N^{(j)}]b_v |H_b(v,s) \rangle +\nn \\
&-&\frac{1}{\pi}{\rm Im}\frac{1}{\Delta_0^2}\langle H_b(v,s)|{\bar b}_v[ \Gamma_M^{(i)\dagger} {\cal P}\gamma^{\mu_1}{\cal P} \Gamma_N^{(j)}](i D_{\mu_1})b_v |H_b(v,s)\rangle +\nn \\
&+&\frac{1}{\pi}{\rm Im}\frac{1}{\Delta_0^3}\langle H_b(v,s)|{\bar b}_v[ \Gamma_M^{(i)\dagger} {\cal P}\gamma^{\mu_1}
{\cal P}\gamma^{\mu_2}{\cal P} \Gamma_N^{(j)}](i D_{\mu_1})(i D_{\mu_2})b_v |H_b(v,s)\rangle +\label{expansion}
\\
&-&\frac{1}{\pi}{\rm Im}\frac{1}{\Delta_0^4} \langle H_b(v,s)|{\bar b}_v[ \Gamma_M^{(i)\dagger} {\cal P}\gamma^{\mu_1}
{\cal P}\gamma^{\mu_2}{\cal P} \gamma^{\mu_3}{\cal P}\Gamma_N^{(j)}](i D_{\mu_1})(i D_{\mu_2})(i D_{\mu_3})b_v |H_b(v,s) \rangle \,\,.\nn 
\eea
This expression involves $H_b$ matrix elements of QCD operators of increasing dimensions,  written as
\bea
&&
\langle H_b(v,s)|{\bar b}_v[ \Gamma_M^{(i)\dagger} {\cal P}\gamma^{\mu_1}\dots  \gamma^{\mu_n}{\cal P}\Gamma_N^{(j)}](i D_{\mu_1})\dots(i D_{\mu_n})b_v |H_b(v,s)\rangle =
\label{trace-form} \\
&&={\rm Tr} \left[(\Gamma_M^{(i)\dagger} {\cal P}\gamma^{\mu_1}\dots  \gamma^{\mu_n}{\cal P}\Gamma_N^{(j)})_{ba} \langle H_b(v,s)|({\bar b}_v)_a(i D_{\mu_1})\dots(i D_{\mu_n})(b_v)_b |H_b(v,s)\rangle \right] \,  \nn 
\eea
with $a,b$  Dirac indices. 
The hadron matrix elements 
\be
{\cal M}_{\mu_1 \dots \mu_n}=\langle H_b(v,s)|({\bar b}_v)_a(i D_{\mu_1})\dots(i D_{\mu_n})(b_v)_b |H_b(v,s)\rangle \label{matel}
\ee  
can be expressed  in terms of  nonperturbative parameters, the number of which increases with the dimension of the  operators.
The expansion to order ${\cal O}(1/m_b^3)$ requires   
\bea
\langle H_b(v,s)|{\bar b}_v (iD)^2 b_v|H_b(v,s)\rangle&=&-2m_H\,{\hat \mu}_\pi^2 \label{mupi}\\
\langle H_b(v,s)|{\bar b}_v (iD_\mu)(iD_\nu)(-i \sigma^{\mu \nu})b_v|H_b(v,s)\rangle&=&2m_H\,{\hat \mu}_G^2 \label{mug} 
\\
\langle H_b(v,s)|{\bar b}_v (iD_\mu)(i v \cdot D) (i D^\mu) b_v|H_b(v,s)\rangle&=&2m_H\,{\hat \rho}_D^3\label{rd} \\
\langle H_b(v,s)|{\bar b}_v (iD_\mu)(i v \cdot D) (i D_\nu) (-i \sigma^{\mu \nu})b_v|H_b(v,s)\rangle &=&2m_H\,{\hat \rho}_{LS}^3 \,\, .\label{rls}
 \eea
 
A method to compute  ${\cal M}_{\mu_1 \dots \mu_n}$  is exploited in \cite{Dassinger:2006md} for $B$ meson, and   more parameters  than those listed in (\ref{mupi}-\ref{rls}) are needed for $n=4$. The order $n=5$ has also been analyzed \cite{Mannel:2010wj}. For a heavy baryon,  the dependence on the spin four-vector $s_\mu$ must be kept in \eqref{matel}. This is important since,  for hadrons with spin, considering the hadron polarization leads to interesting observables to analyze. 

In  Appendix \ref{appA} we collect the expressions of the matrix elements needed for the expansion at ${\cal O}(1/m_b^3)$ keeping the  $s_\mu$ dependence. The computation procedure  is  described in  \cite{Dassinger:2006md}. One starts from the highest dimension operator,  which in our  case $n=3$  has  dimension 6, and determines it in the static heavy quark limit, replacing $b_v(x) \to h_v(x)$, the heavy quark field defined in the heavy quark effective theory (HQET). $h_v(x)$ is related to the  QCD $b(x)$  field: $h_v(x)=e^{i m_b v \cdot x}P_+ b(x)$. $h_v$ satisfies the equations $P_+h_v(x)=h_v(x)$ and $v \cdot D \,h_v(x)=0$.
In principle, the matrix element 
${\cal M}_{\mu_1 \mu_2 \mu_3}=\langle H_b(v,s)|({\bar h}_v)_a(i D_{\mu_1})(i D_{\mu_2})(i D_{\mu_3})(h_v)_b |H_b(v,s)\rangle$    
can be expanded over the set of 16 independent Dirac matrices. However,  in HQET it is given in terms of only  two Dirac structures,  $P_+$ and ${\hat S}^\mu=P_+\gamma^\mu  \gamma_5 P_+$, an observation which simplifies the parametrization  \cite{Mannel:1994kv}.  On the other hand, the  matrix elements of  lower dimension operators are  computed in QCD  expanding over the full set of Dirac matrices. The coefficients of  Dirac structures in the  $d$ dimension matrix element are recursively computed from the  $d+1$ terms, and   Eqs.(\ref{mupi})-(\ref{rls}) are used.

The parameters in Eqs.(\ref{mupi})-(\ref{rls}) are denoted by a hat   to distinguish them  from the  corresponding parameters  defined in HQET, with $b_v$ replaced by $h_v$.  For ${\hat \mu}_\pi^2,\,{\hat \rho}_D^3,\,{\hat \rho}_{LS}^3$ the difference between the two definitions involves terms appearing at ${\cal O}(1/m_b^4)$, hence  in our case ${\hat \mu}_\pi^2=\mu_\pi^2$, ${\hat \rho}_D^3=\rho_D^3$ and ${\hat \rho}_{LS}^3=\rho_{LS}^3$.
For ${\hat \mu}_G^2$ the relation between the two definitions  is ${\hat \mu}_G^2=\mu_G^2-\displaystyle\frac{1}{m_b}\left(\rho_D^3+\rho_{LS}^3 \right)$, a combination  often present in our expressions.
 
 The formalism is suitable for the analysis in the Standard Model and in NP  with  the Hamiltonian \eqref{hamil}.
Our results are obtained for non-vanishing charged lepton mass, at order $1/m_b^3$ in the HQE,  in the case of a polarized baryon and with all  operators in (\ref{hamil}) taken into account. In the existing literature one or more  of the above points are relaxed. 
For the inclusive semileptonic $B$ decays and non vanishing lepton masses,  NP operators  have been considered at order $1/m_b^2$
 in \cite{Kamali:2018bdp}, and we agree with those results at that order after taking the spin-average in our expressions.
 $V+A$ and $S-P$ operators have been studied at the leading order in the $1/m_b$ expansion in  \cite{Mannel:2017jfk}, while a  $V+A$ operator  has been  considered for the mode $B \to X \tau {\bar \nu}_\tau$ performing the HQE at ${\cal O}(1/m_b^{2})$ in \cite{Grossman:1994ax}.
The hadronic tensor has been  computed by an OPE in terms of operators
comprising  the HQET field $h_v$  in \cite{Manohar:1993qn}. In this analysis the polarized $\Lambda_b$  inclusive semileptonic decay  is considered  at order $1/m_b^2$ in SM for massless leptons, the case of massive leptons at the same order in $1/m_b$ is studied in \cite{Balk:1997fg}.  We agree with such results at that order in the $b$ mass expansion.
 As a last remark,  in  $b \to u$ semileptonic transition
 we neglect weak annihilation contributions, which mainly affect the endpoint region of the charged lepton energy spectrum \cite{Bigi:1993bh}. 
 
Using the matrix elements ${\cal M}_{\mu_1 \dots \mu_n}$ collected  in Appendix \ref{appA} the hadronic tensor  can be computed. It is  expanded  in Lorentz structures depending on $v$, $q$ and $s$. The related invariant  functions are given in Appendix \ref{appB}  for   the Standard Model and for  the effective Hamiltonian Eq.~\eqref{hamil}. 

\section{Decay distributions} \label{distributions}
For the  $H_b(v,s) \to X(p_X) \ell^-(p_\ell) {\bar \nu}_\ell(p_\nu)$ transition the four-fold differential decay distribution  is given by
\be
\frac{d^4 \Gamma}{dq^2 \,d(v \cdot q) \,dE_\ell  \,d \cos \theta_P}=\frac{G_F^2 |V_{Ub}|^2}{32 (2 \pi)^3 m_H} \sum_{i,j}C_i^* C_j \frac{1}{\pi}{\rm Im}(T^{ij})_{MN}(L^{ij})^{MN} \,\, ,\label{full}
\ee
with  $p_\ell=(E_\ell,{\vec p}_\ell)$ and $\theta_P$  the angle between ${\vec p}_\ell$ and ${\vec s}$  in the $H_b$ rest frame.
The structure functions in which the hadronic tensor is expanded depend on $q^2$ and $v \cdot q$.
The various  decay distributions are obtained  integrating (\ref{full}) over the phase space \cite{Jezabek:1996ia}.
To compute the spectrum in $q^2$ and in the charged lepton energy $E_\ell$   the order in the integration  must be specified. 
Integrating first  in $E_\ell$,  the  integration limits are 
\bea
E_1^*\le E_\ell \le E_2^*  \,\, ,\hskip 1cm E_{1,2}^*=\frac{v \cdot q (q^2+m_\ell^2) \pm \sqrt{(v \cdot q)^2 -q^2} \,(q^2-m_\ell^2)}{2 q^2}\,.
\label{elfirst}
\eea
 The replacement
\be
\frac{1}{\pi}{\rm Im}\frac{1}{\Delta_0^n} \to \frac{(-1)^{n-1}}{(n-1)!}\delta^{(n-1)}(\Delta_0) \label{delta}
\ee
in the hadronic tensor can be used to integrate over $v \cdot q$.  The last   $q^2$ integration is  for
\be
m_\ell^2 \le q^2 \le (m_b -m_U)^2 \,\,.
\ee
To compute the charged lepton energy spectrum one integrates in a different order \cite{Gremm:1995rv}. 
The first integration over $v \cdot q$ is in the range  
\be 
E_\ell  + \frac{(q^2 - m_\ell^2)}{2 m_\ell^2} E_{\ell \,-} \le v \cdot q \le E_\ell  + \frac{(q^2 - m_\ell^2)}{2 m_\ell^2} E_{\ell \,+} \,\,\, ,
\ee
where $E_{\ell \pm}=E_\ell \pm \sqrt{E_\ell^2-m_\ell^2}$. 
Then one integrates over $q^2$ with  integration limits
\be
\frac{E_{\ell \,-}}{m_b-E_{\ell \,-}} \left(m_b^2-m_U^2-m_b E_{\ell \,-} \right) \le q^2 \le \frac{E_{\ell \,+}}{m_b-E_{\ell \,+}} \left(m_b^2-m_U^2-m_b E_{\ell \,+} \right) \,.\label{q2first}
\ee
The range for the last integration in $E_\ell$ is
\be
m_\ell \le E_\ell \le \frac{m_b^2-m_U^2+m_\ell^2}{2 m_b} \,.
\ee

Keeping the dependence on $\cos \theta_P$,  the corresponding double decay distributions are obtained.
Notice that the kinematics involves the quark masses,   the dependence on the decaying hadron is contained in the matrix elements of the OPE operators. However,  the OPE breaks down in the endpoint region of the spectra, as signaled by  singularities  as the derivatives of the $\delta$ function. Such terms must be resummed in a $H_b$   shape function. Convolution of the distributions with such a function  smears the spectra at the endpoint  and transforms the phase space boundaries from the partonic  to the hadronic kinematics:
 $q^2_{max}=(m_{H_b}-m_X)^2$ and $\dd (E_\ell)_{max}=\frac{m_{H_b}^2-m_X^2+m_\ell^2}{2 m_{H_b}}$, with $m_X$ the mass of  the lightest hadron containing the  $U$ quark produced in the  decay.
We do not include the effects of the shape function, the profile  of which is not known in the baryon case, keeping in mind that the OPE results loose reliability in  the endpoint region.

Expanding the tensor $T^{ij}$ in invariant functions, as provided in Appendix \ref{appB},  the fully differential distribution is obtained upon contraction with the leptonic tensor.
We express the distribution as 
\be
\frac{d^4 \Gamma}{dq^2 \,d(v \cdot q) \,dE_\ell  \,d \cos \theta_P}=\sum_{i,j}
\frac{d^4 \Gamma^{ij}}{dq^2 \,d(v \cdot q) \,dE_\ell  \,d \cos \theta_P} \,. \label{d4G}
\ee
In this expression the first term is
\bea
&&\frac{d^4 \Gamma^{11}}{dq^2 \,d(v \cdot q) \,dE_\ell  \,d \cos \theta_P}={\cal N}|\left(1+\epsilon_V \right)|^2 \nn \\
&&
\Bigg\{8(q^2-m_\ell^2)W_1+4\left[-(q^2-m_\ell^2)+4E_\ell(v \cdot q - E_\ell) \right]W_2\nn \\
&&+8\left[(q^2+m_\ell^2)v \cdot q -2q^2 E_\ell) \right]W_3 
+4m_\ell^2(q^2-m_\ell^2)W_4+16m_\ell^2(v \cdot q-E_\ell)W_5 \nn \\
&&
-2 \frac{ \cos \theta_P}{\sqrt{E_\ell^2-m_\ell^2}} \left(q^2+m_\ell^2-2 (v \cdot q) E_\ell \right) \Bigg[2G_1(q^2-m_\ell^2)+G_2\big[-(q^2-m_\ell^2)\nn \\
&&+4 E_\ell(v \cdot q -E_\ell) \big]  +2 G_3\left[(q^2+m_\ell^2)v \cdot q -2q^2 E_\ell) \right]+4G_5m_\ell^2(v \cdot q -E_\ell) \nn \\
&&-4E_\ell G_6 - 4 m_\ell^2 G_7 -4E_\ell G_8-2(q^2+m_\ell^2)G_9\Bigg]
 \label{dG11} \\ 
&&-16  \cos \theta_P \sqrt{E_\ell^2-m_\ell^2}\Big[ (v \cdot q - 2 E_\ell) G_6 -m_\ell^2G_7-v \cdot q\,G_8-q^2\,G_9 \Big] \Bigg\} \nn
\eea
where ${\cal N}=\displaystyle\frac{G_F^2 |V_{Ub}|^2}{32 (2 \pi)^3 m_H}$,
 $W_a=\displaystyle\frac{1}{\pi}{\rm Im}T_a$ and $G_a=\displaystyle\frac{1}{\pi}{\rm Im}S_a$ with  the index $a=1,2,\dots$ corresponding to the invariant functions $T_{1-5}$ and $S_{1-9}$ in \eqref{T1}-\eqref{T5} and \eqref{S1}-\eqref{S9}.
 For $\epsilon_V^\ell=0$ this term corresponds to the SM distribution. 

Let us consider   the other terms in Eq.~\eqref{d4G} for the NP contributions.
Considering the scalar and pseudoscalar operators, we have:
\bea
\frac{d^4 \Gamma^{22(33)}}{dq^2 \,d(v \cdot q) \,dE_\ell  \,d \cos \theta_P}&=&{\cal N}|\epsilon_{S(P)}|^2 4 (q^2-m_\ell^2) W_{S(P),1} \\ 
\nn \\
\frac{d^4 \Gamma^{23+32}}{dq^2 \,d(v \cdot q) \,dE_\ell  \,d \cos \theta_P}&=&{\cal N}
\left(-2{\rm Re}[\epsilon_S \epsilon_P^*]\right)
 \frac{2 \cos \theta_P}{\sqrt{E_\ell^2-m_\ell^2}}(q^2-m_\ell^2)  (m_\ell^2 + q^2 - 2 v \cdot q E_\ell) G_{SP,1} \nn \\
\eea
with $W_{S(P),1}$ and $G_{SP,1}$ obtained from the imaginary parts of the  functions $T$ and $S$  in \eqref{TS}-\eqref{TP}.

From the interference terms, we have:
\bea
&&\frac{d^4 \Gamma^{12+21(13+31)}}{dq^2 \,d(v \cdot q) \,dE_\ell  \,d \cos \theta_P}={\cal N}\,
2 \, {\rm Re}[(1+\epsilon_V)\epsilon_{S(P)}^*] m_\ell  \nn \\ 
&\Bigg\{4&\Big[2(v \cdot q-E_\ell)W_{SMS(SMP),1}+(q^2-m_\ell^2)W_{SMS(SMP),2} \Big] \nn \label{dG12}\\
&-& \frac{2 \cos \theta_P}{\sqrt{E_\ell^2-m_\ell^2}}\left(q^2+m_\ell^2-2 (v \cdot q) E_\ell \right)  
\Big[2  (v \cdot q -E_\ell)G_{SMS(SMP),1} \\
&+&(q^2-m_\ell^2)G_{SMS(SMP),2} -2G_{SMS(SMP),3}\Big] 
+8\cos \theta_P \sqrt{E_\ell^2-m_\ell^2} G_{SMS(SMP),3} \Bigg\}\nn  
\eea
with $W_{SMS(SMP),i}$ and $G_{SMS(SMP),i}$ obtained  from the imaginary parts of the  functions $T$ and $S$  in \eqref{TSMS1},\eqref{TSMS2} and  \eqref{SMS1}-\eqref{SMS4}.

Continuing with the distributions, we have:

\bea
&&\frac{d^4 \Gamma^{44}}{dq^2 \,d(v \cdot q) \,dE_\ell  \,d \cos \theta_P}={\cal N} |\epsilon_T|^2\nn \\ 
&& \Bigg\{16(q^2-m_\ell^2)(q^2+2m_\ell^2) \left(W_{T4}+W_{T9} \right)\nn \\
&&+16 \left[-(q^2-m_\ell^2)+8E_\ell(v \cdot q -E_\ell) \right] \left(W_{T2}+W_{T6}-W_{T10} \right) \nn \\
&&+16 \left[(q^2-m_\ell^2)v \cdot q+4m_\ell^2(v \cdot q -E_\ell) \right] \left(2W_{T5}+W_{T7}+W_{T8}-W_{T11}-W_{T12} \right) \nn \\
&&+16 \big[m_\ell^4+q^2(v \cdot q -2 E_\ell)^2-m_\ell^2\big(q^2+v \cdot q(-3v \cdot q+4 E_\ell) \big) \big] \left(W_{T14}-W_{T15} \right) \nn \\
&&-8  \frac{\cos \theta_P}{\sqrt{E_\ell^2-m_\ell^2}}\left( m_\ell^2+q^2-2E_\ell  \,v \cdot q \right)
  \Big[\left[-(q^2-m_\ell^2)+8E_\ell(v \cdot q -E_\ell) \right] \left(G_{T2}+G_{T6}\right) \nn \\
&&+ \left[(q^2-m_\ell^2)v \cdot q+4m_\ell^2(v \cdot q -E_\ell) \right] \left(2G_{T5}+G_{T7}+G_{T8}-G_{T11}-G_{T12} \right) \nn \\
&&+2\big[m_\ell^2(v \cdot q -2 E_\ell)+v \cdot q [q^2-4E_\ell(v \cdot q -E_\ell)] \big]G_{T22} \Big]  \\
&&-4E_\ell\left(2G_{T14}+G_{T23}+v \cdot q \,G_{T24}+G_{T30}+G_{T32}-G_{T34}-G_{T36} \right) \nn \\
&&
-(3m_\ell^2+q^2) \left(2G_{T15}+G_{T31}+G_{T33}-G_{T35}-G_{T37}+G_{T27A}+G_{T27B} \right) \nn \\
&&+2[m_\ell^2+E_\ell(v \cdot q -2E_\ell)]\left(G_{T27A}+G_{T27B}+G_{T28}-G_{T29} \right) \Big]
\nn \\
&&-32 \cos \theta_P\sqrt{E_\ell^2-m_\ell^2}\Big[2(v \cdot q-2 E_\ell)\left(2G_{T14}+G_{T23}+v \cdot q \,G_{T24}+G_{T30}+G_{T32}-G_{T34}-G_{T36} \right)
\nn \\
&& -2m_\ell^2 \left(2G_{T15}+G_{T31}+G_{T33}-G_{T35}-G_{T37}+G_{T27A}+G_{T27B} \right) \nn \\
&&
+\big[ m_\ell^2 + v \cdot q (v \cdot q - 2 E_\ell)\big]\left(G_{T27A}+G_{T27B}+G_{T28}-G_{T29} \right)\Big] \Bigg\} \nn
\eea
with $W_{Ti}$ and $G_{Ti}$  from the imaginary parts of the functions $T$ and $S$  in \eqref{ST0}-\eqref{ST34};
\bea
&&
\frac{d^4 \Gamma^{14+41}}{dq^2 \,d(v \cdot q) \,dE_\ell  \,d \cos \theta_P}={\cal N} 
2 {\rm Re}[(1+\epsilon_V)\epsilon_T^*] m_\ell
\nn \\ &&
\Bigg\{16(v \cdot q -E_\ell) \big[-3W_{SMT,1}+3W_{SMT,3}-(v \cdot q)(W_{SMT,5}+W_{SMT,7} )\big]\nn \\
&&+8(q^2-m_\ell^2)\big(-3W_{SMT,2}+3W_{SMT,4}+W_{SMT,5}+W_{SMT,7} \big)\nn \\
&&+8 m_\ell \big[2 q^2 E_\ell-(m_\ell^2+q^2) \big] \big(W_{SMT,6}+W_{SMT,8}\big) \nn \\
&&+8 \frac{\cos \theta_P}{\sqrt{E_\ell^2-m_\ell^2}}\left(q^2+m_\ell^2-2 (v \cdot q) E_\ell \right) \Big[(v \cdot q -E_\ell)\big(3G_{SMT,1}-3G_{SMT,3}-G_{SMT,11}+G_{SMT,25} \big) \nn \\
&& 
-3G_{SMT,9}+3G_{SMT,10}-G_{SMT,12}+G_{SMT,16}-v \cdot q \,\left(G_{SMT,13}-G_{SMT,17} \right) \label{dG14}\\
&&+E_\ell \left(-G_{SMT,14}+G_{SMT,18} \right)\Big]
-16  \cos \theta_P \sqrt{E_\ell^2-m_\ell^2}  \Big[3G_{SMT,9}-3G_{SMT,10}\nn \\
&&+G_{SMT,12}-G_{SMT,16}
+v \cdot q \,\left(G_{SMT,13}+G_{SMT,14} -G_{SMT,17}-G_{SMT,18} \right)\Big]\Bigg\} \nn
\eea
with $W_{SMT,i}$ and $G_{SMT,i}$  from the imaginary parts of the  functions $T$ and $S$  in \eqref{TSMT1}-\eqref{SSMTall};
\bea
&&
\frac{d^4 \Gamma^{24+42(34+43)}}{dq^2 \,d(v \cdot q) \,dE_\ell  \,d \cos \theta_P}={\cal N} 
2 {\rm Re}[\epsilon_T\epsilon_{S(P)}^*]
\nn \\ &&
\Bigg\{-8[(q^2+m_\ell^2)(v \cdot q)-2 q^2 E_\ell]\left(W_{ST(PT),1}+W_{ST(PT),2}\right)
\nn \\
&&+4\frac{\cos \theta_P }{\sqrt{E_\ell^2-m_\ell^2}} \left(q^2+m_\ell^2-2 (v \cdot q) E_\ell \right)\Bigg[[(q^2+m_\ell^2)(v \cdot q)-2 q^2 E_\ell]\left(G_{ST(PT),1}+G_{ST(PT),2}\right) \nn \\
&& +(q^2+m_\ell^2)\left(G_{ST(PT),3}-G_{ST(PT),4}\right)+2E_\ell\left(G_{ST(PT),5}+G_{ST(PT),6} \right) \Bigg]
\nn \\
&&+16\cos \theta_P \sqrt{E_\ell^2-m_\ell^2}
\Big[q^2\left(G_{ST(PT),3}-G_{ST(PT),4}\right)+ v \cdot q \left(G_{ST(PT),5}+G_{ST(PT),6} \right) \Big] \Bigg\} . \nn \\
\eea
In this last case  $W_{ST,i}$ and $G_{ST,i}$ are obtained  from the imaginary parts of the  functions  $T$ and $S$  in \eqref{TST1}-\eqref{STTall}.

The distributions related to the $O_R$ operator
\be
 \frac{d^4\Gamma^{55}}{dq^2 \,d(v \cdot q) \,dE_\ell  \,d \cos \theta_P} \quad \quad {\rm and } \quad \quad \frac{d^4 \Gamma^{15+51}}{dq^2 \,d(v \cdot q) \,dE_\ell  \,d \cos \theta_P}
\ee
have the same form  of Eq.~\eqref{dG11} with suitable substitutions:   $d^4\Gamma^{55}$ is obtained from Eq.~\eqref{dG11} replacing
$|\left(1+\epsilon_V \right)|^2 \to |\epsilon_R|^2$,   $W_a \to W_{Ra}=\displaystyle\frac{1}{\pi}{\rm Im}T_{Ra}$ and $G_a \to G_{Ra}=\displaystyle\frac{1}{\pi}{\rm Im}S_{Ra}$ with the functions $T_R$ and $S_R$ collected in \eqref{TSR}.
In the case of $d^4 \Gamma^{15+51}$ the replacements are: $|\left(1+\epsilon_V \right)|^2 \to 2{\rm Re}[(1+\epsilon_V)\epsilon_R^*] $,  $W_a \to W_{SMRa}=\displaystyle\frac{1}{\pi}{\rm Im}T_{SMRa}$ and $G_a \to G_{SMRa}=\displaystyle\frac{1}{\pi}{\rm Im}S_{SMRa}$, with the functions $T_{SMR}$ and $S_{SMR}$ collected in \eqref{TSMA1}-\eqref{SMRzero}.

Analogously, the distributions 
\be
 \frac{d^4\Gamma^{25+52(35+53)}}{dq^2 \,d(v \cdot q) \,dE_\ell  \,d \cos \theta_P} 
\ee
have the same form  of Eq.~\eqref{dG12} substituting   
 ${\rm Re}[(1+\epsilon_V)\epsilon_{S(P)}^*]  \to {\rm Re}[\epsilon_R\epsilon_{S(P)}^*]$,  $W_{SMS(SMP)a} \to W_{RS(P)a}=\displaystyle\frac{1}{\pi}{\rm Im}T_{RS(RP)a}$, and $G_{SMS(SMP)a} \to G_{RS(RP)a}=\displaystyle\frac{1}{\pi}{\rm Im}S_{RS(RP)a}$. The functions $T_{RS(RP)}$ and $S_{RS(RP)}$ are collected in \eqref{TRS}-\eqref{TRP}.

Finally, the distributions
\be
 \frac{d^4\Gamma^{45+54}}{dq^2 \,d(v \cdot q) \,dE_\ell  \,d \cos \theta_P} 
\ee
have the same form of Eq.~\eqref{dG14} with the substitutions:   
 ${\rm Re}[(1+\epsilon_V)\epsilon_T^*]  \to {\rm Re}[\epsilon_R\epsilon_T^*] $,  $W_{SMTa} \to W_{RTa}=\displaystyle\frac{1}{\pi}{\rm Im}T_{RTa}$ and $G_{SMTa} \to G_{RS(RP)a}=\displaystyle\frac{1}{\pi}{\rm Im}S_{RTa}$, and the functions $T_{RT}$ and $S_{RT}$ collected in \eqref{TAT1}-\eqref{SAT25}.

The above expressions can be used to compute all double and single  decay distributions. We do not present the lengthy formulae here, but only give  the full decay width, which can be  cast in the form:

\be
\Gamma(H_b \to X \ell^- {\bar \nu}_\ell)=\Gamma_b\sum_i  \left\{C_0^{(i)}+\frac{\mu^2_\pi}{m_b^2} C_{\mu_\pi^2}^{(i)}+\frac{\mu^2_G}{m_b^2} C_{\mu_G^2}^{(i)}+\frac{\rho_D^3}{m_b^3} C_{\rho_D^3}^{(i)}+\frac{\rho_{LS}^3}{m_b^3} C_{\rho_{LS}^3}^{(i)} \right\} , \label{fullwidth}
\ee
with $\Gamma_b=\displaystyle\frac{G_F^2 m_b^5 V_{Ub}^2}{192 \pi^3}$. The index $i$ indicates the contribution of the various operators and of the interferences:   $i=SM, S, P, T, R$,  $SP, SMS, SMP$, $SP, ST, SMT$, and  $SMR, SR, PR, TR$.
The   coefficients  $C^{(i)}$ are   collected  in Appendix \ref{appC} and contain  the couplings  $\epsilon^\ell_{V,S,P,T,R}$  in the effective Hamiltonian.  All $C_{\rho_{LS}^3}^{(i)}$ vanish.

 For the SM terms in Eq.~\eqref{fullwidth} various perturbative corrections are known.
The leading electroweak correction $A_{ew}=1.014$ is a multiplying factor.   QCD corrections are known at ${\cal O}(\alpha_s^2)$ for the leading term 
and at  ${\cal O}(\alpha_s)$ for the $1/m_b^2$ terms in   \eqref{fullwidth}, and can be included  following, e.g.,  
\cite{Czarnecki:1994bn,Jezabek:1996db,DeFazio:1999ptt,Trott:2004xc,Aquila:2005hq,Alberti:2014yda}. In ratios of decay widths involving different lepton species such corrections largely cancel out.
We do not include  QCD corrections in the decay distributions  analyzed in the following.

\section{Numerical results}\label{results}
In our numerical study we use the heavy quark masses in the kinetic scheme $m_b^{kin}(\mu=0.75 \, {\rm GeV})=4.62$ GeV, $m_c^{kin}(\mu=0.75 \, {\rm GeV})=1.20$ GeV, and the up quark mass in the ${\overline {\rm MS}}$ scheme  ${\overline m_u}(2 \,{\rm GeV})=2.16 \pm^{0.49}_{0.26}$ MeV  \cite{Zyla:2020}.
In the case of $B$ mesons the HQE parameters are constrained fitting  the measured  lepton energy and the hadronic mass  distributions  and their moments in  $B \to X_c \ell {\bar \nu}_\ell$ decay \cite{Alberti:2014yda}.
For $\Lambda_b$ only few theoretical estimates of $\mu_\pi^2(\Lambda_b)$ exist \cite{Colangelo:1995qp}.
A relation between $\mu_\pi^2(\Lambda_b)$ and $\mu_\pi^2(B)$ in terms of the measured mass differences between beauty and charmed mesons and baryons can be exploited  
\cite{Bigi:1995jr}:
 \be
\mu_\pi^2(B)-\mu_\pi^2(\Lambda_b)=\frac{2m_b m_c}{m_b-m_c}\left[(m_{\Lambda_b}-m_{\Lambda_c})-({\overline m}_B-{\overline m}_D)\right]\left(1+{\cal O}(1/m_{b,c}^2) \right)
\ee
 (${\overline m}_{B,D}$ is the spin-averaged $B^{(*)}$ and $D^{(*)}$   mass), 
to obtain $\mu_\pi^2(\Lambda_b)$ from  the value of $\mu_\pi^2(B)$. Moreover,
 the approximation $\rho_D^3(\Lambda_b) \simeq \rho_D^3(B)$ can be adopted, increasing  the uncertainty on $\rho_D^3(\Lambda_b)$  with respect to  the value for $B$.  
In our analysis we use:
$ \mu_\pi^2(\Lambda_b)=(0.50 \pm 0.10) \, {\rm GeV}^2$ and 
$\rho_D^3(\Lambda_b) = (0.17 \pm 0.08)\, {\rm GeV}^3$.
The HQE parameters $\mu_G^2$ and $\rho_{LS}^3$  are sensitive to  the total angular momentum of the light degrees of freedom in the hadron. 
For   $\Lambda_b$ they vanish since the  light degrees of freedom  have spin zero, but  for other baryons they are different from zero. For this reason we include the contributions involving such parameters in the various expressions in the Appendices, so that they can be used  for different heavy hadrons.

The description of  NP effects requires input values for the  couplings  $\epsilon^\ell_{V,S,P,T,R}$ in the Hamiltonian (\ref{hamil}).
 As anticipated, in the phenomenological analysis we do not consider the contribution of the operator $O_R$, hence we set $\epsilon_R^\ell=0$.
For $U=u$,  allowed regions   for the other couplings  have been  determined  from the analysis of purely leptonic $B$ decays and of semileptonic $B$ transitions to $\pi$ and $\rho(770)$   \cite{Colangelo:2019axi}. Accordingly, for  $b \to u\, \mu \, {\bar \nu}_\mu$ we set the benchmark point (BP): $({\rm Re}[\epsilon_V^\mu],\,{\rm Im}[\epsilon_V^\mu])=(0,\,0)$, 
$({\rm Re}[\epsilon_P^\mu],\,{\rm Im}[\epsilon_P^\mu])=(-0.03,\,-0.02)$, $({\rm Re}[\epsilon_T^\mu],\,{\rm Im}[\epsilon_T^\mu])=(0.12,\,0)$ and $({\rm Re}[\epsilon_S^\mu],\,{\rm Im}[\epsilon_S^\mu])=(-0.04,\,0)$. For $b \to u \, \tau \,{\bar \nu}_\tau$ the BP is:  $\epsilon_V^\tau=0$,  $\epsilon_S^\tau=0$,
$({\rm Re}[\epsilon_P^\tau],\,{\rm Im}[\epsilon_P^\tau])=(0.01,\,0)$ and $({\rm Re}[\epsilon_T^\tau],\,{\rm Im}[\epsilon_T^\tau])=(0.12,\,0)$.
For $U=c$ we discuss NP effects i) considering only the tensor  operator,  with $({\rm Re}( \epsilon_T^\mu),\,{\rm Im} (\epsilon_T^\mu))=(0.115,\,-0.005)$ and $({\rm Re}( \epsilon_T^\tau),\,{\rm Im}( \epsilon_T^\tau))=(-0.067,\, 0)$, as  fixed in \cite{Colangelo:2018cnj}. For one observable we also consider ii) non vanishing  couplings only for the $\tau$ mode, with $\rm Re [\epsilon_V^\tau]=0.16$, ${\rm Re}[\epsilon_S^\tau]=-0.235$, ${\rm Re}[\epsilon_P^\tau]=-0.095$ and ${\rm Re}[\epsilon_T^\tau]=0.05$ fixed in  \cite{Shi:2019gxi}; the values
iii) $\rm Re [\epsilon_V^\tau]=0.07$ and  iv) $\rm Re [\epsilon_S^\tau]=0.025$, \,  $\rm Re [\epsilon_P^\tau]=0.535$, also set in \cite{Shi:2019gxi}.

At odds with the $B$ case,  at present there is not enough experimental information on $\Lambda_b$ decays to restrict the ranges of the effective couplings.
For $b \to c$ modes, semileptonic $\Lambda_b $ transitions to $\Lambda_c^+$, $\Lambda_c^+\pi^+ \pi^-$, $\Lambda_c(2595)$, $\Lambda_c(2625)$, $\Sigma_c(2455)^0 \pi^+$ and
$\Sigma_c(2455)^{++} \pi^-$ have been observed.  The branching fractions  are  measured for the modes into  $\Lambda_c$ baryons, and  the result 
 ${\cal B}(\Lambda_b \to \Lambda_c \ell^- {\bar \nu}_\ell + {\rm anything} )=(10.9 \pm 2.2)\times 10^{-2}$ 
  is quoted,  with $\ell=e,\,\mu$ \cite{Zyla:2020}.
For $b \to u$,  the exclusive branching ratio  ${\cal B}(\Lambda_b \to p \mu^- {\bar \nu})=(4.1 \pm 1.0) \times 10^{-4} $ is measured   \cite{Zyla:2020}.

Using $|V_{cb}|=0.042$ and $|V_{ub}|=0.0037$, together with  $\tau_{\Lambda_b}=(1.471\pm 0.009) \times 10^{-12}$ s \cite{Zyla:2020}, we obtain  the inclusive $\Lambda_b$ branching fractions for the two quark transitions and for final $\tau$ and $\mu$ lepton. The results for the Standard Model, for the central value of the parameters and neglecting QCD corrections, are collected in Table \ref{tab:br}.
\begin{table}[t]
\centering
\renewcommand{\arraystretch}{1.3}
\begin{tabular}{|l|c|}
\hline
 & SM  \\
\hline  
${\cal B}(\Lambda_b \to X_c  \mu {\bar \nu}_\mu)$ \hfill &
$11.0 \times 10^{-2}$ \hfill \\
\hline
${\cal B}(\Lambda_b \to X_c  \tau {\bar \nu}_\tau)$ \hfill &
$\, 2.4 \times 10^{-2}$ \hfill\\
\hline
${\cal B}(\Lambda_b \to X_u  \mu {\bar \nu}_\mu)$ \hfill &
$11.65 \times 10^{-4}$ \hfill \\
\hline
${\cal B}(\Lambda_b \to X_u  \tau {\bar \nu}_\tau)$ \hfill &
$\,2.75 \times 10^{-4}$ \hfill\\
\hline
\end{tabular}
\caption{Inclusive semileptonic $\Lambda_b$ branching fractions in SM, obtained for the central values of the parameters. \label{tab:br}}
\end{table}

 A remark about the various sources of uncertainties is in order.  In a first-principle computation, such as the one we have described here,  the theoretical uncertainties are connected to the quark masses, to the hadronic parameters, to the perturbative corrections and to the size of next-order terms in the heavy quark expansion. All such uncertainties can be reduced in a systematic way. This is the case, in particular, of the values of the  hadronic parameters, the knowledge of which can be improved using  nonperturbative QCD methods, such as QCD sum rules and lattice QCD.  For example, for $\mu_\pi^2(\Lambda_b)$ and  $\rho_D^3(\Lambda_b)$  we have quoted an uncertainty of $20\%$ and $50\%$, respectively, which could be reduced by dedicated QCD analyses. For other baryons, such parameters are even  less known and deserve new studies. On the other hand, the sensitivity to NP effects of the observables we have described needs to be assessed in the actual experimental conditions. In this respect, the analytic formulae we have provided can be used, e.g., to scrutinize by appropriate  simulations the individual effects of the various  low-energy operators in \eqref{hamil},  when  the experimental analyses are planned.

\subsection{Observables in the $\Lambda_b \to X_c \ell {\bar \nu}$ mode}
\begin{figure}
\begin{center}
\includegraphics[width = 0.47\textwidth]{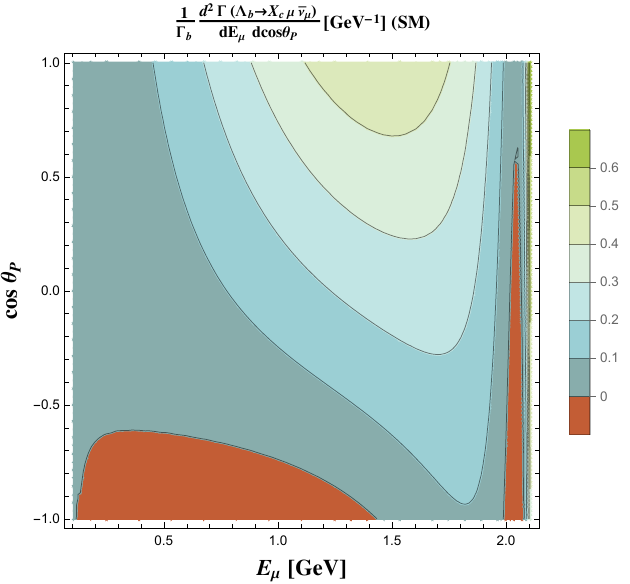}
\includegraphics[width = 0.47\textwidth]{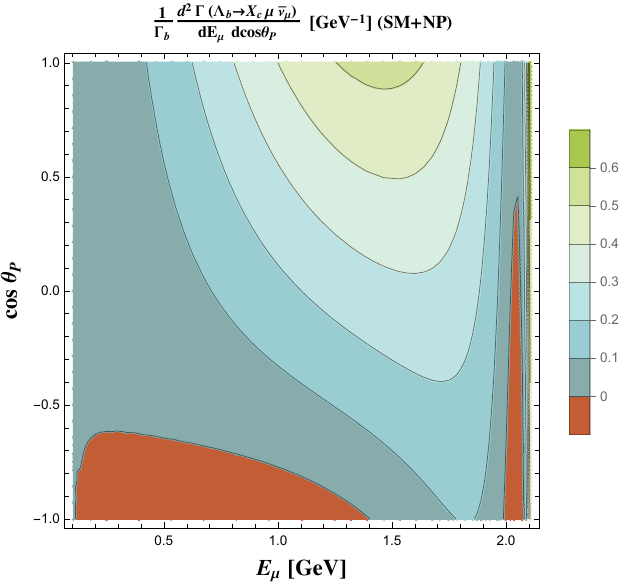}\vspace{0.3cm}\\
\includegraphics[width = 0.47\textwidth]{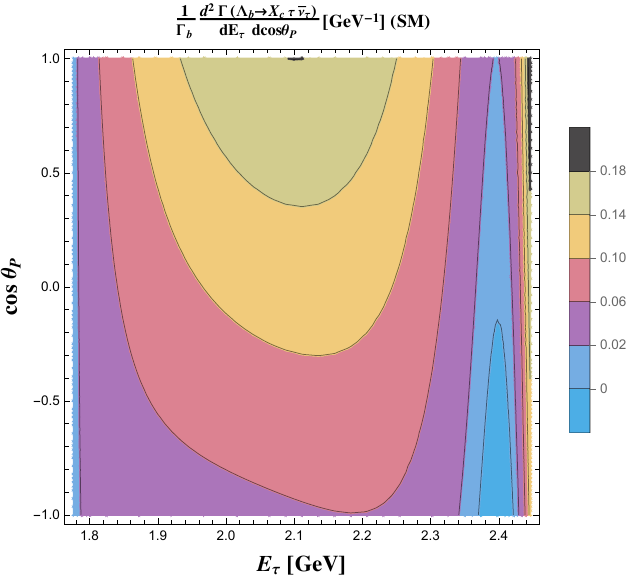}
\includegraphics[width = 0.47\textwidth]{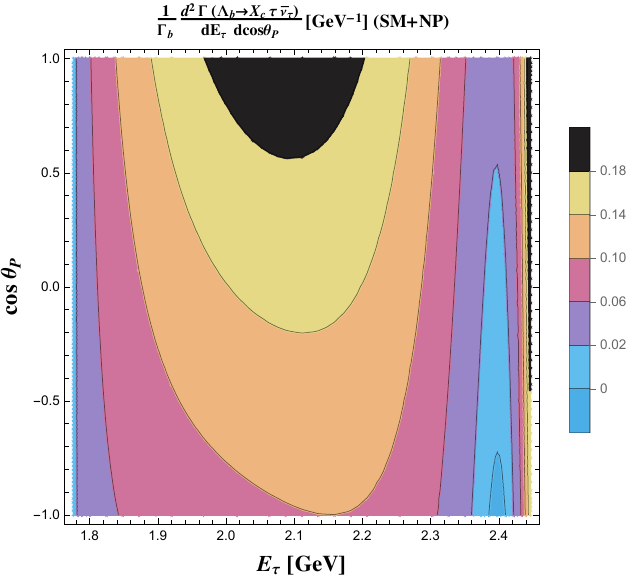}
    \caption{\baselineskip 12pt  \small Contour plots of the  distribution $\dd \frac{1}{\Gamma_b} \frac{d^2 \Gamma}{ d E_\ell \, d\cos \theta_P}$ for $\Lambda_b \to X_c \ell \bar \nu_\ell$.    The top and bottom panels refer to  $\ell=\mu$ and $\ell=\tau$, respectively,  the left and right plots to the Standard Model  and to  NP at the benchmark point.  }\label{contourplotCharm}
\end{center}
\end{figure}

The double differential distribution $\dd \frac{1}{\Gamma_b} \frac{d^2 \Gamma}{ d E_\ell \, d\cos \theta_P}$ for the SM and  NP at the chosen benchmark point is shown in Fig.~~\ref{contourplotCharm} for the  muon and for the  $\tau$  final state. In the case of NP  there is an enhancement of the distribution, more pronounced in the $\tau$ case, 
 for charged lepton energy $E_\mu\simeq 1.7$ GeV and $E_\tau\simeq 2.1$ GeV.

\begin{figure}
\begin{center}
\includegraphics[width = 0.45\textwidth]{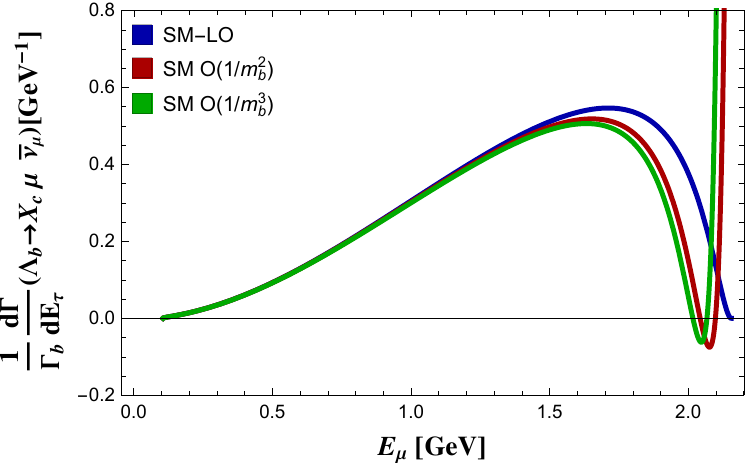} 
\hskip 0.4cm 
\includegraphics[width = 0.45\textwidth]{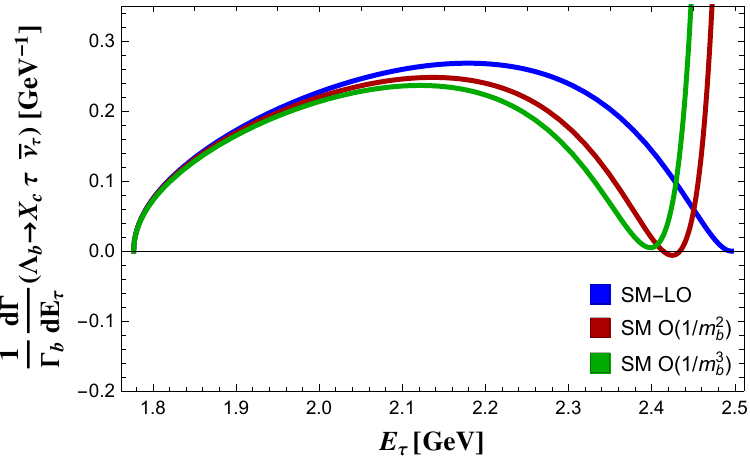} 
\caption{ \baselineskip 12pt  \small Charged lepton energy spectrum in  SM for $\Lambda_b \to X_c \ell \bar \nu_\ell$, with $\ell=\mu$ (left) and  $\ell=\tau$ (right).   The result at leading order  in the HQE (blue line),  at ${\cal O}(1/m_b^2)$ (red line) and at  ${\cal O}(1/m_b^3)$ (green line) are displayed.    }\label{fig:ElspectrumCharmSM}
\end{center}
\end{figure}

The   charged lepton energy spectrum is useful to assess the role of the various terms in the $1/m_b$ expansion. In
Fig.~\ref{fig:ElspectrumCharmSM} we show the result for the muon and the $\tau$ case. The impact of the next-to-leading and next-to-next-to-leading corrections in the HQE is higher for large values of $E_\ell$ ($\ell=\mu,\,\tau$),  excluding the end-point region where  the expansion  breaks down. In the case of $\tau$ the corrections affects a wider energy range. The  parametric hierarchy between $1/m_b^2$ and $1/m_b^3$ corrections is numerically confirmed.

\begin{figure}
\begin{center}
\includegraphics[width = 0.45\textwidth]{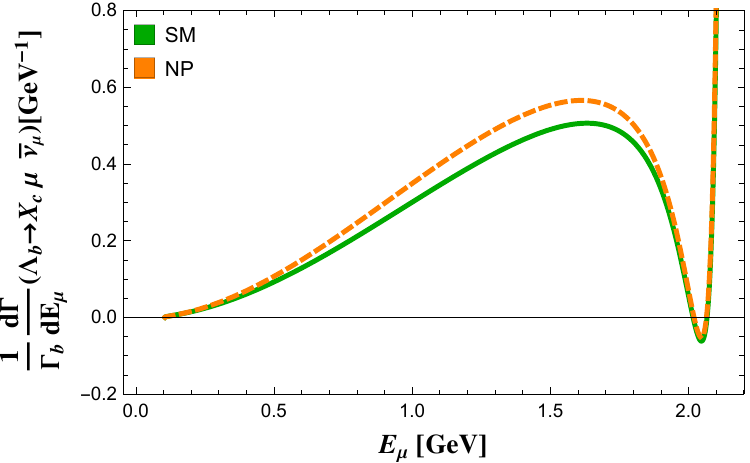} 
\hskip 0.4cm 
\includegraphics[width = 0.45\textwidth]{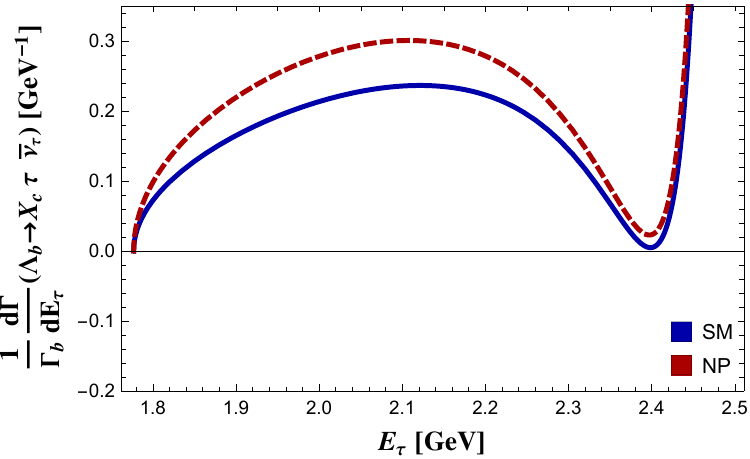} 
\caption{ \baselineskip 12pt  \small Charged lepton energy spectrum for $\Lambda_b \to X_c \ell \bar \nu_\ell$, with $\ell=\mu$ (left) and  $\ell=\tau$ (right). The solid line is the SM result, the dashed line the result for NP at the benchmark point.  }\label{fig:ElspectrumCharm}
\end{center}
\end{figure}

\begin{figure}
\begin{center}
\includegraphics[width = 0.45\textwidth]{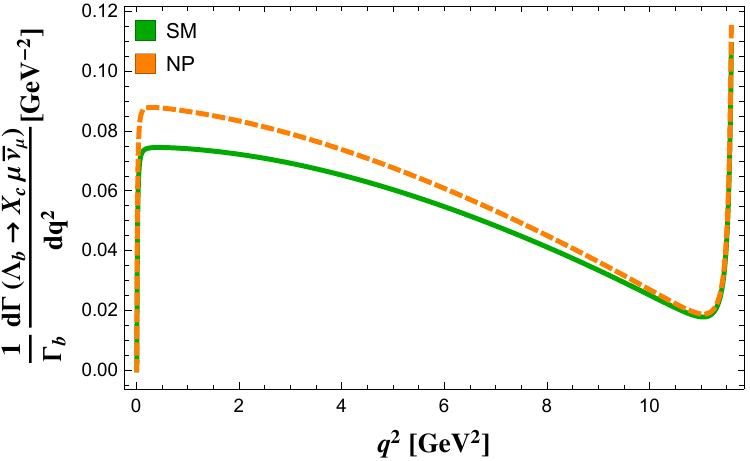} 
\hskip 0.4cm 
\includegraphics[width = 0.45\textwidth]{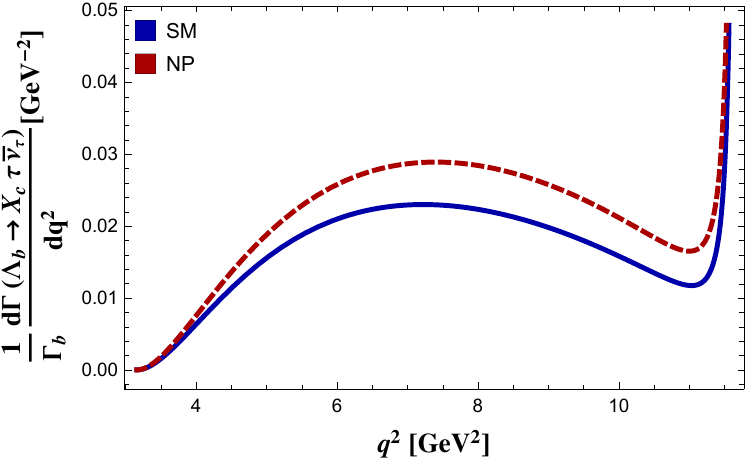} 
\caption{ \baselineskip 12pt  \small Decay distribution in the dilepton invariant mass  $q^2$ for $\Lambda_b \to X_c \ell \bar \nu_\ell$, with $\ell=\mu$ (left) and  $\ell=\tau$ (right).  The solid line is the SM result, the dashed line the result for NP at the benchmark point. }\label{fig:q2spectrumCharm}
\end{center}
\end{figure}

Comparison of the SM prediction (at ${\cal O}(1/m_b^3)$) to NP at the benchmark point is provided in Fig.~\ref{fig:ElspectrumCharm}, where the NP  enhancement already observed in the double distribution is  evident, in particular for the  $\tau$ mode.
The enhancement due to NP can also be observed  in the $q^2$ spectrum, Fig.~\ref{fig:q2spectrumCharm}: in the muon mode the impact is larger for smaller values of $q^2$,  while in the $\tau$ modes the spectrum displays an enhancement in almost all the $q^2$ range.

\begin{figure}
\begin{center}
\includegraphics[width = 0.45\textwidth]{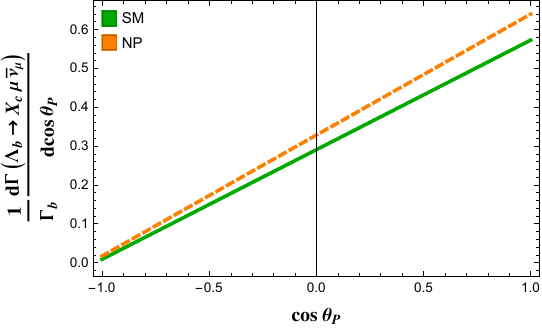} 
\hskip 0.4cm 
\includegraphics[width = 0.45\textwidth]{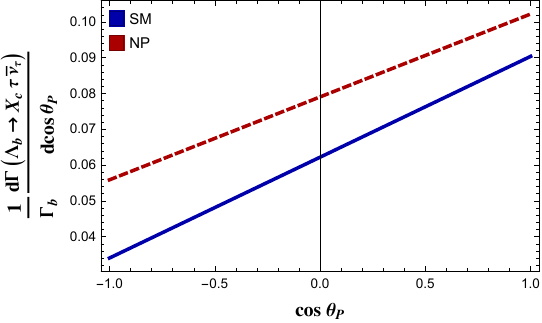} 
\caption{ \baselineskip 12pt  \small   $\dd \frac{1}{\Gamma_b} \frac{d \Gamma}{d\cos \theta_P}$ distribution for $\Lambda_b \to X_c \ell \bar \nu_\ell$, with $\ell=\mu$ (left) and  $\ell=\tau$ (right). The solid line is the  SM result,  the dashed line the NP result at the benchmark point.  }\label{fig:cosPspecrumCharm}
\end{center}
\end{figure}

\begin{figure}
\begin{center}
\includegraphics[width = 0.6\textwidth]{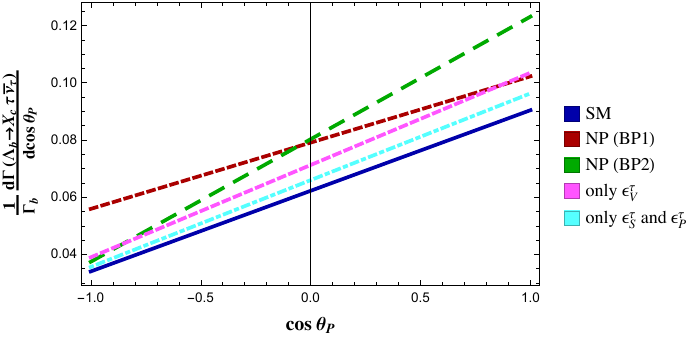} 
\caption{ \baselineskip 12pt  \small  
 $\dd \frac{1}{\Gamma_b} \frac{d \Gamma}{d\cos \theta_P}$  distribution for $\Lambda_b \to X_c \tau \bar \nu_\tau$ in SM and in  NP scenarios with different  values of the effective couplings.  The line NP(BP1) corresponds to  $\epsilon^\tau_{T}$ chosen in this paper.  The lines  NP(BP2),  "only $\epsilon^\tau_{V}$" and "only $\epsilon^\tau_{S}$ and $\epsilon^\tau_{P}$" correspond to the values of the couplings  fixed in ref.\cite{Shi:2019gxi}, see the text.}\label{fig:kamalic}
\end{center}
\end{figure}

A significant sensitivity to NP is found in the $\cos \, \theta_P$  distribution displayed in Fig.~\ref{fig:cosPspecrumCharm}. The dependence of $\dd  \frac{d \Gamma}{d\cos \theta_P}$ on $\cos \, \theta_P$ is linear, and NP contributions modify both the slope and the  intercept of the curve. In principle,  a measurement of few points in the distribution  would allow to access  NP.  This is confirmed by the comparison of different scenarios, corresponding to different benchmark points. Fig.~\ref{fig:kamalic} shows how the various operators have a different impact on the intercept and slope of the distribution.  In particular,  the tensor operator  produces  a large deviation from SM. 

\subsection{ $\Lambda_b \to X_u \ell {\bar \nu}$ mode}
\begin{figure}
\begin{center}
\includegraphics[width = 0.47\textwidth]{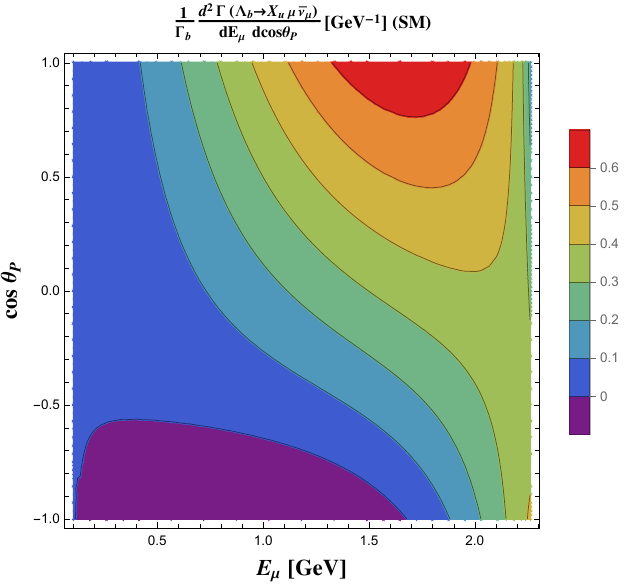}
\includegraphics[width = 0.47\textwidth]{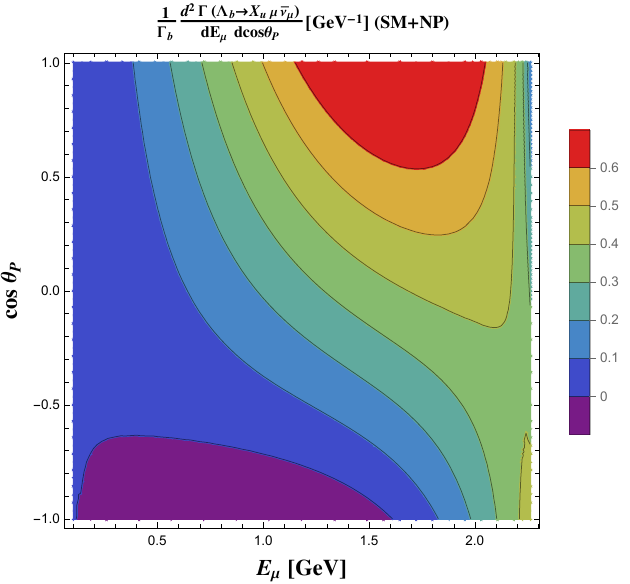} \vspace{0.3cm}\\
\includegraphics[width = 0.47\textwidth]{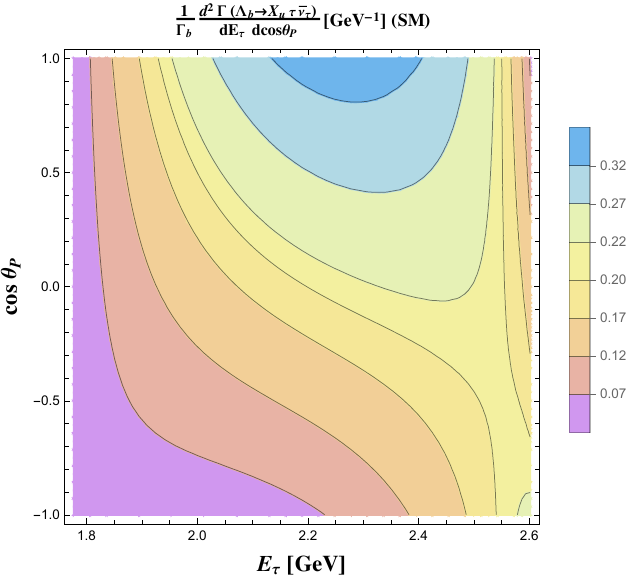}
\includegraphics[width = 0.47\textwidth]{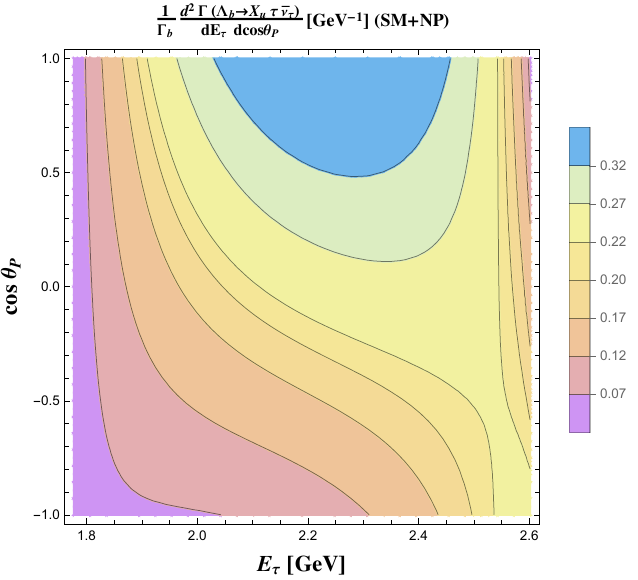}
\caption{\baselineskip 12pt  \small Contour plots of the  distribution $\dd \frac{1}{\Gamma_b} \frac{d^2 \Gamma}{ d E_\ell \, d\cos \theta_P}$ for  $\Lambda_b \to X_u \ell \bar \nu_\ell$. 
The top and bottom panels refer to  $\ell=\mu$ and $\ell=\tau$, respectively,  the left and right panels to the  Standard Model and to NP at the benchmark point.  }\label{contourplotUp}
\end{center}
\end{figure}

\begin{figure}
\begin{center}
\includegraphics[width = 0.45\textwidth]{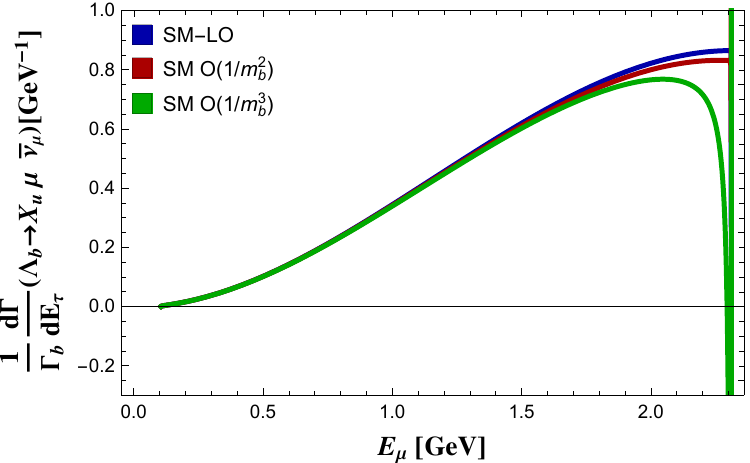} 
\hskip 0.4cm 
\includegraphics[width = 0.45\textwidth]{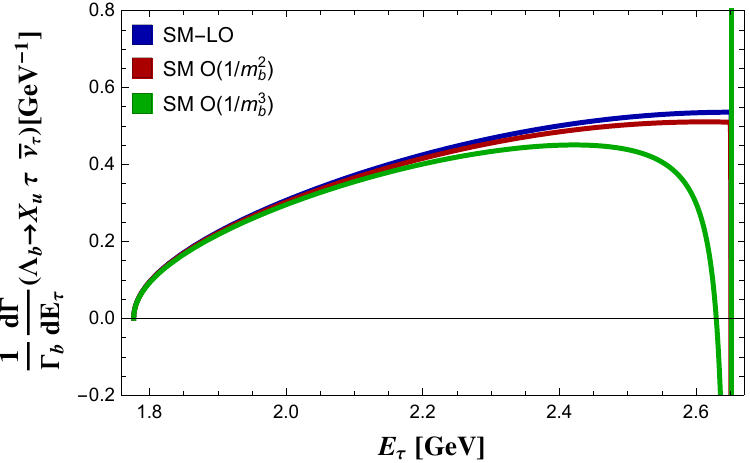} 
\caption{ \baselineskip 12pt  \small Charged lepton energy spectrum in  SM for  $\Lambda_b \to X_u \ell \bar \nu_\ell$, with $\ell=\mu$ (left) and  $\ell=\tau$ (right). The leading order result in the HQE is the blue line, the  ${\cal O}(1/m_b^2)$ the red line, the  ${\cal O}(1/m_b^3)$ the green line.  }\label{fig:ElspectrumUpSM}
\end{center}
\end{figure}

\begin{figure}
\begin{center}
\includegraphics[width = 0.45\textwidth]{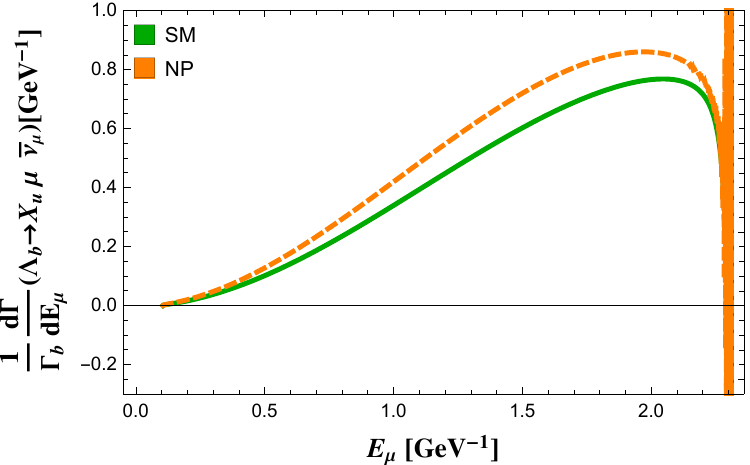} 
\hskip 0.4cm 
\includegraphics[width = 0.45\textwidth]{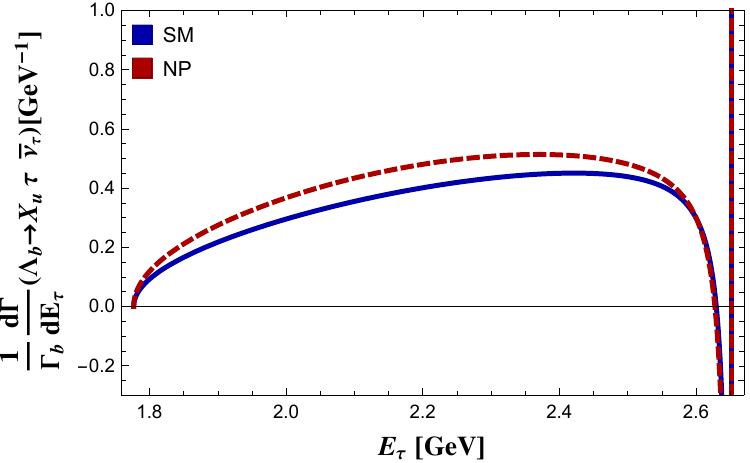} 
\caption{ \baselineskip 12pt  \small Charged lepton energy spectrum for  $\Lambda_b \to X_u \ell \bar \nu_\ell$,  with $\ell=\mu$ (left) and  $\ell=\tau$ (right). The solid line corresponds to  SM,  the dashed line to NP  at the benchmark point.   }\label{fig:ElspectrumUp}
\end{center}
\end{figure}

\begin{figure}
\begin{center}
\includegraphics[width = 0.45\textwidth]{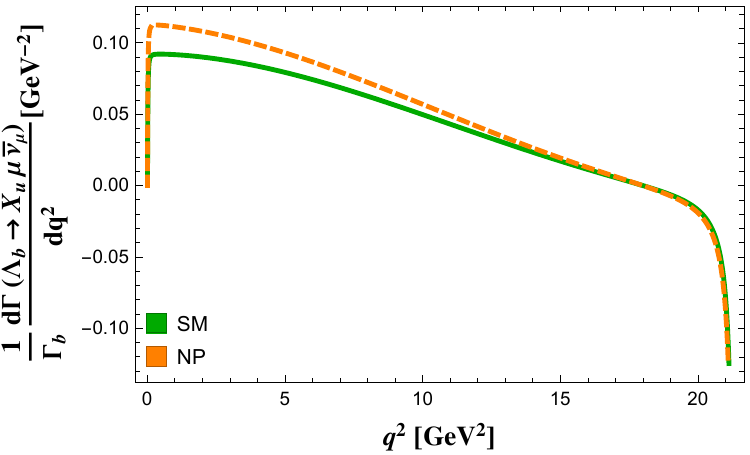} 
\hskip 0.4cm 
\includegraphics[width = 0.45\textwidth]{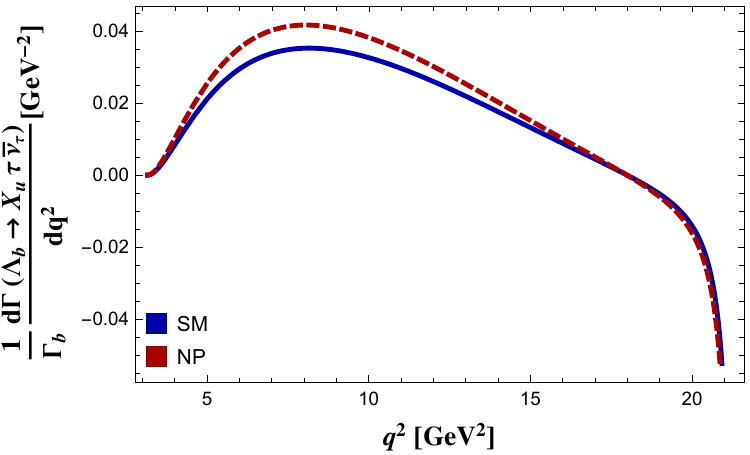} 
\caption{ \baselineskip 12pt  \small $q^2$ distribution for $\Lambda_b \to X_u \ell \bar \nu_\ell$, with $\ell=\mu$ (left) and  $\ell=\tau$ (right). The solid line corresponds to  SM,  the dashed line to NP  at the benchmark point.    }\label{fig:q2spectrumUp}
\end{center}
\end{figure}

$b \to u$ transition displays similar features. The enhancement due to NP appears in the double differential spectra in Fig.~\ref{contourplotUp}, although in this case  it is similar in the $\mu$ and $\tau$ mode.
The various terms in the HQE alter the lepton energy spectrum for large energy, as shown in Fig.~\ref{fig:ElspectrumUpSM}.   NP affects  a wide $E_\ell$ range, with a similar impact for the muon and $\tau$ modes, Fig.~\ref{fig:ElspectrumUp}. The enhancement in the $q^2$ spectrum, displayed in Fig.~\ref{fig:q2spectrumUp}, is lower than in the decay to charm. 
The distribution in $\cos \, \theta_P$ is  sensitive to NP also in this mode. 
\begin{figure}
\begin{center}
\includegraphics[width = 0.45\textwidth]{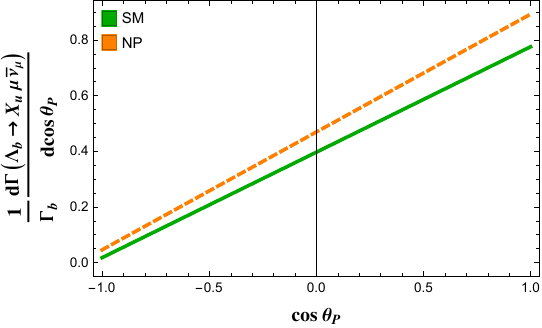} 
\hskip 0.4cm 
\includegraphics[width = 0.45\textwidth]{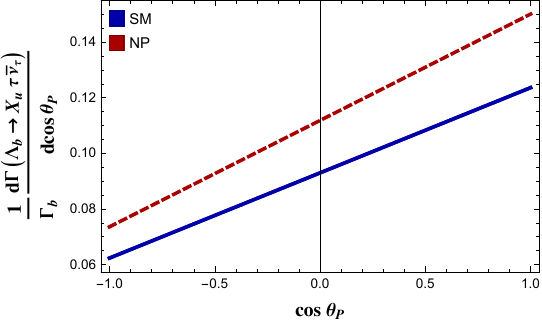} 
\caption{ \baselineskip 12pt  \small   Decay distribution $\dd \frac{1}{\Gamma_b} \frac{d \Gamma}{d\cos \theta_P}$  for  $\Lambda_b \to X_u \ell \bar \nu_\ell$, with $\ell=\mu$ (left) and  $\ell=\tau$ (right). The solid line corresponds to  SM,  the dashed line to NP  at the benchmark point.  }\label{fig:cosPspecrumUp}
\end{center}
\end{figure}
\subsection{Ratio $R_{\Lambda_b}(X_U)$}
For inclusive semileptonic $\Lambda_b$ decays it is interesting to consider  a ratio analogous to $R(D^{(*)})$ for $B$ meson, to
compare the $\tau$ and the muon mode  using a quantity in which  several theoretical uncertainties are canceled:
\be
R_{\Lambda_b}(X_U)=\frac{\Gamma(\Lambda_b \to X_U \, \tau \, {\bar \nu}_\tau)}{\Gamma(\Lambda_b \to X_U \, \mu \, {\bar \nu}_\mu)} \hskip 1.5 cm (U=u,\,c) \,\, .
\label{ratioR}
\ee
For this ratio we obtain:
\bea
R_{\Lambda_b}(X_u)^{SM} &=& 0.234 \,,  \hspace{1cm} R_{\Lambda_b}(X_u)^{NP}  = 0.238 \,\,\,, \label{Rup} \\
R_{\Lambda_b}(X_c)^{SM} &=& 0.214 \,,   \hspace{1cm} R_{\Lambda_b}(X_c)^{NP} = 0.240 \,\,\,.\label{Rcharm} 
\eea
 As with the other quantities in this study, the ratios (\ref{Rup})-(\ref{Rcharm}) are obtained at leading order in $\alpha_s$. ${\cal O}(\alpha_s^2)$  corrections have been included in the ratio $R_b(X_c)$  \cite{Biswas:2009rb}, showing that they are small and supporting the expectation that perturbative corrections cancel in the ratios to a large extent.
Our results  suggest a higher sensitivity of the  charm mode to NP.  It would be  important to observe the correlation of  this  measurement with the  results for $B$ mesons.

\begin{figure}[ht]
\begin{center}
\includegraphics[width = 0.5\textwidth]{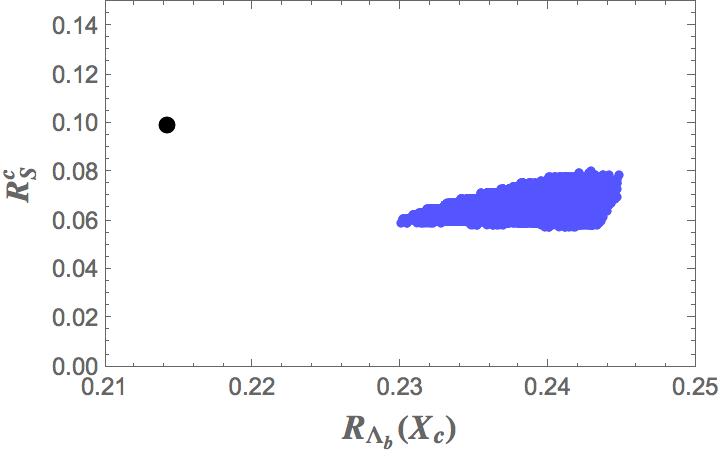} 
\caption{ \baselineskip 12pt  \small   $\Lambda_b \to X_c \tau \bar \nu_\tau$:
 correlation between $R_{\Lambda_b}(X_c)$ and the ratio $R_S^c$ of the slopes of the $\dd  \frac{d \Gamma}{d\cos \theta_P}$  distribution. The dot corresponds to SM, the broad region to NP with the effective couplings varied as specified in the text.}\label{fig:lpslope}
\end{center}
\end{figure}

As a last observable, we define another ratio  sensitive to lepton flavour universality violating NP effects. It can be constructed from 
the  distribution
$\dd \frac{d \Gamma (\Lambda_b \to X_U \ell {\bar \nu}_\ell)}{d\cos \theta_P}=A_\ell^U+B_\ell^U \, \cos \theta_P$. The intercept of the distribution  is  $A_\ell^U=\dd\frac{1}{2} \Gamma (\Lambda_b \to X_U \ell {\bar \nu}_\ell)$, hence  $R_{\Lambda_b}(X_U)=\dd \frac{A_\tau^U}{A_\mu^U}$. The ratio of the slopes $R_S^U=\dd \frac{B_\tau^U}{B_\mu^U}$ has a definite value in  SM, and can deviate  from it due to NP.  In  SM we find: $R_S^c=0.1$ and $R_S^u=0.08$. A correlation between $R_{\Lambda_b}(X_U)$ and  $R_S^U$ can be costructed. 
As an example, for $\Lambda_b \to X_c \tau \bar \nu_\tau$ with the effective Hamiltonian extended including a  tensor operator, we vary the couplings $({\rm Re}(\epsilon_T^\mu),\,{\rm Im}(\epsilon_T^\mu))$ and $({\rm Re}(\epsilon_T^\tau),\,{\rm Im}(\epsilon_T^\tau))$ in the regions determined in \cite{Colangelo:2018cnj}. The  correlation plot 
 in  Fig.~\ref{fig:lpslope} shows that the (challenging) measurement of the two ratios would separate  SM from NP.
 
\section{Conclusions}
We have presented a reappraisal of  the calculation of the inclusive semileptonic decay width of a heavy hadron, focusing  on  polarized $\Lambda_b$. We present the expressions for the full differential decay distribution  and for the fully integrated width at order ${\cal O}(1/m_b^3)$ in the HQE, at leading order in $\alpha_s$ and for non vanishing charged lepton mass. The computation is done extending
 the SM effective Hamiltonian by the inclusion of  the full set of $D=6$ semileptonic operators, each one weighted by a lepton-flavour dependent  coefficient. 

Our study improves the SM result, previously known at order ${\cal O}(1/m_b^2)$ for a polarized hadron,  providing the expressions of the hadronic matrix elements. 
This allows to analyze  other  $b$-flavoured baryon modes, as well as other inclusive processes. Moreover, the study supplies the  elements for  analyzing different operators in the effective weak Hamiltonian density. 

Possible NP effects are systematically  investigated in  various distributions. In particular,
in view of the  tension in the ratios $R(D^{(*)})$ for $B$ mesons,  we have studied the analogous ratios $R_{\Lambda_b}(X_{c,u})$.
Among other results, we have found that the $\dd\frac{d \Gamma}{ d  \cos \, \theta_P}$ distribution,  linear in $\cos \, \theta_P$,  is sensitive to NP.
The slope of the distribution,  for a  hadronic final state  $X_c$ or $X_u$,   depends on the final lepton species, hence the ratio $R_S^{c,u}$  of the slopes,  for  $\ell=\tau$ vs $\ell=\mu$, is  sensitive to  possible lepton flavour universality violation. For   $\Lambda_b \to X_c \ell {\bar \nu}_\ell$  when the effective Hamiltonian  includes a  tensor operator, we have shown that  a deviation  from SM in  $R_{\Lambda_b}(X_c)$ is related to a deviation in $R_S^c$, an interesting, although challenging,  correlation to investigate.

\vspace*{1cm}
\noindent {\bf Acknowledgements.} We thank J. Aebischer for discussions.
FDF thanks A. Buttaro for advice.
This study has been  carried out within the INFN project (Iniziativa Specifica) QFT-HEP.

\newpage
\appendix
\numberwithin{equation}{section}
\section{Hadronic Matrix Elements}\label{appA}

In this appendix we collect the hadronic matrix elements involved in the heavy quark expansion to ${\cal O}(1/m_b^3)$.
The relations are employed:
\be
i\,\epsilon^{\mu \nu \alpha \beta} v_\alpha P_+ \gamma_\beta \gamma_5 P_+  = -P_+ (-i\, \sigma^{\mu \nu}) P_+ \,\,\, ,
\hspace*{1cm}
\sigma^{\mu \nu} = -\frac{i }{2}\epsilon^{\mu \nu \alpha \beta} \sigma_{\alpha \beta}\gamma_5 \,\,\ .
\ee
The  terms independent of the  spin four-vector $s_\mu$ agree with  Ref.\cite{Dassinger:2006md}. \\

{\bf \noindent Dimension 6 operator}
\par
The matrix element is computed in HQET:
 \bea
\langle H_b(v,s)|({\bar h}_v)_a (i D)^{\tau}(i D)^{\lambda}(i D)^{\sigma}(h_v)_b|H_b(v,s)\rangle=
(A^{D6})^{\tau \lambda \sigma }(P_+)_{ba} + (B^{D6})^{\tau \lambda \sigma \mu} \left[ P_+\gamma_\mu  \gamma_5 P_+ \right]_{ba}\nn\\
\label{dim6}
\eea
with $a, b$ Dirac indices.
 $A^{(D6)}$ and  $B^{(D6)}$ are  parity-even and parity-odd, respectively.
Using the expansion
\bea
 (A^{(D6)})^{\tau \lambda \sigma }&=& A^{(D6)}_1 \Pi^{\tau  \sigma } v^\lambda+A^{(D6)}_2 \, i \, \epsilon^{\tau  \sigma \alpha \beta }v_\alpha s_\beta v^\lambda \,\, \nn \\
(B^{(D6)})^{\tau \lambda \sigma \mu}&=& B^{(D6)}_1 \Pi^{\tau  \sigma } v^\lambda s^\mu +B^{(D6)}_2 \, i \, \epsilon^{\tau  \sigma \alpha \mu }v_\alpha v^\lambda \,\, 
 \eea
we find:
\bea
A^{(D6)}_1&=&-B^{(D6)}_1=\displaystyle\frac{m_H}{3}{\hat \rho}_D^3 \nn \\
A^{(D6)}_2&=&\displaystyle\frac{m_H}{2}{\hat \rho}_{LS}^3  \\
B^{(D6)}_2&=&\displaystyle\frac{m_H}{6}{\hat \rho}_{LS}^3 \,\,\, . \nn \eea
This gives the expression of the matrix element:
\bea
&&\langle H_b(v,s)|({\bar h}_v)_a (i D)^{\tau}(i D)^{\lambda}(i D)^{\sigma}(h_v)_b|H_b(v,s)\rangle =\nn\\
&&
\left(\frac{m_H}{3}{\hat \rho}_D^3\Pi^{\tau  \sigma } v^\lambda+\frac{m_H}{2}{\hat \rho}_{LS}^3 \, i \, \epsilon^{\tau  \sigma \alpha \beta }v_\alpha s_\beta v^\lambda \right) [P_+]_{ba} \nn \\ &&
+\left(
-\frac{m_H}{3}{\hat \rho}_D^3\Pi^{\tau  \sigma } v^\lambda s^\mu +\frac{m_H}{6}{\hat \rho}_{LS}^3 \, i \, \epsilon^{\tau  \sigma \alpha \mu } v_\alpha v^\lambda \right)
\left[ P_+\gamma_\mu  \gamma_5 P_+ \right]_{ba}  \,\,.
\label{dim6ris}
\eea

{\bf  \noindent Dimension 5 operator}
\par
The matrix element, computed in QCD,  can be expressed as:
 \bea
\langle H_b(v,s)|({\bar b}_v)_a (i D)^{\tau}(i D)^{\sigma}(b_v)_b|H_b(v,s) \rangle
& =&(A^{(D5)})^{\tau  \sigma }g_{ba} + (B^{(D5)})^{\tau  \sigma }(\gamma_5)_{ba}+(C^{(D5)})^{\tau \sigma \mu} (\gamma_\mu)_{ba}\nn \\
&+&(D^{(D5)})^{\tau \sigma \mu} (\gamma_\mu \gamma_5)_{ba}+   (E^{(D5)})^{\tau \sigma \mu \nu }(-i \sigma_{\mu \nu})_{ba}  \,\,\, ,\nn \\
\label{dim5} 
\eea
 with $A^{(D5)}, C^{(D5)}, E^{(D5)}$ parity-even and   $B^{(D5)}, D^{(D5)}$ parity-odd.
 They can be expanded as:
\bea
 (A^{(D5)})^{\tau  \sigma }&=&A^{(D5)}_1 g^{\tau \sigma} +A^{(D5)}_2 v^\tau v^\sigma +A^{(D5)}_3 i \, \epsilon^{\tau \sigma \alpha \beta }v_\alpha s_\beta  \nn \\
 (B^{(D5)})^{\tau  \sigma }&=& B^{(D5)}_1 v^\tau s^\sigma + B^{(D5)}_2 s^\tau v^\sigma \nn 
 \\
 (C^{(D5)})^{\tau \sigma \mu}&=&C^{(D5)}_1 g^{\tau \sigma} v^\mu+C^{(D5)}_2 g^{\tau \mu} v^\sigma+C^{(D5)}_3 g^{\mu \sigma} v^\tau+C^{(D5)}_4 v^\tau v^\sigma v^\mu+\nn \\
 &
 +& C^{(D5)}_5 \, i \, \epsilon^{\tau \sigma \alpha \beta} v_\alpha s_\beta v^\mu+C^{(D5)}_6 \, i \, \epsilon^{ \sigma \alpha \beta  \mu} v_\alpha s_\beta v^\tau+C^{(D5)}_7 \, i \, \epsilon^{ \tau \alpha \beta  \mu} v_\alpha s_\beta v^\sigma
 \nn  \\
 (D^{(D5)})^{\tau \sigma \mu}&=&D^{(D5)}_1 g^{\tau \sigma} s^\mu+D^{(D5)}_2 v^\tau v^\sigma s^\mu+D^{(D5)}_3\, i\, \epsilon^{\tau \sigma \theta \mu}v_\theta
\nn \\
&&+D^{(D5)}_4 g^{\sigma \mu}s^\tau+D^{(D5)}_5 g^{\tau \mu} s^\sigma+D^{(D5)}_6 v^\mu v^\sigma s^\tau +D^{(D5)}_7 v^\mu v^\tau s^\sigma
 \\
(E^{(D5)})^{\tau \sigma \mu \nu}=&&E^{(D5)}_1(g^{\mu \tau} g^{\nu \sigma}-g^{\nu \tau} g^{\mu \sigma})+
\nn \\ &+&E^{(D5)}_2 (g^{\mu \tau} v^{\nu}-g^{\nu \tau} v^{\mu }) v^\sigma+E^{(D5)}_3 (g^{\mu \sigma} v^{\nu}-g^{\nu \sigma} v^{\mu }) v^\tau +\nn \\
&+&
E^{(D5)}_4 g^{\tau \sigma}\,i\,\epsilon^{\mu \nu \alpha \beta}v_\alpha s_\beta +E^{(D5)}_5 v^\tau v^\sigma\, i\,\epsilon^{\mu \nu \alpha \beta}v_\alpha s_\beta+
\nn \\
&+&
E^{(D5)}_{6}\,i\, \epsilon^{\mu \nu \sigma \alpha} s_\alpha v_\tau+E^{(D5)}_{7}\,i\, \epsilon^{\mu \nu \tau \alpha} s_\alpha v_\sigma + \nn \\
 &+&
  E^{(D5)}_{8}\,i\, \epsilon^{\mu \nu \sigma \alpha } v_\alpha s_\tau +E^{(D5)}_{9}\,i\, \epsilon^{\mu \nu \tau \alpha } v_\alpha s_\sigma  \,\, .\nn
 \eea
We obtain:

\begin{center}
\begin{tabular}{l l}
\noalign{\bigskip}
$\dd A^{(D5)}_1=-A^{(D5)}_2=-\frac{m_H}{6}{\hat \mu}^2_\pi $  & $ \dd A^{(D5)}_3= \frac{m_H}{4}\Big[{\hat \mu}^2_G  +\frac{ {\hat \rho}_D^3+ {\hat \rho}_{LS}^3 }{m_b} \Big]$\smallskip \\
$\dd B^{(D5)}_1=-B^{(D5)}_2=\frac{m_H}{12m_b}{\hat \rho}_D^3$\smallskip \\
$\dd C^{(D5)}_1= -\frac{m_H }{6}{\hat \mu}_\pi^2$ &
$\dd C^{(D5)}_2=C^{(D5)}_3=\frac{m_H }{12m_b}\left( {\hat \rho}_D^3+ {\hat \rho}_{LS}^3 \right)$\smallskip \\
$\dd C^{(D5)}_4=\frac{m_H }{6}{\hat \mu}_\pi^2-\frac{m_H }{6m_b}\left( {\hat \rho}_D^3+ {\hat \rho}_{LS}^3 \right)$&
$\dd C^{(D5)}_5=\frac{m_H}{4}\Big[{\hat \mu}^2_G  +\frac{ {\hat \rho}_D^3+ {\hat \rho}_{LS}^3 }{m_b} \Big]$\smallskip \\
$\dd C^{(D5)}_6=-C^{(D5)}_7=-\frac{m_H}{24m_b}\left( 2{\hat \rho}_D^3+3 {\hat \rho}_{LS}^3 \right)$\smallskip \\
$\dd D^{(D5)}_1=-D^{(D5)}_2=\frac{m_H}{6}{\hat \mu}_\pi^2 $&
$\dd D^{(D5)}_3=\frac{m_H}{12}\Big[{\hat \mu}^2_G  +\frac{ {\hat \rho}_D^3+ {\hat \rho}_{LS}^3 }{m_b} \Big]$ \smallskip  \\
$\dd D^{(D5)}_6=D^{(D5)}_7= \frac{m_H }{12m_b} {\hat \rho}_D^3$\smallskip \\
$\dd E^{(D5)}_1=-\frac{m_H }{24}\Big[{\hat \mu}^2_G  +\frac{ {\hat \rho}_D^3+ {\hat \rho}_{LS}^3 }{m_b} \Big] $&
$\dd E^{(D5)}_2 = -E^{(D5)}_3=\frac{m_H}{24}{\hat \mu}_G^2+\frac{m_H }{12m_b}\left( {\hat \rho}_D^3+ {\hat \rho}_{LS}^3 \right)$\smallskip \\
$\dd E^{(D5)}_4=-\frac{m_H }{12}{\hat \mu}_\pi^2 $&
$\dd E^{(D5)}_5= \frac{m_H }{12}{\hat \mu}_\pi^2-\frac{m_H }{24m_b}\left( 2{\hat \rho}_D^3+ 3{\hat \rho}_{LS}^3 \right)$\smallskip \\
$\dd E^{(D5)}_6=E^{(D5)}_{7}= \frac{m_H }{48m_b}\left( 2{\hat \rho}_D^3+ 3{\hat \rho}_{LS}^3 \right)$&
$\dd D^{(D5)}_4=D^{(D5)}_5= E^{(D5)}_{8}=E^{(D5)}_{9}=0$  . \bigskip \\
\end{tabular}
\end{center}
\noindent
The previous expressions allow us to write the   matrix element:
 \bea
&&\langle H_b(v,s)|({\bar b}_v)_a (i D)^{\tau}(i D)^{\sigma}(b_v)_b|H_b(v,s)\rangle= -\frac{m_H }{3}{\hat \mu}_\pi^2\, \Pi^{\tau \sigma} \Big[P_+-s_\mu{\hat S}^\mu \Big]_{ba} \nn \\
&+&\frac{m_H }{2}\Big({\hat \mu}^2_G  +\frac{ {\hat \rho}_D^3+ {\hat \rho}_{LS}^3 }{m_b} \Big)\,i\,\epsilon^{\tau \sigma \alpha \beta}v_\alpha \Big[s_\beta \, P_++\frac{1}{3} {\hat S}_\beta \Big]_{ba}
+\frac{m_H }{6m_b}{\hat \rho}_D^3 \left[v^\tau s^\sigma P_+ \gamma_5-v^\sigma s^\tau\gamma_5 P_+ \right]_{ba} \label{dim5fin} 
\nn \\
&-&\frac{m_H }{6m_b}\left( {\hat \rho}_D^3+ {\hat \rho}_{LS}^3 \right)\Big[-v^\tau s^\sigma P_+ \gamma_5-v^\sigma s^\tau \gamma_5 P_+ +(2v^\tau v^\sigma P_+-v^\sigma \gamma^\tau P_+ -v^\tau P_+ \gamma^\sigma )(1-{\spur s} \gamma_5)\Big]_{ba} \nn 
\\
&-&\frac{m_H }{12m_b}{\hat \rho}_{LS}^3\Big[-v^\tau s^\sigma P_+ \gamma_5-v^\sigma s^\tau \gamma_5 P_+ -(2v^\tau v^\sigma P_+-v^\sigma \gamma^\tau P_+ -v^\tau P_+ \gamma^\sigma ){\spur s} \gamma_5\Big]_{ba} \,\,\, . 
\eea
\\
{\bf \noindent Dimension 4 operator}
\par
The procedure for computing the matrix element in QCD, using the expansion in Dirac matrices, is analogous to the $D=5$ case. The results is:
\bea 
\langle H_b(v,s)|({\bar b}_v)_a (i D)^{\tau}(b_v)_b|H_b(v,s)\rangle&=&
\frac{m_H }{2m_b}\left( {\hat \mu}^2_\pi-{\hat \mu}^2_G \right)\left[\left( v^\tau P_+ \, -\frac{1}{3}(\gamma^\tau-v^\tau {\spur v}) \right) (1-{\spur s}\gamma_5)\right]_{ba}
\nn \\ 
&-&\frac{m_H}{3m_b} {\hat \mu}^2_\pi s^\tau  [P_+ \gamma_5]_{ba}\nn \\
&+&\frac{m_H }{12m_b}{\hat \mu}^2_G \Big[-(\gamma^\tau-v^\tau {\spur v}) {\spur s}\gamma_5+3s^\tau \gamma_5 \Big]_{ba} \nn \\
&+&\frac{m_H }{12m_b^2}\left( {\hat \rho}_D^3+ {\hat \rho}_{LS}^3 \right)[\gamma^\tau-4v^\tau {\spur v}]_{ba} \nn \\
&+&\frac{m_H }{12m_b^2} {\hat \rho}_D^3
\left[v^\tau {\spur s}\gamma_5 -s^\tau {\spur v} \gamma_5 \right]_{ba} \label{dim4ris} \\
&+&
\frac{m_H }{6m_b^2}{\hat \rho}_D^3 \Big[-(\gamma^\tau-2v^\tau {\spur v}) {\spur s}\gamma_5+s^\tau \gamma_5 \Big]_{ba} \nn \\
&+&
\frac{m_H }{8m_b^2}{\hat \rho}_{LS}^3 \Big[-(\gamma^\tau-3v^\tau {\spur v}) {\spur s}\gamma_5+s^\tau \gamma_5 \Big]_{ba} .  \nn
\eea
{\bf Dimension 3 operator}
\par
The matrix element computed in QCD reads:

 \bea
 \langle H_b(v,s)|({\bar b}_v)_a (b_v)_b|H_b(v,s)\rangle&=& 
 \left[\left(m_H P_+ -\frac{m_H}{4m_b^2}
 \left( {\hat \mu}^2_\pi-{\hat \mu}^2_G \right) \right) (1-{\spur s}\gamma_5)\right]_{ba} \label{d3ris} \nn \\
 &+& \frac{m_H}{4m_b^2}
 \left( {\hat \mu}^2_\pi-{\hat \mu}^2_G \right)\left[P_+ {\spur s}\gamma_5\right]_{ba}-\frac{m_H}{6m_b^2}
 \left( {\hat \mu}^2_\pi+\frac{{\hat \rho}_D^3}{m_b} \right) \left[P_- {\spur s}\gamma_5 \right]_{ba} . \nn \\
 \eea
 
The  matrix elements can be related  using the equation of motion for  $b_v$:
\bea 
\langle H_b(v,s)| {\bar b}_v(i D)^{\mu_1} \dots (iD)^{\mu_n} \Gamma b_v|H_b(v,s)\rangle&=&\frac{1}{2}\langle H_b(v,s)| {\bar b}_v(i D)^{\mu_1} \dots (iD)^{\mu_n} \{\Gamma,\, \spur v\} b_v|H_b(v,s)\rangle\nn \\
&&\hskip -3.5cm+\frac{1}{2m_b}\langle H_b(v,s)| {\bar b}_v\{(i \spur D),\,(i D)^{\mu_1} \dots (iD)^{\mu_n} \Gamma \} b_v|H_b(v,s)\rangle \,\,\, \hspace*{1cm}
\eea
 for a generic Dirac matrix $\Gamma$.  This allows to relate the coefficients of  matrix elements of operators of different dimensions,  
providing a check of the results \cite{Mannel:2018mqv}.
%
\section{Hadronic tensor for the Standard Model and for the extended Hamiltonian }\label{appB}
We provide the tensor $T^{ij}$ for the $b \to U$ modes ($U=u,\,c$)   for the Standard Model and for the effective Hamiltonian in Eq.~\eqref{hamil}, expanded  in  invariant functions. We provide their expressions for the single operators (Standard Model, S, P, T) and for the interferences.
 \begin{itemize}
 \item Standard Model
 \par
This case amounts to choosing  $i=j=1$ in Eq.~\eqref{Tij-gen} and  $  J^{(1)}_\mu=\bar U \gamma_\mu(1-\gamma_5) b$.
For a polarized baryon the corresponding tensor  $T^{\mu \nu}_{SM}$ can be expanded in terms of the  functions $T_{1,\dots,5}$ and $S_{1,\dots,13}$ \cite{Manohar:1993qn}:
 \bea
 T^{\mu \nu}_{SM} &=&-g^{\mu \nu}T_1+v^\mu v^\nu T_2-i\, \epsilon^{\mu \nu \alpha \beta}v_\alpha q_\beta T_3+q^\mu q^\nu T_4 +(q^\mu v^\nu+q^\nu v^\mu)T_5 \nn \\
 &-&(q \cdot s)\Big[-g^{\mu \nu}S_1+v^\mu v^\nu S_2-i\, \epsilon^{\mu \nu \alpha \beta}v_\alpha q_\beta S_3+q^\mu q^\nu S_4 +(q^\mu v^\nu+q^\nu v^\mu)S_5 \Big] \hspace*{1cm} \nn \\
 &+&(s^\mu v^\nu+s^\nu v^\mu)S_6+(s^\mu q^\nu+s^\nu q^\mu)S_7+i\, \epsilon^{\mu \nu \alpha \beta}v_\alpha s_\beta \, S_8+i\, \epsilon^{\mu \nu \alpha \beta}q_\alpha s_\beta \, S_9 \nn \\
 &+&(s^\mu v^\nu-s^\nu v^\mu)S_{10}+(s^\mu q^\nu-s^\nu q^\mu)S_{11} \label{TSMris}  \\
 &+&\left(v^\mu \, \epsilon^{\nu \alpha \beta \delta}q_\alpha v_\beta s_\delta +v^\nu \, \epsilon^{\mu \alpha \beta \delta}q_\alpha v_\beta s_\delta \right) i\,S_{12} \nn \\
&+& \left(q^\mu \, \epsilon^{\nu \alpha \beta \delta}q_\alpha v_\beta s_\delta +q^\nu \, \epsilon^{\mu \alpha \beta \delta}q_\alpha v_\beta s_\delta \right)i\, S_{13} \,\, . \nn
 \eea
 The  $1/m_b$ expansion of the these functions  reads:
 \bea
 T_1 &=& 2 m_H \Bigg\{\frac{1}{\Delta_0} \left[2(m_b-v \cdot q)+ \frac{\left( {\hat \mu}^2_\pi-{\hat \mu}^2_G \right)}{3m_b}-2 \frac{\left( {\hat \rho}_D^3+ {\hat \rho}_{LS}^3 \right) }{3m_b^2} \right]  \nn \\
 &+&\frac{2}{3m_b\Delta_0^2}\Big[ \left( {\hat \mu}^2_\pi-{\hat \mu}^2_G \right) \big[2 (q^2-(v \cdot q)^2)\nn \\
 &+&3(v \cdot q)(m_b-v \cdot q) \big]+{\hat \mu}^2_G 4m_b(m_b- v \cdot q) \nn \\
 &+& \frac{\left( {\hat \rho}_D^3+ {\hat \rho}_{LS}^3 \right) }{m_b}\left[6m_b(m_b-v \cdot q)-q^2+4 (v \cdot q)^2 \right]-4m_b{\hat \rho}_{LS}^3\Big]   \nn\\
 &-&\frac{8}{3\Delta_0^3}\left[q^2-(v \cdot q)^2\right](m_b- v \cdot q) \Big[ {\hat \mu}^2_\pi-\frac{{\hat \rho}_D^3+{\hat \rho}_{LS}^3}{m_b}\Big] \hspace*{1cm} \label{T1} \\
 &-&\frac{16}{3\Delta_0^4}{\hat \rho}_D^3\left[q^2-(v \cdot q)^2\right](m_b- v \cdot q)^2
 \Bigg\} \nn
 \eea
 \bea
 T_2 &=& 2 m_H \Bigg\{\frac{2}{\Delta_0} \Big[2m_b+\frac{5}{3m_b}\left( {\hat \mu}^2_\pi-{\hat \mu}^2_G \right)-\frac{4}{3m_b^2}\left( {\hat \rho}_D^3+ {\hat \rho}_{LS}^3 \right) \Big] \nn \\
 &+&\frac{4}{3m_b\Delta_0^2}\Big[ 7m_b v \cdot q \, {\hat \mu}^2_\pi+m_b(2m_b-5v \cdot q)\,{\hat \mu}^2_G \nn \\
& +& 6(m_b - v \cdot q)\, {\hat \rho}_D^3+2(2m_b - 3v \cdot q)\,{\hat \rho}_{LS}^3 \Big]  \label{T2}  \\
 &-&\frac{8}{3\Delta_0^3}\Big[2m_b \left[q^2-(v \cdot q)^2 \right] \,{\hat \mu}^2_\pi-2v \cdot q(m_b-v \cdot q)\left( 2{\hat \rho}_D^3+ {\hat \rho}_{LS}^3 \right)+q^2{\hat \rho}_{LS}^3 \Big] \nn \\
 &-&\frac{32}{3\Delta_0^4}{\hat \rho}_D^3 \left[q^2-(v \cdot q)^2\right]m_b(m_b- v \cdot q)
 \Bigg\} \nn
 \eea
 \bea
 T_3 &=& - 2 m_H \Bigg\{\frac{2}{\Delta_0} +\frac{2}{3m_b^2\Delta_0^2}\Big[5m_b v \cdot q \left( {\hat \mu}^2_\pi-{\hat \mu}^2_G \right)+6m_b^2 {\hat \mu}^2_G\nn \\
&+&2(3m_b-2v \cdot q)\left( {\hat \rho}_D^3+  {\hat \rho}_{LS}^3  \right) \Big] \nn \\
 &+&\frac{8}{3\Delta_0^3}\Big[-[q^2-(v \cdot q)^2]{\hat \mu}^2_\pi+v \cdot q \,(m_b-v \cdot q)\frac{{\hat \rho}_D^3}{m_b}-(m_b-v \cdot q)^2 \,\frac{{\hat \rho}_{LS}^3}{m_b} \Big] 
 \nn \\
 &-&\frac{16}{3\Delta_0^4}{\hat \rho}_D^3[q^2-(v \cdot q)^2](m_b- v \cdot q)
 \Bigg\}  \label{T3}
 \eea
 \bea
 T_4 &=& 2m_H \Bigg\{\frac{4}{3m_b\Delta_0^2} \Big[2\left( {\hat \mu}^2_\pi-{\hat \mu}^2_G \right)- \frac{\left( {\hat \rho}_D^3+ {\hat \rho}_{LS}^3 \right)}{m_b}  \Big] \nn \\
 &+&\frac{8}{3m_b\Delta_0^3}\Big[2(m_b-v \cdot q){\hat \rho}_D^3+(m_b-2 v \cdot q){\hat \rho}_{LS}^3 \Big]
 \Bigg\} \hspace*{3cm} \label{T4} 
 \eea
 \bea
 T_5 &=& 2m_H \Bigg\{-\frac{2}{\Delta_0}+\frac{2}{3m_b\Delta_0^2}\Big[-4m_b  {\hat \mu}^2_\pi-5 v \cdot q \left( {\hat \mu}^2_\pi-{\hat \mu}^2_G \right)+4 \frac{v \cdot q}{m_b}\left( {\hat \rho}_D^3+ {\hat \rho}_{LS}^3 \right) \Big] \nn \\
 &+&\frac{8}{3\Delta_0^3}\Bigg[ \left[q^2-(v \cdot q)^2\right]{\hat \mu}^2_\pi + \left[-2m_b^2+m_b v \cdot q +(v \cdot q)^2\right]\frac{{\hat \rho}_D^3}{m_b}  \label{T5}\\
 &+&\left[-m_b^2+m_b v \cdot q +(v \cdot q)^2\right]\frac{{\hat \rho}_{LS}^3}{m_b} \Bigg]  
 +\frac{16}{3\Delta_0^4}{\hat \rho}_D^3 \left[q^2-(v \cdot q)^2\right](m_b- v \cdot q)
 \Bigg\} \nn
 \eea
 \bea
 S_1&=& 2m_H \Bigg\{-\frac{2}{\Delta_0}\left[1-\frac{7{\hat \mu}^2_\pi-9{\hat \mu}^2_G}{12 m_b^2}+\frac{{\hat \rho}_D^3}{6m_b^3}  \right] \nn \\
 &+&
 \frac{2}{3m_b\Delta_0^2}\left[-5v \cdot q \,  {\hat \mu}^2_\pi+3(v \cdot q -2m_b){\hat \mu}^2_G-4{\hat \rho}_D^3-3{\hat \rho}_{LS}^3 \right] \nn \\
 &+&\frac{8}{3\Delta_0^3}\left[[q^2-(v \cdot q)^2]\,  {\hat \mu}^2_\pi-v \cdot q \,(m_b-v \cdot q)\frac{{\hat \rho}_D^3}{m_b} \right]   \label{S1}\\
 &+& \frac{16}{3\Delta_0^4} {\hat \rho}_D^3(m_b-v \cdot q)\left[q^2-(v \cdot q)^2\right] \Bigg\}\nn
\eea
\bea
 S_2&=& 2m_H \Bigg\{\frac{2}{3\Delta_0^2}\left[4m_b {\hat \mu}^2_\pi-6m_b{\hat \mu}^2_G-8{\hat \rho_D^3}-9{\hat \rho}_{LS}^3 \right] \hspace*{2cm} \nn\\
 &+&  \frac{8}{3\Delta_0^3} \left[2(m_b-v \cdot q){\hat \rho}_D^3-3(v \cdot q) \, {\hat \rho}_{LS}^3 \right] \Bigg\}  \label{S2}
\eea
\bea
 S_3&=&- 2m_H \Bigg\{\frac{2}{3m_b\Delta_0^2}\Big[2{\hat \mu}^2_\pi+\frac{{\hat \rho}_D^3}{m_b} \Big]+\frac{4}{3m_b\Delta_0^3}\Big[2(m_b-v \cdot q){\hat \rho}_D^3 -3m_b {\hat \rho}_{LS}^3 \Big]\Bigg\}  \hspace*{2cm}
 \eea
with $S_4= 0$  and $S_5=S_3$,
 \bea
 S_6&=& 2m_H \Bigg\{\frac{1}{\Delta_0}\Big[-2m_b-\frac{1}{2m_b}\left( {\hat \mu}^2_\pi+{\hat \mu}^2_G \right)-\frac{{\hat \rho}_D^3}{3m_b^2} \Big]\nn \\
 &+&
 \frac{1}{3\Delta_0^2}\Big[-10v\cdot q \,{\hat \mu}^2_\pi-4(m_b+ v \cdot q) \frac{{\hat \rho}_D^3}{m_b}-9 v \cdot q \frac{{\hat \rho}_{LS}^3}{m_b}\Big] \nn \\
  &+&\frac{4}{3\Delta_0^3}\Big[2m_b[q^2-(v \cdot q)^2]\,  {\hat \mu}^2_\pi-2v \cdot q \,(m_b-v \cdot q){\hat \rho}_D^3+3[q^2-(v \cdot q)^2]\, {\hat \rho}_{LS}^3 \Big] \nn \\
  &+&
  \frac{16m_b}{3\Delta_0^4} [q^2-(v \cdot q)^2](m_b-v \cdot q){\hat \rho}_D^3 \Bigg\} \,   \label{S6}
 \eea
   \bea
 S_7&=& 2m_H \Bigg\{ \frac{2}{\Delta_0}\Big[1-\frac{7 {\hat \mu}^2_\pi-9{\hat \mu}^2_G}{12m_b^2} +\frac{{\hat \rho}_D^3}{6m_b^3} \Big] \nn \\
 &+&
 \frac{1}{3m_b\Delta_0^2}\Big[2(2m_b+3v \cdot q) {\hat \mu}^2_\pi+6(m_b-v \cdot q){\hat \mu}^2_G+2(2m_b-v \cdot q)\frac{{\hat \rho}_D^3}{m_b}+3 {\hat \rho}_{LS}^3 \Big]  \nn \\
 &+&\frac{8}{3\Delta_0^3}\Big[-[q^2-(v \cdot q)^2]{\hat \mu}^2_\pi+(m_b-v \cdot q){\hat \rho}_D^3 \Big]   \label{S7}\\
  &-&
  \frac{16}{3\Delta_0^4} [q^2-(v \cdot q)^2](m_b-v \cdot q){\hat \rho}_D^3 \Bigg\} \, \nn
 \eea
   \bea
 S_8&=& -2m_H \Bigg\{ -\frac{2m_b}{\Delta_0}\Big[1-\frac{5{\hat \mu}^2_\pi-3{\hat \mu}^2_G}{12m_b^2}-\frac{{\hat \rho}_D^3}{6m_b^3} \Big] \nn \\
  &+&
 \frac{1}{3\Delta_0^2}\Big[-10 v \cdot q \,{\hat \mu}^2_\pi-12(m_b-v \cdot q) \,{\hat \mu}^2_G-4(3m_b-2 v \cdot q)\frac{{\hat \rho}_D^3}{m_b}+9 v \cdot q \, \frac{{\hat \rho}_{LS}^3}{m_b} \Big] \hspace*{1,cm}\nn \\
  &+&\frac{4}{3\Delta_0^3}\Big[[q^2-(v \cdot q)^2] \,\left[2m_b{\hat \mu}^2_\pi-3{\hat \rho}_{LS}^3 \right]-2v \cdot q \,(m_b -v \cdot q){\hat \rho}_D^3 \Big]   \label{S8}\\
  &+& \frac{16}{3\Delta_0^4}{\hat \rho}_D^3 m_b (m_b -v \cdot q)[q^2-(v \cdot q)^2]
  \Bigg\} \nn
 \eea
   \bea
 S_9&=& -2m_H \Bigg\{ \frac{2}{\Delta_0}\Big[1-\frac{7{\hat \mu}^2_\pi-9{\hat \mu}^2_G}{12m_b^2}+\frac{{\hat \rho}_D^3}{6m_b^3} \Big] \nn \\
 &+&
 \frac{1}{3m_b\Delta_0^2}\Big[2(2m_b+3v \cdot q)\,{\hat \mu}^2_\pi+6(m_b - v \cdot q)\,{\hat \mu}^2_G-2v \cdot q \,\frac{{\hat \rho}_D^3}{m_b}-3{\hat \rho}_{LS}^3 \Big] \nn \\
  &+&\frac{8}{3\Delta_0^3}\Big[-[q^2-(v \cdot q)^2] \,{\hat \mu}^2_\pi +(m_b-v \cdot q){\hat \rho}_D^3\Big]  \label{S9}
  \\
  &-&\frac{16}{3\Delta_0^4}{\hat \rho}_D^3  (m_b -v \cdot q)[q^2-(v \cdot q)^2]\Bigg\} \nn
 \eea
 and $S_{10,11,12,13} = 0$.
  \item Scalar operator in $H_{eff}$
 \par
 This case amounts to choosing  $i=j=2$ in Eq.~\eqref{Tij-gen}.  $T_S$ is expanded as
 \be
 T_S=T_{S1} +(q \cdot s) S_{S1} \label{TS}
 \ee
 with
 \bea
 T_{S1}&=&2m_H\Bigg\{\frac{1}{\Delta_0} \Big[(m_b+m_U-v \cdot q)-\frac{m_b+m_U}{2m_b^2}\left( {\hat \mu}^2_\pi-{\hat \mu}^2_G \right) \Big] \nn \\
 &+&\frac{1}{3m_b\Delta_0^2} \Big[\Big(2[q^2-(v \cdot q)^2]+3v \cdot q\,(m_b+m_U-v \cdot q) \Big)
 \left( {\hat \mu}^2_\pi-{\hat \mu}^2_G \right) \nn \\
 &-&[q^2-4(v \cdot q)^2]\frac{{\hat \rho}_D^3+{\hat \rho}_{LS}^3}{m_b}\Big] \label{TS1} \\
 &+&\frac{4}{3\Delta_0^3} [q^2-(v \cdot q)^2]\Big[-(m_b+m_U-v \cdot q){\hat \mu}^2_\pi+(m_b-v \cdot q)\frac{{\hat \rho}_D^3}{m_b}-v \cdot q \,\frac{{\hat \rho}_{LS}^3}{m_b} \Big] \nn \\
  &-&\frac{8}{3\Delta_0^4} [q^2-(v \cdot q)^2](m_b-v \cdot q)(m_b+m_U-v \cdot q){\hat \rho}_D^3  \Bigg\} \nn
 \eea
and
$ S_{S1}=0$.
   \item Pseudoscalar operator in $H_{eff}$
 \par
 The tensor is obtained choosing  $i=j=3$ in Eq.~\eqref{Tij-gen}, and is expanded as
 \be
 T_P=T_{P1} +(q \cdot s) \, S_{P1} \,\,.\label{TP}
 \ee
The two functions in (\ref{TP}) are given by the corresponding ones in (\ref{TS})  replacing $m_U \to -m_U$.
\item Interference between the SM and the scalar operator in $H_{eff}$
 \par
The  tensor is obtained when $(i,j)=(1,2)$ and $(2,1)$ in Eq.~\eqref{Tij-gen}. We denote the two contributions as $T_{SMS}$ and $T_{SSM}$, respectively.
Using  the  expansion
  \bea
T_{SMS}^{\mu}&=&T_{SMS,1}\, v^\mu+T_{SMS,2} \, q^\mu\nn \\
  &-&(q \cdot s) \Big[S_{SMS,1} \, v^\mu+S_{SMS,2} \, q^\mu \Big]+S_{SMS,3}\,s^\mu+S_{SMS,4}\,i\,\epsilon^{\mu \alpha \beta \delta}q_\alpha v_\beta s_\delta   \hspace{1.5cm} \label{TSMS}
\eea
 and the analogous one for $T_{SSM}$,
we find:
 \bea
 T_{SMS,1} &=&2m_H\Bigg\{\frac{1}{\Delta_0}(m_b+m_U)\nn \\
 &-&\frac{v \cdot q}{3 m_b^2 \Delta_0^2}\Big[-5m_b(m_b+m_U){\hat \mu}^2_\pi+m_b(m_b+5m_U){\hat \mu}^2_G
 -4(m_b-m_U)\left( {\hat \rho}_D^3+ {\hat \rho}_{LS}^3 \right)\Big] \nn \\
 &-&\frac{4}{3 m_b \Delta_0^3}\Big[m_b(m_b+m_U) \left[q^2-(v \cdot q)^2\right]{\hat \mu}^2_\pi-(m_b+m_U)v \cdot q \,(m_b-v \cdot q){\hat \rho}_D^3 \nn \\
 &+&\left[m_b \left[q^2-(v \cdot q)^2 \right]+m_U (v \cdot q)^2 \right] {\hat \rho}_{LS}^3 \Big] \label{TSMS1} \\
 &-&\frac{8}{3  \Delta_0^4}{\hat \rho}_D^3(m_b+m_U)(m_b-v \cdot q) \left[q^2-(v \cdot q)^2\right]  \Bigg\} \nn
 \eea
 \bea
 T_{SMS,2} &=&2m_H\Bigg\{-\frac{1}{\Delta_0}\left[1-\frac{\left( {\hat \mu}^2_\pi-{\hat \mu}^2_G \right) }{2m_b^2} \right]
 \nn \\
 &-&\frac{1}{3 m_b^2 \Delta_0^2}\Big[m_b(2m_U+3 v \cdot q)\left( {\hat \mu}^2_\pi-{\hat \mu}^2_G \right) +2m_b^2\left( {\hat \mu}^2_\pi+{\hat \mu}^2_G \right)  \nn\\
 &+& (m_b-m_U)\left( {\hat \rho}_D^3+ {\hat \rho}_{LS}^3 \right) \Big]  \label{TSMS2}\\
 &-&\frac{4}{3 m_b \Delta_0^3}\Big[-m_b[q^2-(v \cdot q)^2]{\hat \mu}^2_\pi+{\hat \rho}_D^3(m_b+m_U)(m_b-v \cdot q)-m_U v\cdot q
  {\hat \rho}_{LS}^3 \Big]\nn \\
 &+&\frac{8}{3  \Delta_0^4}{\hat \rho}_D^3(m_b-v \cdot q)\left[q^2-(v \cdot q)^2\right]  \Bigg\}\nn
 \eea
 \bea
 S_{SMS,1}&=&2m_H\Bigg\{\frac{1}{\Delta_0}\left[1-\frac{\left(5 {\hat \mu}^2_\pi-3{\hat \mu}^2_G \right) }{12m_b^2}-\frac{{\hat \rho}_D^3}{6m_b^3} \right] \nn \\
 &+&\frac{1}{3 m_b^2 \Delta_0^2}\Big[m_b(2m_b+2m_U+5v \cdot q){\hat \mu}^2_\pi-6m_b v \cdot q \,{\hat \mu}^2_G \nn \\
 &+&(-m_b+m_U-4 v \cdot q){\hat \rho}_D^3-\frac{9}{2}(v \cdot q)\, \rho_{LS}^3 \Big]  \nn \\
 &+&\frac{2}{3 m_b \Delta_0^3}\Big[-2m_b[q^2-(v \cdot q)^2] {\hat \mu}^2_\pi \label{SMS1} \\
&+&2(m_b-v \cdot q)(m_b+m_U+v \cdot q) {\hat \rho}_D^3-3 (v \cdot q)^2  {\hat \rho}_{LS}^3 \Big]\nn \hspace*{3cm}\\
  &-&\frac{8}{3  \Delta_0^4}{\hat \rho}_D^3(m_b-v \cdot q)\left[q^2-(v \cdot q)^2\right]  \Bigg\}\nn 
\eea
\bea
 S_{SMS,2}&=&2m_H\Bigg\{\frac{1}{6m_b^2\Delta_0^2}\Big[-4m_b{\hat \mu}^2_\pi+6m_b{\hat \mu}^2_G+4{\hat \rho}_D^3+3 {\hat \rho}_{LS}^3 \Big] \nn  \hspace*{2cm} \\
 &+&\frac{2}{3 m_b \Delta_0^3}\Big[-2(m_b - v \cdot q) {\hat \rho}_D^3+3 v \cdot q {\hat \rho}_{LS}^3 \Big] \Bigg\}  \label{SMS2}
 \eea
 \bea
&& S_{SMS,3}=2m_H \Bigg\{\frac{1}{\Delta_0}\Big[-m_b-m_U+v \cdot q+\frac{7m_b+7m_U-5 v \cdot q}{12 m_b^2}{\hat \mu}^2_\pi \nn \\ 
 &&-\frac{3m_b+3m_U- v \cdot q}{4 m_b^2}{\hat \mu}^2_G-\frac{m_b+m_U+ v \cdot q}{6 m_b^3} {\hat \rho}_D^3 \Big] \nn \\
 &&+\frac{1}{3 m_b \Delta_0^2}\Big[-\left(2q^2+v \cdot q(3m_b+3m_U-5 v \cdot q)\right){\hat \mu}^2_\pi\nn\\
 &&+3 \left(q^2+v \cdot q (m_b+m_U-2 v \cdot q) \right) {\hat \mu}^2_G + \left(2 q^2+v \cdot q(-m_b+m_U-4 v \cdot q)\right) \frac{{\hat \rho}_D^3}{m_b} \nn \\
 &&+3[q^2-3 (v \cdot q)^2] \frac{{\hat \rho}_{LS}^3}{2m_b} \Big]  \label{SMS3} \\
&& +\frac{2}{3 m_b \Delta_0^3}[q^2-(v \cdot q)^2] \Big[2m_b(m_b+m_U-v \cdot q){\hat \mu}^2_\pi-2(m_b- v \cdot q) {\hat \rho}_D^3+3 v \cdot q {\hat \rho}_{LS}^3 \Big] \hspace*{1cm}\nn \\
&&+\frac{8}{3  \Delta_0^4}{\hat \rho}_D^3(m_b-v \cdot q)[q^2-(v \cdot q)^2] (m_b+m_U-v \cdot q) \Bigg\}\nn
 \eea
 \bea
&& S_{SMS,4} =2m_H\Bigg\{\frac{1}{\Delta_0}\left[1-\frac{\left(5 {\hat \mu}^2_\pi-3{\hat \mu}^2_G \right) }{12m_b^2}-\frac{{\hat \rho}_D^3}{6m_b^3} \right] \nn \\
 &&+\frac{1}{3 m_b^2 \Delta_0^2}
 \Big[5m_b v \cdot q{\hat \mu}^2_\pi+6m_b(m_b- v \cdot q ){\hat \mu}^2_G+2(3m_b-2 v \cdot q){\hat \rho}_D^3 +\frac{3}{2}(4m_b-3 v \cdot q){\hat \rho}_{LS}^3 \Big] \nn \\
 &&-\frac{2}{3 m_b \Delta_0^3}\Big[2m_b[q^2-(v \cdot q)^2] {\hat \mu}^2_\pi-2 v \cdot q (m_b - v \cdot q) {\hat \rho}_D^3+3 \left[(m_b- v \cdot q)^2+m_b m_U \right] {\hat \rho}_{LS}^3 \Big] \nn \\
 &&-\frac{8}{3  \Delta_0^4}{\hat \rho}_D^3(m_b-v \cdot q)[q^2-(v \cdot q)^2]  \Bigg\} \label{SMS4}
 \eea
 and 
$T_{SSM,i}=T_{SMS,i}$ (for $i=1,2),$
$ S_{SSM,i}=S_{SMS,i}$ (for $i=1,2,3),$ 
$S_{SSM,4}=-S_{SMS,4}$.
 
  \item Interference between the SM and the pseudoscalar operators in $H_{eff}$
  \par
 The  tensor is obtained for $(i,j)=(1,3)$ and $(i,j)=(3,1)$ in Eq.~\eqref{Tij-gen}. We denote the two contributions as $T_{SMP}$ and $T_{PSM}$, respectively, 
 with the  expansion
  \bea
T_{SMP}^{\mu}&=&T_{SMP,1} \, v^\mu+T_{SMP,2} \, q^\mu\nn \\
&-&(q \cdot s) \Big[S_{SMP,1}\, v^\mu+S_{SMP,2}\, q^\mu \Big]+S_{SMP,3}\, s^\mu+S_{SMP,4}\, \,i\,\epsilon^{\mu \alpha \beta \delta}q_\alpha v_\beta s_\delta \label{TSMP} \,\, ,
 \hspace{2cm}
 \eea
and the analogous one for $T_{PSM}$.
The  functions in (\ref{TSMP}) are given by the corresponding ones in (\ref{TSMS}) replacing  $m_U \to -m_U$.

\item Interference between the scalar and pseudoscalar operators in $H_{eff}$
\par
This case amounts to choosing $(i,j)=(2,3)$ and $(i,j)=(3,2)$ in Eq.~\eqref{Tij-gen}. We denote the two terms  as $T_{SP}$ and $T_{PS}$, respectively.
Writing
\be
T_{SP}=T_{SP,1}-(q \cdot s) \, S_{SP,1} \label{TSP}
\ee
and  analogously for $T_{PS}$, we have
$T_{SP,1}=T_{PS,1}=0$ and
\bea
S_{SP,1}&=&S_{PS,1}=2m_H\Bigg\{\frac{1}{\Delta_0}\left[1-m_b\frac{\left(7 {\hat \mu}^2_\pi-9{\hat \mu}^2_G \right) }{12m_b^2}+\frac{{\hat \rho}_D^3}{6m_b^3}\right] 
+\frac{v \cdot q}{3 m_b \Delta_0^2}\left(5 {\hat \mu}^2_\pi-3{\hat \mu}^2_G \right)
\nn \\
&+&\frac{4}{3 m_b \Delta_0^3}\left[-m_b[q^2-(v \cdot q)^2]{\hat \mu}^2_\pi+v \cdot q (m_b-v \cdot q){\hat \rho}_D^3 \right]  \label{SP1}\\
 &-&\frac{8}{3  \Delta_0^4}{\hat \rho}_D^3(m_b-v \cdot q)[q^2-(v \cdot q)^2]  \Bigg\} \,\,\, .\nn
\eea
  \item Tensor operator in $H_{eff}$
 \par
 This case amounts to choosing $i=j=4$ in Eq.~\eqref{Tij-gen}. The corresponding tensor $T_T$ can be expanded as:
  \bea
&&   T_T^{\mup \nup \mu \nu}=i\, \epsilon^{\mu \nu \mup \nup}\, [T_{T0}-(q \cdot s) S_{T0}]+ \Big[g^{\mu \mup}g^{\nu \nup}
  -g^{\mu \nup}g^{\nu \mup} \Big]\left[T_{T1}-(q \cdot s) S_{T1} \right] \nn \\
 &&+\Big\{-g^{\mu \mup} \Big[v^\nu v^{\nup} \left[T_{T2}-(q \cdot s) S_{T2} \right] \nn \\
&&-i \, \epsilon^{\nu \nup \alpha \beta} v_\alpha q_\beta \left[T_{T3}-(q \cdot s) S_{T3} \right]
  +q^\nu q^\nup \left[T_{T4}-(q \cdot s) S_{T4} \right] \nn \\
  &&+(q^\nu v^\nup +q^\nup v^\nu) \left[T_{T5}-(q \cdot s) S_{T5} \right] \Big] +\big(\mu \leftrightarrow \nu  \land \mup \leftrightarrow \nup \big)- \big(\mu \leftrightarrow \nu  \big) - \big(\mup \leftrightarrow \nup  \big) \Big\}
 \nn \\
 &&  +\Big\{i\, v^\mu \epsilon^{\alpha \nu \mup \nup} v_\alpha \, \left[T_{T6}-(q \cdot s) S_{T6} \right] +i\, q^\mu \epsilon^{\alpha \nu \mup \nup} v_\alpha \, \left[T_{T7}-(q \cdot s) S_{T7} \right]\nn \\
 && +i\, v^\mu \epsilon^{\alpha \nu \mup \nup} q_\alpha \, \left[T_{T8}-(q \cdot s)S_{T8} \right]
 +i\, q^\mu \epsilon^{\alpha \nu \mup \nup} q_\alpha \, \left[T_{T9}-(q \cdot s) S_{T9}\right] -
   \big(\mu \leftrightarrow \nu  \big) \Big\} \nn \\
&&  +\Big\{i\, v^\mup \epsilon^{\alpha \nup \mu \nu} v_\alpha \, \left[T_{T10}-(q \cdot s) S_{T10} \right]+i\, q^\mup \epsilon^{\alpha \nup \mu \nu} v_\alpha \, \left[T_{T11}-(q \cdot s) S_{T11}\right]\nn \\
&&  +i\, v^\mup \epsilon^{\alpha \nup \mu \nu} q_\alpha \, \left[T_{T12}-(q \cdot s) S_{T12} \right]
+i\, q^\mup \epsilon^{\alpha \nup \mu \nu}q_\alpha \, \left[T_{T13}-(q \cdot s)S_{T13} \right] 
-\big(\mup \leftrightarrow \nup  \big) \Big\}\nn \\
 &&+\Big\{
-g^{\mu \mup} \Big[(s^\nu v^\nup+s^\nup v^\nu)\, S_{T14}+(q^\nu s^\nup+q^\nup s^\nu)\, S_{T15}+i\, \epsilon^{\nu \nup \alpha \beta}v_\alpha s_\beta \, S_{T16}\nn \\ 
  &&+i\, \epsilon^{\nu \nup \alpha \beta}q_\alpha s_\beta \, S_{T17} 
  +(s^\nu v^\nup-s^\nup v^\nu)\, S_{T18}+(q^\nu s^\nup-q^\nup s^\nu)\, S_{T19} \nn \\
  &&+i v^\nu \, \epsilon^{\nup \alpha \beta \delta}q_\alpha v_\beta s_\delta \, S_{T20 A}+i v^\nup \, \epsilon^{\nu \alpha \beta \delta}q_\alpha v_\beta s_\delta \, S_{T20 B} \nn \\ 
   &&+i q^\nu \, \epsilon^{\nup \alpha \beta \delta}q_\alpha v_\beta s_\delta \, S_{T21 A}+i q^\nup \, \epsilon^{\nu \alpha \beta \delta}q_\alpha v_\beta s_\delta \, S_{T21 B} \Big]  \\ 
   &&
+v^\mu v^\mup \Big[(q \cdot s) \, i\,  \epsilon^{\nu \nup \alpha \beta}q_\alpha v_\beta \, S_{T22}+ i\, \epsilon^{\nu \nup \alpha \beta}v_\alpha s_\beta \, S_{T23}+i\, \epsilon^{\nu \nup \alpha \beta}q_\alpha s_\beta \, S_{T24}
 \nn \\ &&
 +i q^\nu \, \epsilon^{\nup \alpha \beta \delta}q_\alpha v_\beta s_\delta \, S_{T25 A}+i q^\nup \, \epsilon^{\nu \alpha \beta \delta}q_\alpha v_\beta s_\delta \, S_{T25 B} \Big] \nn \\ 
   &&
   -i\,\epsilon^{\mu \mup \alpha \beta} v_\alpha q_\beta\, v^\nu s^\nup \, S_{T26}
   +q^\mu v^\mup \, \Big[i \, \epsilon^{\nu \nup \alpha \beta} v_\alpha s_\beta \, S_{T27A} \Big]
    +q^\mup v^\mu \, \Big[i \, \epsilon^{\nu \nup \alpha \beta} v_\alpha s_\beta \, S_{T27B}
 \Big]
  \nn 
 \\ &&+\big(\mu \leftrightarrow \nu  \land \mup \leftrightarrow \nup \big)- \big(\mu \leftrightarrow \nu  \big) - \big(\mup \leftrightarrow \nup  \big) \Big\} \nn \\ 
   &&
   +i\, (q^\mu v^\nu-q^\nu v^\mu) \epsilon^{\mup \nup \alpha \beta}v_\alpha q_\beta \, T_{T14}+i\, (q^\mup v^\nup-q^\nup v^\mup) \epsilon^{\mu \nu \alpha \beta}v_\alpha q_\beta \, T_{T15} \nn \\ 
   &&
   +i\, (q^\mu v^\nu-q^\nu v^\mu) \epsilon^{\mup \nup \alpha \beta}v_\alpha s_\beta \, S_{T28}+i\, (q^\mup v^\nup-q^\nup v^\mup) \epsilon^{\mu \nu \alpha \beta}v_\alpha s_\beta \, S_{T29}
  \nn \\ 
   &&
  +\Big\{i\, s^\mu \epsilon^{\alpha \nu \mup \nup} v_\alpha \,S_{T30}
 +i\, s^\mu \epsilon^{\alpha \nu \mup \nup} q_\alpha \,S_{T31}
   +i\, v^\mu \epsilon^{\alpha \nu \mup \nup} s_\alpha \, S_{T32}+i\, q^\mu \epsilon^{\alpha \nu \mup \nup} s_\alpha \, S_{T33} \nn \\
&&-   \big(\mu \leftrightarrow \nu  \big) \Big\}
 \nn   \\
  &&+\Big\{i\, s^\mup \epsilon^{\alpha \nup \mu \nu} v_\alpha \, S_{T34}
   +i\, s^\mup \epsilon^{\alpha \nup \mu \nu} q_\alpha \, S_{T35}
 +i\, v^\mup \epsilon^{\alpha \nup \mu \nu}s_\alpha \, S_{T36}\nn \\
&&+i\, q^\mup \epsilon^{\alpha \nup \mu \nu}s_\alpha \, S_{T37}-\big(\mup \leftrightarrow \nup  \big) \Big\} \nn
  \eea
  The various functions are given by:
\bea
T_{T0}&=&2m_H\Bigg\{\frac{1}{\Delta_0}\Big[-2(m_b- v \cdot q)-\frac{5}{3m_b}\left({\hat \mu}_\pi^2-{\hat \mu}_G^2 \right)+\frac{4}{3m_b^2}\left({\hat \rho}_D^3+{\hat \rho}_{LS}^3 \right) \Big]\nn \\
&+&\frac{2}{3m_b\Delta_0^2}\Big[[-2q^2-3m_bv \cdot q +5( v \cdot q)^2]{\hat \mu}_\pi^2+[-3m_b^2+2q^2+6m_b v \cdot q -5 (v \cdot q)^2]{\hat \mu}_G^2 \nn \\
&+&\frac{-7 m_b(m_b- v \cdot q)+q^2-4 (v \cdot q)^2}{m_b}\left({\hat \rho}_D^3+{\hat \rho}_{LS}^3 \right)+3 m_b {\hat \rho}_{LS}^3 \Big] \label{ST0}\\
&+&\frac{8}{3m_b\Delta_0^3}(m_b- v \cdot q)[q^2-(v \cdot q)^2]\big[m_b{\hat \mu}_\pi^2-\left({\hat \rho}_D^3+{\hat \rho}_{LS}^3 \right)\big] \nn\\
&+&\frac{16}{3\Delta_0^4}(m_b- v \cdot q)^2[q^2-(v \cdot q)^2]{\hat \rho}_D^3 \Bigg\}
\nn
\eea
\bea
S_{T0}&=&2m_H\Bigg\{\frac{2}{\Delta_0}\Big[1-\frac{\left(7 {\hat \mu}^2_\pi-9{\hat \mu}^2_G \right)}{12m_b^2}+\frac{{\hat \rho}_D^3}{6m_b^3} \Big]\nn \\
&+&\frac{2}{3m_b\Delta_0^2}\Big[5 v \cdot q \,{\hat \mu}_\pi^2 +3(m_b - v \cdot q){\hat \mu}_G^2 + {\hat \rho}_D^3 \Big] \nn \\
&+&\frac{8}{3m_b\Delta_0^3}\Big[-m_b[q^2-(v \cdot q)^2] \,{\hat \mu}_\pi^2+v \cdot q (m_b - v \cdot q) {\hat \rho}_D^3 \Big]  \hspace*{2cm}\label{SS0}\\
&-&\frac{16}{3\Delta_0^4}(m_b- v \cdot q)[q^2-(v \cdot q)^2]{\hat \rho}_D^3 \Bigg\} \nn
\eea
\bea
  T_{T1}&=&2m_H\Bigg\{\frac{1}{\Delta_0}
   \Big[2(m_b-v \cdot q)+\frac{5}{3m_b}\left( {\hat \mu}^2_\pi-{\hat \mu}^2_G \right)-\frac{4}{3m_b^2}\left( {\hat \rho}_D^3+ {\hat \rho}_{LS}^3 \right) \Big] \nn \\
   &+&\frac{2}{3m_b\Delta_0^2}\Big[[2q^2+(3m_b-5v \cdot q)v \cdot q ]{\hat \mu}^2_\pi+[4m_b^2-2q^2-7m_b v \cdot q+5 (v \cdot q)^2]{\hat \mu}^2_G
   \nn \\
   &+&4m_b{\hat \rho}_D^3+\frac{4(m_b-v \cdot q)^2-q^2}{m_b}\left( {\hat \rho}_D^3+ {\hat \rho}_{LS}^3 \right) \Big]  \label{TT1}\\
   &-&\frac{8}{3m_b\Delta_0^3}[q^2-(v \cdot q)^2]\Big[(m_b - v \cdot q)\left(m_b {\hat \mu}^2_\pi-{\hat \rho}_D^3 \right)-(2m_b-v \cdot q) {\hat \rho}_{LS}^3 \Big] \nn \\
   &-&\frac{16}{3\Delta_0^4}(m_b- v \cdot q)^2[q^2-(v \cdot q)^2]{\hat \rho}_D^3 \Bigg\} \nn
   \eea
  \bea
T_{T2}&=&2m_H\Bigg\{\frac{2}{\Delta_0} \Big[2m_b+\frac{5}{3m_b}\left( {\hat \mu}^2_\pi-{\hat \mu}^2_G \right)-\frac{4}{3m_b^2}\left( {\hat \rho}_D^3+ {\hat \rho}_{LS}^3 \right) \Big] \nn \\
&+&\frac{4}{3m_b\Delta_0^2}\Big[7m_b (v \cdot q){\hat \mu}^2_\pi+m_b(4m_b-5 v \cdot q) {\hat \mu}^2_G\nn \\
&+&2(4m_b-3v \cdot q) {\hat \rho}_D^3+2(2m_b-3 v \cdot q){\hat \rho}_{LS}^3  \Big]  \nn \\
&+&\frac{8}{3\Delta_0^3}\Big[-2m_b[q^2-(v \cdot q)^2]{\hat \mu}^2_\pi+4v \cdot q (m_b-v \cdot q){\hat \rho}_D^3  \label{TT2} \\    
&+&[q^2+2v \cdot q (m_b-v \cdot q) ]{ \rho}_{LS}^3  \Big] \nn \\
      &-&\frac{32}{3\Delta_0^4}m_b(m_b- v \cdot q)[q^2-(v \cdot q)^2]{\hat \rho}_D^3 \Bigg\} \nn
\eea
\bea
   T_{T3}&=&2m_H\Bigg\{-\frac{2}{\Delta_0} 
  -\frac{2}{3m_b^2\Delta_0^2}\Big[5m_b v \cdot q\left( {\hat \mu}^2_\pi-{\hat \mu}^2_G \right)+6 m_b^2{\hat \mu}^2_G\nn \\
&+&2(3m_b-2 v \cdot q)\left( {\hat \rho}_D^3+ {\hat \rho}_{LS}^3 \right)\Big] \nn \\
   &+&\frac{8}{3m_b\Delta_0^3}\Big[m_b [q^2-(v \cdot q)^2]{\hat \mu}^2_\pi-v \cdot q(m_b - v \cdot q){\hat \rho}_D^3+
   (m_b - v \cdot q)^2{ \rho}_{LS}^3  \Big]  \hspace*{2cm} \nn \\
      &+&\frac{16}{3\Delta_0^4}(m_b- v \cdot q)[q^2-(v \cdot q)^2]{\hat \rho}_D^3 \Bigg\}  \label{TT3}
\eea
\bea
   T_{T4}&=&2m_H\Bigg\{
  \frac{4}{3m_b^2\Delta_0^2}\Big[2m_b\left( {\hat \mu}^2_\pi-{\hat \mu}^2_G \right)-\left( {\hat \rho}_D^3+ {\hat \rho}_{LS}^3 \right)\Big] \nn \\
  &+&
  \frac{8}{3m_b\Delta_0^3}\Big[2(m_b - v \cdot q){\hat \rho}_D^3 +(3m_b -2 v \cdot q){ \rho}_{LS}^3  \Big]  \Bigg\} \label{TT4}
 \\ \nn \\
   T_{T5}&=&2m_H\Bigg\{-\frac{2}{\Delta_0} 
  -\frac{2}{3m_b^2\Delta_0^2}\Big[m_b(4m_b+5 v \cdot q){\hat \mu}^2_\pi+m_b(2m_b-5 v \cdot q)
  {\hat \mu}^2_G  \nn \\
  &+&2(m_b-2 v \cdot q)\left( {\hat \rho}_D^3+
   {\hat \rho}_{LS}^3 \right)\Big] \nn \\
    &+&\frac{8}{3m_b\Delta_0^3}\Big[m_b [q^2-(v \cdot q)^2]{\hat \mu}^2_\pi+[-2m_b^2+m_b v \cdot q +(v \cdot q)^2]{\hat \rho}_D^3 \nn \\
    &+&[-m_b^2-m_b v \cdot q +(v \cdot q)^2]{\hat \rho}_{LS}^3 \Big] \label{TT5} \\
          &+&\frac{16}{3\Delta_0^4}(m_b- v \cdot q)[q^2-(v \cdot q)^2]{\hat \rho}_D^3 \Bigg\} \nn
 \eea
 and
 \bea
   S_{T1}&=&2m_H\Bigg\{ -\frac{2}{\Delta_0} \Big[1-\frac{\left(7 {\hat \mu}^2_\pi-9{\hat \mu}^2_G \right)}{12m_b^2}+\frac{{\hat \rho}_D^3}{6m_b^3} \Big] \nn \\
  & +& \frac{2}{3m_b\Delta_0^2}\Big[- v \cdot q \left(5  {\hat \mu}^2_\pi-3{\hat \mu}^2_G \right)+2\left(2{\hat \rho}_D^3+3{\hat \rho}_{LS}^3\right) \Big] \nn \\
   &+&\frac{8}{3m_b\Delta_0^3}\Big[m_b [q^2-(v \cdot q)^2]{\hat \mu}^2_\pi- v \cdot q(m_b- v \cdot q){\hat \rho}_D^3 \Big] \hspace*{2cm} \label{ST1}\\
   &+&\frac{16}{3\Delta_0^4}(m_b- v \cdot q)[q^2-(v \cdot q)^2]{\hat \rho}_D^3 \Bigg\} \nn \\
\nn \\
   S_{T2}&=&2m_H\Bigg\{ \frac{2}{3m_b\Delta_0^2}\Big[2m_b\left(2 {\hat \mu}^2_\pi+3{\hat \mu}^2_G \right)+8 {\hat \rho}_D^3+9
   {\hat \rho}_{LS}^3  \Big] \nn \\
  & +&\frac{8}{3\Delta_0^3}\Big[2(m_b-v \cdot q){\hat \rho}_D^3 +3 v \cdot q {\hat \rho}_{LS}^3 \Big] \Bigg\} \label{ST2}
\eea
\be
   S_{T3}=2m_H\Bigg\{ -\frac{2}{3m_b^2\Delta_0^2}\left[2m_b {\hat \mu}^2_\pi+{\hat \rho}_D^3 \right]-\frac{4}{3m_b\Delta_0^3}\Big[
  2 (m_b- v \cdot q){\hat \rho}_D^3-3m_b {\hat \rho}_{LS}^3 \Big] \Bigg\} \hspace*{1cm} \label{ST3}
\ee
with $S_{T4}=0$,
\bea
   S_{T5}=2m_H\Bigg\{ -\frac{2}{3m_b^2\Delta_0^2}\left[2m_b {\hat \mu}^2_\pi+{\hat \rho}_D^3 \right]
   -\frac{4}{3m_b\Delta_0^3}
  \left[ 2(m_b- v \cdot q){\hat \rho}_D^3+3m_b {\hat \rho}_{LS}^3\right]\Bigg\} \hspace*{1.3cm}\label{ST5}
\eea
\bea
 T_{T6}&=&2m_H \Bigg\{\frac{4m_b}{\Delta_0}\Big[1+\frac{5}{6m_b^2}\left({\hat \mu}_\pi^2-{\hat \mu}_G^2 \right)-\frac{2}{3m_b^3}\left({\hat \rho}_D^3+{\hat \rho}_{LS}^3 \right) \nn \\
 &+& \frac{2}{3m_b\Delta_0^2}\Big[14m_b (v \cdot q) {\hat \mu}_\pi^2+m_b(7m_b-10 v \cdot q){\hat \mu}_G^2 \nn \\
& +&
 3(5m_b-4v \cdot q){\hat \rho}_D^3+4(2m_b-3 v \cdot q) {\hat \rho}_{LS}^3 \Big] \label{TT6} \\
  &-&\frac{16}{3\Delta_0^3}\Big[m_b[q^2-(v \cdot q)^2] \,{\hat \mu}_\pi^2-v \cdot q \,(m_b - v \cdot q)\left(2 {\hat \rho}_D^3 +
  {\hat \rho}_{LS}^3 \right) \Big] \nn \\
  &-&\frac{32}{3\Delta_0^4}m_b(m_b- v \cdot q)[q^2-(v \cdot q)^2]{\hat \rho}_D^3 \Big\} \nn
\eea
\bea
  S_{T6}&=&2m_H \Bigg\{\frac{8}{3\Delta_0^2} {\hat \mu}_\pi^2+\frac{16}{3\Delta_0^3}(m_b - v \cdot q){\hat \rho}_D^3 \Bigg\} \hspace*{2cm} \label{ST6}
\eea
  \bea
  T_{T7}&=&2m_H \Bigg\{-\frac{2}{\Delta_0}
  \nn \\
  &-&\frac{2}{3m_b^2\Delta_0^2}\Big[m_b(4m_b+5 v \cdot q){\hat \mu}_\pi^2
  +m_b (2m_b-5 v \cdot q){\hat \mu}_G^2 +2(m_b-2 v \cdot q)\left({\hat \rho}_D^3+{\hat \rho}_{LS}^3 \right) \Big]\nn \\
  &+&\frac{8}{3m_b\Delta_0^3}\Big[m_b[q^2-(v \cdot q)^2] \,{\hat \mu}_\pi^2+[-2m_b^2+m_b v \cdot q +(v \cdot q)^2]{\hat \rho}_D^3-[m_b^2-(v \cdot q)^2]{\hat \rho}_{LS}^3 \Big] \nn \\
 &+&\frac{16}{3\Delta_0^4}(m_b- v \cdot q)[q^2-(v \cdot q)^2]{\hat \rho}_D^3 \Bigg\} \label{TT7}
 \eea
\bea
 S_{T7}&=& S_{T8}= -S_{T11}=S_{T12}=2m_H \Bigg\{-\frac{2}{3m_b^2\Delta_0^2}\big(2m_b {\hat \mu}_\pi^2+{\hat \rho}_D^3 \big)-\frac{8}{3m_b\Delta_0^3}(m_b - v \cdot q){\hat \rho}_D^3\Bigg\} \nn \\ \label{ST7}
\eea
\bea
  T_{T8}&=&2m_H \Bigg\{-\frac{2}{\Delta_0}\nn \\
  &-&\frac{2}{3m_b^2\Delta_0^2}\Big[m_b (4m_b+5 v \cdot q) {\hat \mu}_\pi^2+m_b(m_b-5 v \cdot q){\hat \mu}_G^2 +(m_b-4 v \cdot q)\left({\hat \rho}_D^3+{\hat \rho}_{LS}^3 \right) \Big]\nn \\
   &+&\frac{8}{3m_b\Delta_0^3}\Big[m_b[q^2-(v \cdot q)^2] \,{\hat \mu}_\pi^2
  +[-2m_b^2+m_b v \cdot q +(v \cdot q)^2]{\hat \rho}_D^3-[m_b^2 -( v \cdot q)^2] \,{\hat \rho}_{LS}^3 \Big]\nn \\
   &+&\frac{16}{3\Delta_0^4}(m_b- v \cdot q)[q^2-(v \cdot q)^2]{\hat \rho}_D^3 \Bigg\} \label{TT8}
    \eea
    \bea
 T_{T9}&=&2m_H \Bigg\{\frac{4}{3m_b^2\Delta_0^2}\Big[2m_b\left({\hat \mu}_\pi^2-{\hat \mu}_G^2 \right)-\left({\hat \rho}_D^3+{\hat \rho}_{LS}^3 \right) \Big]
  +\frac{16}{3m_b\Delta_0^3}(m_b - v \cdot q)\left({\hat \rho}_D^3+{\hat \rho}_{LS}^3 \right) \Bigg\}
 \nn \\ \label{TT9}
 \eea
\bea
 T_{T10}&=& 2m_H \Bigg\{-\frac{2}{3 \Delta_0^2} \big(m_b {\hat \mu}_G^2+{\hat \rho}_D^3\big) \Bigg\}  \hspace*{6cm} \label{TT10} 
\eea
 \bea
  T_{T11}&=& 2m_H \Bigg\{\frac{2}{\Delta_0}+
  \frac{2}{3m_b^2\Delta_0^2}\Big[5m_b (v \cdot q) \left({\hat \mu}_\pi^2-{\hat \mu}_G^2 \right)+6m_b^2 {\hat \mu}_G^2\nn \\
&+&2(3m_b-2 v \cdot q)\left({\hat \rho}_D^3+{\hat \rho}_{LS}^3 \right) \Big]\nn \\
   &-&\frac{8}{3m_b\Delta_0^3}\Big[m_b[q^2-(v \cdot q)^2] \,{\hat \mu}_\pi^2-v \cdot q (m_b - v \cdot q){\hat \rho}_D^3+(m_b - v \cdot q)^2 \,{\hat \rho}_{LS}^3 \Big]\nn \\
   &-&\frac{16}{3\Delta_0^4}(m_b- v \cdot q)[q^2-(v \cdot q)^2]{\hat \rho}_D^3 \Bigg\} \label{TT11}
  \eea
    \bea
    T_{T12}&=&2m_H \Bigg\{-\frac{2}{\Delta_0}
  \nn \\
  &-&\frac{2}{3m_b^2\Delta_0^2}\Big[5 m_b v \cdot q \left({\hat \mu}_\pi^2-{\hat \mu}_G^2 \right) +5 m_b^2{\hat \mu}_G^2 +(5m_b-4 v \cdot q)\left({\hat \rho}_D^3+{\hat \rho}_{LS}^3 \right) \Big]\nn \\
   &+&\frac{8}{3m_b\Delta_0^3}\Big[m_b[q^2-(v \cdot q)^2] \,{\hat \mu}_\pi^2
  -v \cdot q (m_b -v \cdot q){\hat \rho}_D^3+(m_b- v \cdot q)^2 \,{\hat \rho}_{LS}^3 \Big]\nn \\
   &+&\frac{16}{3\Delta_0^4}(m_b- v \cdot q)[q^2-(v \cdot q)^2]{\hat \rho}_D^3 \Bigg\} \label{TT12}
\eea
with
$ T_{T13} = S_{T9} = S_{T10} = S_{T13} = 0$,
\be  T_{T14}=-T_{T15}=-\frac{16m_H}{3\Delta_0^3}{\hat \rho}_{LS}^3  \label{TT14}
\ee 
\bea
   S_{T14}&=&2m_H\Bigg\{-\frac{2m_b}{\Delta_0} \Big[1+\frac{\left({\hat \mu}^2_\pi+{\hat \mu}^2_G \right)}{4m_b^2}+\frac{{\hat \rho}_D^3}{6m_b^3} \Big] \nn \\
   &-&
    \frac{1}{3\Delta_0^2}\Big[10 v \cdot q \,{\hat \mu}^2_\pi+12(m_b-v \cdot q) {\hat \mu}^2_G+4\frac{(4m_b-3v \cdot q)}{m_b}{\hat \rho}_D^3-9\frac{v \cdot q}{m_b}{\hat \rho}_{LS}^3 \Big] \nn \\
    &+&\frac{4}{3\Delta_0^3}\Big[[q^2-(v \cdot q)^2](2m_b{\hat \mu}^2_\pi -3{\hat \rho}_{LS}^3)-2v \cdot q (m_b-v \cdot q){\hat \rho}_D^3 \Big] \nn \\
    &+&\frac{16}{3\Delta_0^4}m_b(m_b- v \cdot q)[q^2-(v \cdot q)^2]{\hat \rho}_D^3 \Bigg\}  \label{ST14}
   \eea
\bea
   S_{T15}&=&2m_H\Bigg\{\frac{2}{\Delta_0} \Big[1-\frac{7{\hat \mu}^2_\pi-9{\hat \mu}^2_G}{12m_b^2}+\frac{{\hat \rho}_D^3}{6m_b^3} \Big] \nn \\
   &+&
    \frac{1}{3m_b^2\Delta_0^2}\Big[2m_b(2m_b+3 v \cdot q){\hat \mu}^2_\pi+6m_b(m_b-v \cdot q){\hat \mu}^2_G 
    -2v \cdot q \,{\hat \rho}_D^3-3m_b
   {\hat \rho}_{LS}^3  \Big] \nn \\
   &+&\frac{8}{3\Delta_0^3}\Big[- [q^2-(v \cdot q)^2]{\hat \mu}^2_\pi+(m_b - v \cdot q){\hat \rho}_D^3 \Big] \label{ST15} \\
    &-&\frac{16}{3\Delta_0^4}(m_b- v \cdot q)[q^2-(v \cdot q)^2]{\hat \rho}_D^3 \Bigg\} \nn
   \eea
\bea
   S_{T16}&=&2m_H\Bigg\{-\frac{2m_b}{\Delta_0} \Big[1-\frac{\left(5{\hat \mu}^2_\pi-3{\hat \mu}^2_G \right)}{12m_b^2}-
   \frac{{\hat \rho}_D^3}{6m_b^3} \Big] \nn \\
 &+&
    \frac{2}{3m_b\Delta_0^2}\Big[-5m_b v \cdot q {\hat \mu}^2_\pi-6m_b(m_b-v \cdot q){\hat \mu}^2_G
    -2(3m_b-2v \cdot q){\hat \rho}_D^3
   +\frac{9}{2}v \cdot q \,{\hat \rho}_{LS}^3  \Big] \nn \\
  &+&\frac{4}{3\Delta_0^3}\Big[ [q^2-(v \cdot q)^2](2m_b{\hat \mu}^2_\pi-3{\hat \rho}_{LS}^3)-2v \cdot q(m_b - v \cdot q){\hat \rho}_D^3   \Big] \nn \\
    &+&\frac{16}{3\Delta_0^4}m_b(m_b- v \cdot q)[q^2-(v \cdot q)^2]{\hat \rho}_D^3 \Bigg\}  \label{ST16}
   \eea
\bea
   S_{T17}&=&2m_H\Bigg\{\frac{2}{\Delta_0} \Big[1-\frac{\left(7{\hat \mu}^2_\pi-9{\hat \mu}^2_G \right)}{12m_b^2}+
   \frac{{\hat \rho}_D^3}{6m_b^3} \Big] \nn \\
 &+&
    \frac{2}{3m_b^2\Delta_0^2}\Big[m_b(2m_b +3v \cdot q ){\hat \mu}^2_\pi+3m_b(m_b-v \cdot q){\hat \mu}^2_G
    -v \cdot q{\hat \rho}_D^3-\frac{3}{2}m_b{\hat \rho}_{LS}^3
  \Big]  \nn \\
  &-&\frac{8}{3\Delta_0^3}\Big[[q^2-(v \cdot q)^2]{\hat \mu}^2_\pi-(m_b - v \cdot q){\hat \rho}_D^3 \Big]
    -\frac{16}{3\Delta_0^4}(m_b- v \cdot q)[q^2-(v \cdot q)^2]{\hat \rho}_D^3 \Bigg\} \nn \\ \label{ST17}
     \eea
\bea
S_{T22}&=&-16m_H \frac{1}{\Delta_0^3}{\hat \rho}_{LS}^3  \label{ST22}\\
S_{T23}&=& 2m_H\Bigg\{-\frac{4}{\Delta_0^2}\Big(m_b {\hat \mu}^2_G+{\hat \rho}_D^3 \Big)-\frac{8}{\Delta_0^3}q^2 {\hat \rho}_{LS}^3  \Bigg\} \hspace*{2cm} \label{ST23} \\
S_{T24}&=&16m_H\frac{1}{\Delta_0^3}(v \cdot q) {\hat \rho}_{LS}^3    \label{ST24}
\eea
with  
$S_{T18}=S_{T19}=S_{T20A}= S_{T20B}=S_{T21A}=S_{T21B}=S_{T25A}=S_{T25B}=S_{T26}=0$,
\be
S_{T27A}=S_{T27B}=\frac{4m_H}{\Delta_0^2}\Bigg\{{\hat \mu}_G^2+\frac{\left({\hat \rho}_D^3+{\hat \rho}_{LS}^3 \right)}{m_b} \Bigg\} \label{ST27A}
\ee
 \be
 S_{T28}=-S_{T29}=\frac{4m_H}{3m_b \Delta_0^2}\Bigg\{3m_b{\hat \mu}_G^2+5 {\hat \rho}_D^3+6 {\hat \rho}_{LS}^3 \Bigg\} \label{ST28}
\ee
\bea
S_{T30}&=&S_{T32}=2m_H \Bigg\{-\frac{2m_b}{\Delta_0}\Big[1+\frac{1}{4m_b^2}\left({\hat \mu}_\pi^2+{\hat \mu}_G^2 \right)+\frac{1}{6m_b^3}{\hat \rho}_D^3 \Big] \nn \\
&-& \frac{2}{3m_b\Delta_0^2}\Big[5m_b v \cdot q {\hat \mu}_\pi^2  +3m_b(m_b -v \cdot q){\hat \mu}_G^2+(5m_b-2 v \cdot q) 
{\hat \rho}_D^3 \Big] \nn \\
  &+&\frac{8}{3\Delta_0^3}\Big[m_b[q^2-(v \cdot q)^2] \,{\hat \mu}_\pi^2-v \cdot q (m_b - v \cdot q){\hat \rho}_D^3 \Big] \label{ST30}\\
   &+&\frac{16}{3\Delta_0^4}m_b(m_b- v \cdot q)[q^2-(v \cdot q)^2]{\hat \rho}_D^3 \Bigg\}\nn
   \eea
 \bea
 S_{T31}&=&S_{T33}=S_{T35}= -S_{T37}=2m_H \Bigg\{\frac{2}{\Delta_0}\Big[1-\frac{1}{12m_b^2}\left(7{\hat \mu}_\pi^2-9{\hat \mu}_G^2 \right)+\frac{1}{6m_b^3}{\hat \rho}_D^3 \Big] \nn \\
 &+& \frac{2}{3m_b^2\Delta_0^2}\Big[m_b (2m_b+3v \cdot q) {\hat \mu}_\pi^2 
 +3m_b(m_b -v \cdot q){\hat \mu}_G^2
 +(m_b-v \cdot q){\hat \rho}_D^3 \Big] \nn \\
  &-&\frac{8}{3\Delta_0^3}\Big[[q^2-(v \cdot q)^2] \,{\hat \mu}_\pi^2- (m_b - v \cdot q){\hat \rho}_D^3 \Big] \label{ST31}\\
   &-&\frac{16}{3\Delta_0^4}(m_b- v \cdot q)[q^2-(v \cdot q)^2]{\hat \rho}_D^3 \Bigg\} \nn
   \eea
  \bea
S_{T34}&=&-S_{T36}=2m_H \Bigg\{-\frac{2m_b}{\Delta_0}\Big[1-\frac{1}{12m_b^2}\left(5{\hat \mu}_\pi^2-3{\hat \mu}_G^2 \right)-\frac{1}{6m_b^3}{\hat \rho}_D^3 \Big] \nn \\
&-& \frac{2}{3m_b\Delta_0^2}\Big[5m_b v \cdot q {\hat \mu}_\pi^2 +3m_b(m_b -v \cdot q){\hat \mu}_G^2+3m_b 
{\hat \rho}_D^3 \Big] \nn \\
  &+&\frac{8}{3\Delta_0^3}\Big[m_b[q^2-(v \cdot q)^2] \,{\hat \mu}_\pi^2-v \cdot q (m_b - v \cdot q){\hat \rho}_D^3 \Big] \label{ST34} \\
   &+&\frac{16}{3\Delta_0^4}m_b(m_b- v \cdot q)[q^2-(v \cdot q)^2]{\hat \rho}_D^3 \Bigg\} \nn
   \eea

 \item Interference between the SM and the tensor operators in $H_{eff}$
\par
The  tensor is obtained for $(i,j)=(1,4)$ and $(i,j)=(4,1)$ in Eq.~\eqref{Tij-gen}. We denote the two contributions with $T_{SMT}$ and $T_{TSM}$, respectively.
We write:
\begin{equation}
\label{TSMT}
\begin{split}
T_{\sm T}^{\alpha\mu\nu} & =i\, (g^{\alpha\mu} \, v^\nu - g^{\alpha\nu} \, v^\mu) \, T_{\sm T,1} + i\,(g^{\alpha\mu} \, q^\nu - g^{\alpha\nu} \, q^\mu) \, T_{\sm T,2} + \epsilon^{\alpha\mu\nu\beta} \, v_\beta \, T_{\sm T,3}  \\
&  + \epsilon^{\alpha\mu\nu\beta} \, q_\beta \, T_{\sm T,4} +i\, v^\alpha \, ( q^\mu \, v^\nu - q^\nu \, v^\mu) \, T_{\sm T,5} + i\,q^\alpha \, ( q^\mu \, v^\nu - q^\nu \, v^\mu) \, T_{\sm T,6}  \\
&  + v^\alpha \, \epsilon^{\mu\nu\beta\delta} \, v_\beta \, q_\delta \, T_{\sm T,7} + q^\alpha \, \epsilon^{\mu\nu\beta\delta} \, v_\beta \, q_\delta \, T_{\sm T,8}  \\
&  - (q \cdot s) \Big[ i\,(g^{\alpha\mu} \, v^\nu - g^{\alpha\mu} \, v^\mu) \, S_{\sm T,1} + i\,(g^{\alpha\mu} \, q^\nu - g^{\alpha\mu} \, q^\mu) \, S_{\sm T,2} + \epsilon^{\alpha\mu\nu\beta} \, v_\beta \, S_{\sm T,3}  \\
&+ \epsilon^{\alpha\mu\nu\beta} \, q_\beta \, S_{\sm T,4} + i\,v^\alpha \, ( q^\mu \, v^\nu - q^\nu \, v^\mu) \, S_{\sm T,5} +i\, q^\alpha \, ( q^\mu \, v^\nu - q^\nu \, v^\mu) \, S_{\sm T,6}  \\
&+ v^\alpha \, \epsilon^{\mu\nu\beta\delta} \, v_\beta \, q_\delta \, S_{\sm T,7} + q^\alpha \, \epsilon^{\mu\nu\beta\delta} \, v_\beta \, q_\delta \, S_{\sm T,8} \Big]  \\
&+i\, (g^{\alpha\mu} \, s^\nu - g^{\alpha\mu} \, s^\mu) \, S_{\sm T,9} + \epsilon^{\alpha\mu\nu\beta} \, s_\beta \, S_{\sm T,10} +i\, s^\alpha \, (q^\mu \, v^\nu - q^\nu \, v^\mu) \, S_{\sm T,11}  \\
&  + i\,v^\alpha \, (v^\mu \, s^\nu - v^\nu \, s^\mu) \, S_{\sm T,12} + i\,v^\alpha \, (q^\mu \, s^\nu - q^\nu \, s^\mu) \, S_{\sm T,13} \\
&+i\, q^\alpha \, (v^\mu \, s^\nu - v^\nu \, s^\mu) \, S_{\sm T,14}  
 + i\,q^\alpha \, (q^\mu \, s^\nu - q^\nu \, s^\mu) \, S_{\sm T,15} \\
&+ v^\alpha \, \epsilon^{\mu\nu\beta\delta} \, v_\beta \, s_\delta \, S_{\sm T,16} + v^\alpha \, \epsilon^{\mu\nu\beta\delta} \, q_\beta \, s_\delta \, S_{\sm T,17}  \\
&  + q^\alpha \, \epsilon^{\mu\nu\beta\delta} \, v_\beta \, s_\delta \, S_{\sm T,18} + q^\alpha \, \epsilon^{\mu\nu\beta\delta} \, q_\beta \, s_\delta \, S_{\sm T,19}  \\
& + (g^{\alpha\mu} \, \epsilon^{\nu\beta\delta\tau} \, q_\beta \, v_\delta \, s_\tau - g^{\alpha\nu} \, \epsilon^{\mu\beta\delta\tau} \, q_\beta \, v_\delta \, s_\tau) \, S_{\sm T,20}  \\
&  + (v^\mu \, \epsilon^{\alpha\nu\beta\delta} \, v_\beta \, s_\delta - v^\nu \, \epsilon^{\alpha\mu\beta\delta} \, v_\beta \, s_\delta) \, S_{\sm T,21} + (v^\mu \, \epsilon^{\alpha\nu\beta\delta} \, q_\beta \, s_\delta - v^\nu \, \epsilon^{\alpha\mu\beta\delta} \, q_\beta \, s_\delta) \, S_{\sm T,22}  \\
&  + (q^\mu \, \epsilon^{\alpha\nu\beta\delta} \, v_\beta \, s_\delta - q^\nu \, \epsilon^{\alpha\mu\beta\delta} \, v_\beta \, s_\delta) \, S_{\sm T,23} + (q^\mu \, \epsilon^{\alpha\nu\beta\delta} \, q_\beta \, s_\delta - q^\nu \, \epsilon^{\alpha\mu\beta\delta} \, q_\beta \, s_\delta) \, S_{\sm T,24}  \\
& + s^\alpha \, \epsilon^{\mu\nu\beta\delta} \,q_\beta \, v_\delta \, S_{\sm T,25}+(q^\mu v^\nu-q^\nu v^\mu) \, \epsilon_{\alpha \beta \delta \theta} q^\beta v^\delta s^\theta \, S_{SMT,26}
\end{split}
\end{equation}
The results for the various functions read:
\bea
T_{\sm T,1}& =&-T_{\sm T,3}  \nn \\
&=&2 m_H \,  \Bigg\{ \frac{2 m_U}{\Delta_0} + \frac{2 m_U}{3 m_b^2 \, \Delta_0^2} \, \Big[ 5 m_b \, v \cdot q \, (\mupi - \muG) \nn\\
&+& 4 \, m_b^2 \, \muG + 4 \, (m_b - v \cdot q) \, (\rhoD + \rhoLS) \Big]  \nn \\
&  - &\frac{8 m_U}{3 m_b \Delta_0^3} \, \Big[ m_b \, [q^2 - (q \cdot v)^2] \, \mupi - (m_b - v \cdot q) \, v \cdot q \, (\rhoD + \rhoLS) \Big] \hspace*{2cm} \label{TSMT1} \\
&  -& \frac{16 m_U}{3 \Delta_0^4} \, (m_b - v \cdot q) \, [q^2 - (v \cdot q)^2] \, \rhoD \Bigg\} \nn
\eea
\bea
T_{\sm T,2}&=&- T_{\sm T,4} =2m_H  \Bigg\{- \frac{2 m_U}{3 m_b^2 \, \Delta_0^2} \, \Big[ 2 m_b\, (\mupi - \muG) - (\rhoD + \rhoLS) \Big] \nn\\
&-& \frac{8 m_U}{3 m_b \Delta_0^3} \, (m_b - v \cdot q) \, (\rhoD + \rhoLS) \Bigg\}  \label{TSMT2}
\eea
\be
T_{\sm T,5} = T_{\sm T,7} =2 m_H  \Bigg\{ \frac{8 m_U}{3 \Delta_0^3} \, \rhoLS \Bigg\} \hspace*{2cm} \label{TSMT5}
\ee
with
$T_{\sm T,6} = T_{\sm T,8} =0,$
\bea
S_{\sm T,1}& =& -S_{\sm T,3} =2 m_H  \Bigg\{\frac{2 m_U}{3 m_b^2 \, \Delta_0^2} \, ( 2 \, m_b \, \mupi + \rhoD) + \frac{8 m_U}{3 m_b \Delta_0^3} \, (m_b - q \cdot v) \, \rhoD \Bigg\} 
\hspace{1.5cm} \label{SSMT1} \\
S_{\sm T,9} &=&-S_{\sm T,10}= 2 m_H \,  \Bigg\{ - \frac{2 m_U}{\Delta_0} \, \Big[ 1 - \frac{7 \mupi - 9 \muG}{12 m_b^2} + \frac{\rhoD}{6 m_b^3} \Big]  \nn \\
&-& \frac{2 m_U}{3 m_b^2 \, \Delta_0^2} \, \Big[ 3 \, m_b \, v \cdot q \, (\mupi - \muG) + 6 \, m_b^2 \, \muG + (4 m_b - v \cdot q) \, \rhoD + 3 \, m_b \, \rhoLS \Big] \nn \\
&+& \frac{8 m_U}{3 \Delta_0^3} \, [q^2 - (v \cdot q)^2] \, \mupi + \frac{16 m_U}{3 \Delta_0^4} \, (m_b - v \cdot q) \, [q^2 - (v \cdot q)^2]\, \rhoD \Bigg\}  \label{SSMT9}
 \\
S_{\sm T,11}&=& 2 m_H  \Bigg\{ - \frac{4 m_U}{\Delta_0^3} \, \rhoLS \Bigg\} \label{SST11} \\
S_{\sm T,12} &= &2 m_H  \Bigg\{ \frac{4 m_U}{3 m_b \, \Delta_0^2} \, \Big[ 3 \, m_b \, \muG + 4 \, \rhoD + \frac{9}{2} \, \rhoLS \Big] + \frac{8 m_U}{ \Delta_0^3} \, v \cdot q \, \rhoLS \Bigg\}\\ \label{SSMT12} 
%
S_{\sm T,16} &=& 2 m_H  \Bigg\{ - \frac{2 m_U}{3 m_b \, \Delta_0^2} \, \Big[ 6m_b \, \muG + 8 \, \rhoD + 9 \, \rhoLS \Big] - \frac{8 m_U}{ \Delta_0^3} \, v \cdot q \, \rhoLS \Bigg\} \,\, .
\label{SSMT16}
\eea
Several functions vanish: 
$S_{\sm T,(2,4,5,6,7,8)} =0$ and
$S_{\sm T,(15,19,20,21,22,23,24,26)} =0$.
The relations hold: 
$\dd S_{\sm T,13} =S_{\sm T,14} =-\frac{1}{3}S_{\sm T,17}=-S_{\sm T,18}=S_{\sm T,25}  =S_{\sm T,11}$ 
and
\bea
T_{T\sm,i}&=&-T_{\sm T,i} \hskip 1.2 cm (i=1,2,5) \nn \\
T_{T\sm,i}&=&T_{\sm  T,i} \hskip 1.5 cm (i=3,4,7) \nn \\
S_{T\sm,i}&=&-S_{\sm T,i} \hskip 1.2 cm (i=1,9,11,12,13,14) \label{SSMTall} \\
S_{T \sm,i}&=&S_{\sm T,i} \hskip 1.5 cm (i=3,10,16,17,18,25) \,\, .  \nn
\eea
\item Interference between the scalar and  tensor operators in $H_{eff}$
\par
The  tensor is obtained for $(i,j)=(2,4)$ and $(i,j)=(4,2)$ in Eq.~\eqref{Tij-gen}. We denote the two contributions as  $T_{ST}^{\mu \nu}$ and $T_{TS}^{\mu \nu}$, respectively.
Writing
\begin{equation}
\label{intST}
\begin{split}
T_{ST}^{\mu\nu} & = \epsilon^{\mu\nu\alpha\beta} \, v_\alpha \, q_\beta \, T_{ST,1} +i\, (q^\mu \, v^\nu - q^\nu \, v^\mu) \, T_{ST,2} + \\
& - (q \cdot s) \, \big[ \epsilon^{\mu\nu\alpha\beta} \, v_\alpha \, q_\beta \, S_{ST,1} +i\, (q^\mu \, v^\nu - q^\nu \, v^\mu) \, S_{ST,2} \big] + \\
&  + \epsilon^{\mu\nu\alpha\beta} \, q_\alpha \, s_\beta \, S_{ST,3} +i\, (q^\mu \, s^\nu - q^\nu \, s^\mu) \, S_{ST,4} + \\
&+ \epsilon^{\mu\nu\alpha\beta} \, v_\alpha \, s_\beta \, S_{ST,5} + i\,(s^\mu \, v^\nu - v^\nu \, s^\mu) \, S_{ST,6} + \\
&+\left[q^\mu \, \epsilon^{\nu \alpha \beta \delta} q_\alpha v_\beta s_\delta -q^\nu \, \epsilon^{\mu \alpha \beta \delta} q_\alpha v_\beta s_\delta \right] S_{ST,7} + \\
&+\left[v^\mu \, \epsilon^{\nu \alpha \beta \delta} q_\alpha v_\beta s_\delta -v^\nu \, \epsilon^{\mu \alpha \beta \delta} q_\alpha v_\beta s_\delta \right] S_{ST,8}
\end{split}
\end{equation}
we obtain:

\bea
T_{ST,1} &=&T_{ST,2} = 2 \, m_H \,  \Bigg\{ - \frac{1}{\Delta_0} - \frac{1}{3 m_b^2 \, \Delta_0^2} \, \Big[ 5 \, m_b \, v \cdot q \, (\mupi - \muG) + 4 \, m_b^2 \muG  \nn \\
&+& 4 \, (m_b - v \cdot q) \, (\rhoD + \rhoLS) \Big]  \nn \\
& +& \frac{4}{3 m_b \Delta_0^3} \, \Big[ [q^2 - (v \cdot q)^2] \, \mupi - (m_b - v \cdot q) \, v \cdot q \, \rhoD  \label{TST1}\\
&+& [ (m_b - v \cdot q)^2 + m_b \, m_U ] \, \rhoLS \Big]  
+ \frac{8}{3 \Delta_0^4} \, (m_b - v \cdot q) \, [q^2 - (v \cdot q)^2] \, \rhoD \Bigg\} \nn
\eea
\be
S_{ST,1} =S_{ST,2} = 2 \, m_H  \Bigg\{- \frac{1}{3 m_b^2 \, \Delta_0^2} \, ( 2 \, m_b \, \mupi + \rhoD) - \frac{2}{3 m_b \Delta_0^3} \, \Big[ 2(m_b - v \cdot q) \, \rhoD + 3 \, m_b \, \rhoLS \Big] \Bigg\} \label{SST1}
\ee
\bea
S_{ST,3} &=&-S_{ST,4} 
= 2 \, m_H  \Bigg\{ -\frac{1}{\Delta_0} \, \Big[  1 - \frac{7 \, \mupi - 9 \muG}{12 m_b^2} + \frac{\rhoD}{6 m_b^3} \Big] \nn \\
& -& \frac{1}{3 m_b^2 \, \Delta_0^2} \, \Big[ m_b \, [ 2 ( m_b + m_U) + 3 v \cdot q ] \, \mupi + 3 \, m_b \, ( m_b - m_U - v \cdot q) \, \muG \nn \\ 
& +& [2 (m_b - m_U) - v \cdot q] \, \rhoD + \frac{3}{2} (m_b - m_U) \, \rhoLS \Big]  \label{SST3}\\
&+& \frac{4}{3 m_b \Delta_0^3} \, \Big[ m_b \, [q^2 - (v \cdot q)^2] \, \mupi - (m_b - v \cdot q) \, (m_b + m_U) \, \rhoD + \frac{3}{2} \, m_U \, v \cdot q \, \rhoLS \Big] \nn \\
& + &\frac{8}{3 \Delta_0^4} \, (m_b - v \cdot q) \, [q^2 - (v \cdot q)^2] \, \rhoD \Bigg\} \nn
\eea

\bea
S_{ST,5} &=&S_{ST,6} = 2 \, m_H \,  \Bigg\{ \frac{m_b + m_U}{\Delta_0} \, \Big[ 1 - \frac{5 \, \mupi - 3 \muG}{12 m_b^2} - \frac{\rhoD}{6 m_b^3} \Big] \nn \\
&+& \frac{1}{3 m_b^2 \, \Delta_0^2} \, \Big[ 5 \, m_b \, (m_b + m_U) \, v \cdot q \, \mupi - 6 \, m_b \, m_U \, v \cdot q \, \muG \nn \\
&+& (m_b - m_U) \, v \cdot q \, (4 \, \rhoD + \frac{9}{2} \, \rhoLS)  \Big] \nn \\
&-& \frac{4}{3 m_b \Delta_0^3} \, \bigg( m_b \, (m_b + m_U) \, [q^2 - (v \cdot q)^2] \, \mupi - (m_b - v \cdot q) \, (m_b + m_U) \, q \cdot v \, \rhoD  \hspace*{1cm} \nn \\
&+&\frac{3}{2} \, (m_b \, q^2 - (m_b - m_U) \, (v \cdot q)^2) \, \rhoLS \Big]  \label{SST5}\\
&-&\frac{8}{3 \Delta_0^4} \, (m_b + m_U) \, (m_b - v \cdot q) \, [q^2 - (q \cdot v)^2] \, \rhoD \Bigg\} \nn
\eea
and
$S_{ST,7} =S_{ST,8} =0.$
The relations hold: 
\bea
T_{ST,i}&=&T_{TS,i}   \,\,\,\,\,\,\,\,  {\rm and}  \hspace{0.7cm}  S_{ST,i}=S_{TS,i} \,\,\,\,\ (i=1,3,5,7,8) \nn\\
T_{ST,i}&=&-T_{TS,i}  \,\,\,\, {\rm and} \hspace{0.7cm}   S_{ST,i}=-S_{TS,i} \,\,\,\,\ (i=2,4,6) . \hspace*{2cm} \label{STTall}
\eea
\item Interference between the pseudoscalar and tensor operators in $H_{eff}$
\par
This case amounts to choosing $(i,j)=(3,4)$ and $(4,3)$ in Eq.~\eqref{Tij-gen}. We denote the two contributions with $T_{PT}^{\mu \nu}$ and $T_{TP}^{\mu \nu}$, respectively.
We use the same expansion as for the scalar-tensor interference in Eq.~\eqref{intST}. The  functions  $T_{TP,i}$ and $S_{TP,i}$
 are obtained from the corresponding ones in (\ref{intST})  substituting  $T_{TP,i}(m_U)=-T_{TS,i} (-m_U)$ and  $S_{TP,i}(m_U)=-S_{TS,i} (-m_U)$. Analogous relations hold between  $T_{PT,i},\,S_{PT,i}$, and $T_{TS,i},\,S_{ST,i}$.
\item Right-handed operator $O_R$ in $H_{eff}$
 \par
This case amounts to choosing   $i=j=5$ in Eq.~\eqref{Tij-gen}.
The corresponding tensor  $T^{\mu \nu}_{R}$ can be expanded as in  Eq.~(\ref{TSMris}), substituting $T_i \to T_{Ri}$ and $S_i \to S_{Ri}$. The following relations hold:
\bea
T_{Ri}&=& T_i \,\,\,\,\,\, (i=1,2,4,5) \hskip 3.4 cm T_{R3}=-T_3 \nn \\
S_{Ri}&=&S_i \,\,\,\,\,\, (i=3,4,8,9,10,11,12,13) \hskip 1 cm S_{Ri}=-S_i \,\,\,\,\,\, (i=1,2,5,6,7) \hspace*{2cm} \label{TSR}
\eea

 \item Interference between the SM and the $O_R$ operators in $H_{eff}$  
\bea
\label{TSMA1}
T_{SMR,1} &=& 2 \, m_H \, m_U \,  \Bigg\{ - \frac{2}{\Delta_0} \bigg( 1 - \frac{\mupi - \muG}{2 m_b^2} \bigg) \nn \\
& +& \frac{2}{3 m_b \, \Delta_0^2} \, \Big[ - 3 \, v \cdot q \, (\mupi - \muG) - 4 \, m_b \muG - 2 \, (\rhoD + \rhoLS)\Big] \\
& +& \frac{8}{3 \Delta_0^3} \, [q^2 - (v \cdot q)^2 ] \, \mupi + \frac{16}{3 \Delta_0^4} \, (m_b - v \cdot q) \, [q^2 - (v \cdot q)^2] \, \rhoD \Bigg\} \hspace*{2cm} \nn
\eea
\bea
\label{TSMA2}
T_{SMR,2} &=& - 2 \, m_H \, m_U \,  \Bigg\{  \frac{8}{3 \, \Delta_0^2} \, \bigg( \muG + 2 \, \frac{(\rhoD + \rhoLS)}{m_b} \bigg) + \frac{16}{3 \Delta_0^3} \, v \cdot q \, \rhoLS \Bigg\} \hspace*{2cm}
\eea
\bea
T_{SMR,3} &=& T_{SMR,4} = 0
\eea
\bea
\label{TSMA5}
T_{SMR,5} &=& \frac{16 \, m_H \, m_U \, \rhoLS}{3 \, \Delta_0^3}
\eea

\bea
\label{SSMA8}
&&S_{SMR,8} =  2 \, m_H \, m_U \,  \Bigg\{ \frac{2}{\Delta_0} \bigg( 1 - \frac{5 \mupi}{12 m_b^2} + \frac{\muG}{4 m_b^2} - \frac{\rhoD}{6 m_b^3} \bigg) \nn \\
&& + \frac{1}{3 m_b \, \Delta_0^2} \, \bigg[ 10 \, v \cdot q \, \mupi + 12 (m_b - v \cdot q) \, \muG \nn \\
&&+ 4 \, (3 m_b - 2 v \cdot q) \, \frac{\rhoD}{m_b} 
+ 3  (4 m_b -3 v \cdot q) \frac{\rhoLS}{m_b} \bigg] \hspace*{2cm}  \\
& &- \frac{8}{3 \Delta_0^3} \, \bigg[ [ q^2 - (v \cdot q)^2 ] \, \mupi - v \cdot q (m_b - v \cdot q) \frac{\rhoD}{m_b} - 3 \, v \cdot q (2 m_b - v \cdot q) \frac{\rhoLS}{2m_b} \bigg]  \nn \\
& &- \frac{16}{3 \Delta_0^4} \, (m_b - v \cdot q) \, [q^2 - (v \cdot q)^2] \, \rhoD \Bigg\}  \nn
\eea
\bea
\label{SSMA9}
S_{SMR,9} &=& - 2 \, m_H \, m_U \,  \Bigg\{ \frac{1}{3 m_b \, \Delta_0^2} \, \bigg( 4 \, \mupi - 6 \, \muG - 4 \, \frac{\rhoD}{m_b} - 3 \, \frac{\rhoLS}{m_b} \bigg) \hspace*{3cm} \nn \\
& +& \frac{8}{3 m_b \Delta_0^3} \, \bigg( (m_b - v \cdot q) \, \rhoD + \frac{3}{2} \, (2 m_b - v \cdot q) \, \rhoLS \bigg) \Bigg\} 
\eea
\bea
\label{SSMA10}
S_{SMR,10} &=& 2 \, m_H \, m_U \,  \Bigg\{ \frac{1}{\Delta_0} \bigg( 2 - \frac{5 \mupi}{6 m_b^2} + \frac{\muG}{2 m_b^2} - \frac{\rhoD}{3 m_b^3} \bigg) \nn \\
& +& \frac{1}{3 m_b \, \Delta_0^2} \, \bigg( 10 \, v \cdot q \, \mupi - 12 \, v \cdot q \, \muG - 4 \, (m_b + 2 v \cdot q) \, \frac{\rhoD}{m_b} - 9 v \cdot q \, \frac{\rhoLS}{m_b} \bigg) \,\,\,\,\nn \\
& -& \frac{8}{3 \Delta_0^3} \, \bigg( [ q^2 - (v \cdot q)^2 ] \, \mupi - v \cdot q \, (m_b - v \cdot q) \frac{\rhoD}{m_b} + 3 \, (v \cdot q)^2 \, \frac{\rhoLS}{2m_b} \bigg)  \nn \\
& -& \frac{16}{3 \Delta_0^4} \, (m_b - v \cdot q) \, [q^2 - (v \cdot q)^2] \, \rhoD \Bigg\} 
\eea
\bea
\label{SSMA11}
S_{SMR,11} &=& 2 \, m_H \, m_U \,  \Bigg\{ \frac{1}{3 m_b \, \Delta_0^2} \, \bigg( - 4 \, \mupi + 6 \, \muG + 4 \, \frac{\rhoD}{m_b} + 3 \, \frac{\rhoLS}{m_b} \bigg) \hspace*{2cm} \nn \\
& -& \frac{8}{3 m_b \Delta_0^3} \, \bigg( (m_b - v \cdot q) \, \rhoD - \frac{3}{2} \, v \cdot q \, \rhoLS \bigg) \Bigg\} 
\eea
and 
$ S_{SMR,(1,2,3,4,5,6,7,12,13)} = 0$.
%
In addition we have:
\bea
T_{RSM,i} &=& T_{SMR,i} \qquad \,\,\,\,(i=1,2,3,4,5)  \nn \\
S_{RSM,i} &=& S_{SMR,i} \qquad  \,\,\,\,(i=1,2,3,4,5,6,7,8,9,12,13)  \label{SMRzero}\\
S_{RSM,i} &=& - S_{SMR,i} \qquad  (i=10,11) . \nn
\eea
%
\item Interference between the right-handed  and the scalar operators in $H_{eff}$
 \par
The  tensor is obtained when  $(i,j)=(5,2)$ and $(2,5)$ in Eq.~\eqref{Tij-gen}. We denote the two contributions as $T_{RS}$ and $T_{SR}$, respectively.
Using  the  expansion as in Eq.~(\ref{TSMS}) changing $T_{SMSi} \to T_{RSi}$ and $S_{SMSi} \to S_{RSi}$ we find:
\bea
T_{RSi}&=&T_{SMSi} \hskip 1.cm  (i=1,2) \nn \\
S_{RSi}&=&-S_{SMSi} \,\,\,\,\,\,\,\,\,\,\,(i=1,2,3) \hskip 1cm  S_{RS4}=S_{SMS4} \label{TRS}
\eea
and analogous relations in the case of the structures in $T_{SR}$.
\item Interference between the right-handed and the pseudoscalar operators in $H_{eff}$
 \par
The  tensor is obtained when  $(i,j)=(5,3)$ and $(3,5)$ in Eq.~\eqref{Tij-gen}. We denote the two contributions as $T_{RP}$ and $T_{PR}$, respectively.
Using  an  expansion analogous to (\ref{TSMP}), substituting $T_{SMPi} \to T_{RPi}$ and $S_{SMPi} \to S_{RPi}$, we find:
\bea
T_{RPi}&=&-T_{SMPi} \,\,\,(i=1,2)\nn \\
S_{RPi}&=&S_{SMPi} \,\,\,\,\,\, (i=1,2,3) \hskip 1cm  S_{RP4}=-S_{SMP4}\label{TRP}
\eea
and analogous relations in the case of the structures in $T_{PR}$.

 \item Interference between the right handed and the tensor operators in $H_{eff}$
\par
The  tensor is obtained for  $(i,j)=(5,4)$ and $(i,j)=(4,5)$ in Eq.~\eqref{Tij-gen}, with the two contributions denoted as  $T_{RT}$ and $T_{TR}$, respectively.
Using  an  expansion  as in Eq.~(\ref{TSMT}), with $T_{SMTi} \to T_{RTi}$ and $S_{SMTi} \to S_{RTi}$, we find:
\bea
T_{R T,1}& =&-T_{RT, 3}=2 m_H \,  \Bigg\{ -\frac{2 m_b}{\Delta_0} - \frac{2 }{3 \, \Delta_0^2} \, \Big[ 5  v \cdot q \, \mupi +(4m_b-3v \cdot q) \muG+4 \rhoD \Big] \nn\\
&  + &\frac{8 }{3  \Delta_0^3} \, \Big[ m_b \, [q^2 - (q \cdot v)^2] \, \mupi - (m_b - v \cdot q) \, v \cdot q \, \rhoD  \Big] \label{TAT1} \\
&  +& \frac{16 m_b}{3 \Delta_0^4} \, (m_b - v \cdot q) \, [q^2 - (v \cdot q)^2] \, \rhoD \Bigg\} \nn
\eea
\bea
T_{R T,2} &=&-T_{RT,4}=2 m_H \,  \Bigg\{ \frac{2}{\Delta_0} \,  \Big[ 1 - \frac{ \, \mupi -  \muG}{2 m_b^2} \Big] \nn \\
&+&\frac{1 }{\Delta_0^2} \, \Big[
\frac{2(2m_b+3v \cdot q)}{3m_b}\mu_\pi^2+\frac{2(4m_b-3v \cdot q)}{3m_b}\mu_G^2+\frac{2}{3m_b}(\rhoD+\rhoLS) \Big] \nn \\
&-& \frac{8}{3 \Delta_0^3}\Big[[q^2 - (v \cdot q)^2] \, \mu_\pi^2-(m_b-v \cdot q)\rhoD]  \label{TAT2} \\
&-&\frac{16}{3 \Delta_0^4} \, (m_b - v \cdot q)[q^2 - (v \cdot q)^2] \,  \rhoD \Bigg\}\nn
  \eea
\bea
T_{R T,5} &=&T_{R T,7}=2 m_H \,  \Bigg\{ \frac{4}{3  \Delta_0^2} \Big[\mu_G^2 +\frac{2}{m_b}(\rhoD+\rhoLS) \Big]+\frac{8}{3 \Delta_0^3} v \cdot q \rhoLS \Bigg\} \hspace*{2cm} \label{TAT5}\eea
\bea
T_{R T,6} &=&T_{R T,8} =-2 m_H \, \Bigg\{\frac{8}{3 \Delta_0^3}\rhoLS \Bigg\} \hspace*{7cm} \label{TAT6}
 \eea
\bea
&&S_{R T,1} =-2 m_H \,  \Bigg\{ \frac{1}{\Delta_0} \, \Big[2-\frac{5 \mu_\pi^2}{6m_b^2}+\frac{ \mu_G^2}{2m_b^2}-\frac{ \rhoD}{3m_b^3} \Big] \nn \\
&&+ \frac{1 }{\Delta_0^2} \, \Big[\frac{2m_b(2m_b+5 v \cdot q)}{3m_b^2}\mu_\pi^2+\frac{4m_b(m_b-v \cdot q)}{m_b^2}\mu_G^2\nn \\
&&+\frac{2(5m_b-4 v \cdot q)}{3m_b^2}\rhoD +\frac{4m_b-3 v \cdot q}{m_b^2}\rhoLS \Big] \label{SAT1} \\
&&+\frac{1 }{  \Delta_0^3} \, \Big[-\frac{8 }{3}\, [q^2 - (q \cdot v)^2] \mu_\pi^2+\frac{8 }{3m_b}(m_b^2-v \cdot q^2)\rhoD+4 v \cdot q \frac{(2m_b-v \cdot q)}{m_b}\rhoLS \Big] \hspace*{1cm} \nn \\
&&-\frac{16 }{ 3 \Delta_0^4}  \, (m_b - v \cdot q)\,[q^2 - (q \cdot v)^2]  \rhoD \Bigg\} \nn
\eea

  \bea
S_{R T,2} &=&2 m_H \,  \Bigg\{ \frac{1 }{\Delta_0^2} \, \Big[\frac{(4 \mu_\pi^2-6 \mu_G^2)}{3m_b}-\frac{4\rhoD+3\rhoLS}{3m_b^2} \Big]\nn \\
&+& \frac{1 }{\Delta_0^3} \,\Big[ \frac{8(m_b-v \cdot q)}{3m_b}\rhoD+\frac{4(2m_b-v \cdot q)}{m_b}\rhoLS \Big] \Bigg\}  \hspace*{2cm}\label{SAT'2}
\eea
 \bea
&& S_{R T,3} =2 m_H \,  \Bigg\{ \frac{2 }{\Delta_0} \, \Big[1-\frac{5\mu_\pi^2-3 \mu_G^2}{12 m_b^2}-\frac{\rhoD}{6 m_b^3}\Big]\nn \\
&&+\frac{1 }{3 \Delta_0^2} \, \Big[\frac{2(2m_b+5 v \cdot q)}{m_b}\mu_\pi^2+\frac{12(m_b-v \cdot q)}{m_b} \mu_G^2+\frac{2(5m_b-4 v \cdot q)}{m_b^2}\rhoD\nn \\
&+&\frac{3(4m_b-3 v \cdot q)}{m_b^2}\rhoLS \Big] \label{SAT3} \\
&&-\frac{4}{3 \Delta_0^3} \, \Big[2\, [q^2 - (v \cdot q)^2] \, \mu_\pi^2-\frac{2[m_b^2-(v \cdot q)^2]}{m_b}\rhoD-\frac{3 v \cdot q (m_b - v \cdot q)}{m_b}\rhoLS \Big]  \nn \\
&&-\frac{16 }{ 3 \Delta_0^4}  \, (m_b - v \cdot q)\,[q^2 - (q \cdot v)^2]  \rhoD \Bigg\} \nn
\eea
 \bea
 S_{R T,4} &=&2 m_H \,  \Bigg\{ \frac{1}{3m_b \Delta_0^2} \, \Big[-4 \mu_\pi^2+6 \mu_G^2+\frac{4 \rhoD+3 \rhoLS}{m_b} \Big] \hspace*{2cm} \nn \\
&-& \frac{4}{3 m_b \Delta_0^3}(m_b - v \cdot q)(2 \rhoD+3 \rhoLS)\Bigg\} \label{SAT4}
\eea
 \be
 S_{RT,7} = S_{RT,26}=-2m_H \frac{4}{\Delta_0^3}\rhoLS  \hspace*{5cm}
 \ee
\bea
S_{R T,9} &=&2 m_H \, \Big\{\frac{2m_b}{\Delta_0} \Big[\frac{m_b-v \cdot q}{m_b}+\frac{m_b+5 v \cdot q}{12m_b^3}\mu_\pi^2-\frac{m_b+v \cdot q}{4 m_b^3}\mu_G^2 \nn \\
&+&\frac{-3m_b+v \cdot q}{6 m_b^4} \rhoD -\frac{1}{2m_b^3}\rhoLS \Big]\nn \\
&+&\frac{1}{\Delta_0^2} \Big[\frac{2[2q^2+(3m_b-5 v \cdot q) v \cdot q]}{3m_b} \mu_\pi^2+\frac{4m_b^2-6m_b v \cdot q +4 (v \cdot q)^2-2 q^2}{m_b}\mu_G^2  \nn \\
&+&\frac{2(8 m_b^2 - 2 q^2 - 7 m_b v \cdot q + 4 (v \cdot q)^2)}{3 m_b^2} \rhoD+\frac{2 m_b^2 - q^2 - 6 m_b v \cdot q + 3 (v \cdot q)^2}{m_b^2}\rhoLS \Big] \nn \\ 
&-&\frac{4}{3 \Delta_0^3} \,[q^2 - (q \cdot v)^2] \, \Big[2(m_b- v \cdot q)\mu_\pi^2-2\frac{(m_b - v \cdot q)}{m_b} \rhoD-3\frac{2m_b- v \cdot q}{m_b} \rhoLS \Big]   \nn\\
&-&\frac{16}{3\Delta_0^4}\,(m_b- v \cdot q)^2\,[q^2 - (q \cdot v)^2] \,  \rhoD 
\Bigg\} 
\eea
\bea
S_{R T,10} &=& -2 m_H \, \Big\{\frac{2m_b}{\Delta_0} \Big[\frac{m_b-v \cdot q}{m_b}+\frac{m_b+5 v \cdot q}{12m_b^3}\mu_\pi^2-\frac{m_b+v \cdot q}{4 m_b^3}\mu_G^2 \nn\\
&+&\frac{-3m_b+v \cdot q}{6 m_b^4} \rhoD -\frac{1}{2m_b^3}\rhoLS \Big]\nn \\
&+&\frac{1}{\Delta_0^2} \Big[\frac{2[2q^2+(3m_b-5 v \cdot q) v \cdot q]}{3m_b} \mu_\pi^2+\frac{4m_b^2-6m_b v \cdot q +4 (v \cdot q)^2-2 q^2}{m_b}\mu_G^2 \nn \\
&+&\frac{2(8 m_b^2 - 2 q^2 - 7 m_b v \cdot q + 4 (v \cdot q)^2)}{3 m_b^2} \rhoD+\frac{2 m_b^2 - q^2 - 6 m_b v \cdot q + 3 (v \cdot q)^2}{m_b^2}\rhoLS \Big]  \nn \\ 
&-&\frac{4}{3 \Delta_0^3} \,[q^2 - (q \cdot v)^2] \, \Big[2(m_b- v \cdot q)\mu_\pi^2-2\frac{(m_b - v \cdot q)}{m_b} \rhoD-3\frac{m_b- v \cdot q}{m_b} \rhoLS \Big] \nn  \\
&-&\frac{16}{3\Delta_0^4}\,(m_b- v \cdot q)^2\,[q^2 - (q \cdot v)^2] \,  \rhoD 
\Bigg\} 
\eea
\bea
S_{R T,11} &=&2 m_H \, \Bigg\{-\frac{2}{\Delta_0}\Big[1-\frac{5}{12 m_b^2}\mu_\pi^2+\frac{1}{4 m_b^2}\mu_G^2-\frac{1}{6m_b^3}\rhoD \Big] \nn \\
&-&\frac{1}{\Delta_0^2}\Big[\frac{10 v \cdot q}{3m_b}\mu_\pi^2+\frac{4(m_b - v \cdot q)}{m_b}\mu_G^2 +\frac{8(m_b - v \cdot q)}{3m_b^2}\rhoD+\frac{4m_b-3 v \cdot q}{m_b^2} \rhoLS \Big] \nn \\
&+& \frac{4}{3 \Delta_0^3} \Big[ 2\,[q^2 - (q \cdot v)^2] \, \mu_\pi^2-\frac{2v \cdot q (m_b - v \cdot q)}{m_b}\rhoD +\frac{3(m_b- v \cdot q)^2}{m_b}\rhoLS \Big] \nn \\
&+& \frac{16}{3 \Delta_0^4}
\,(m_b- v \cdot q)\,[q^2 - (q \cdot v)^2] \,  \rhoD 
\Bigg\} 
\eea
\bea
S_{R T,12} &=&2 m_H \, \Bigg\{-\frac{4m_b}{\Delta_0}\Big[1+\frac{5\mu_\pi^2-9 \mu_G^2}{12 m_b^2}-\frac{10 \rhoD+9\rhoLS}{12 m_b^3} \Big] \nn\\
&-& \frac{2}{3 \Delta_0^2} \Big[14 v \cdot q \, \mu_\pi^2+6(m_b-2 v \cdot q)\mu_G^2+\frac{12(m_b- v \cdot q)}{m_b}\rhoD+\frac{3(3m_b-5 v \cdot q)}{m_b}\rhoLS \Big] \nn \\ 
&+& \frac{8}{3 \Delta_0^3} \Big[2m_b\,[q^2 - (q \cdot v)^2] \,  \mu_\pi^2-v \cdot q (m_b - v \cdot q)(4\rhoD+3 \rhoLS) \Big] \label{SAT12} \\
&+&\frac{32}{3 \Delta_0^4} m_b\,(m_b- v \cdot q)\,[q^2 - (q \cdot v)^2] \,  \rhoD  \Bigg\} \nn
\eea
\bea
S_{R T,13} &=&2 m_H \, \Bigg\{\frac{2}{\Delta_0}\Big[1-\frac{5}{12 m_b^2}\mu_\pi^2+\frac{1}{4 m_b^2}\mu_G^2-\frac{1}{6m_b^3}\rhoD \Big] \nn \\
&+&\frac{1}{\Delta_0^2}\Big[\frac{2(4m_b+5 v \cdot q)}{3m_b}\mu_\pi^2-\frac{4 v \cdot q}{m_b}\mu_G^2 -\frac{4(m_b+2 v \cdot q)}{3 m_b^2}\rhoD-\frac{3v \cdot q}{m_b^2}\rhoLS \Big] \nn \\
&+& \frac{8}{3 \Delta_0^3} \Big[-[q^2 - (q \cdot v)^2] \, \mu_\pi^2+\frac{2m_b^2-m_b v \cdot q -(v \cdot q)^2}{m_b}\rhoD\nn \\
&+&3\frac{m_b^2-(v \cdot q)^2}{2m_b}\rhoLS \Big]  \label{SAT13} \\
&-& \frac{16}{3 \Delta_0^4}\,(m_b- v \cdot q)\,[q^2 - (q \cdot v)^2] \,  \rhoD \Bigg\} \nn
\eea
\bea
S_{R T,14} &=&2 m_H \, \Bigg\{\frac{2}{\Delta_0}\Big[1-\frac{5}{12 m_b^2}\mu_\pi^2+\frac{1}{4 m_b^2}\mu_G^2-\frac{1}{6m_b^3}\rhoD \Big] \nn \\
&+&\frac{1}{\Delta_0^2}\Big[\frac{2(4m_b+5 v \cdot q)}{3m_b}\mu_\pi^2-\frac{4 v \cdot q}{m_b}\mu_G^2 -\frac{8 v \cdot q}{3 m_b^2}\rhoD-\frac{3v \cdot q}{m_b^2}\rhoLS \Big] \nn \\
&+& \frac{8}{3 \Delta_0^3} \Big[-[q^2 - (q \cdot v)^2] \, \mu_\pi^2+\frac{2m_b^2-m_b v \cdot q -(v \cdot q)^2}{m_b}\rhoD+\frac{m_b^2-(v \cdot q)^2}{2m_b}\rhoLS \Big] \nn \\
&-& \frac{16}{3 \Delta_0^4}
\,(m_b- v \cdot q)\,[q^2 - (q \cdot v)^2] \,  \rhoD 
\Bigg\} \label{SAT14}
\eea
\be
S_{R T,15} =2 m_H \,  \Big\{- \frac{2 }{\Delta_0^2} \, \Big[\frac{(4 \mu_\pi^2-6 \mu_G^2)}{3m_b}-\frac{4\rhoD+3\rhoLS}{3m_b^2} \Big]
-\frac{8(m_b-v \cdot q)}{3m_b\Delta_0^3}(2\rhoD+3\rhoLS)\Big\} \,\,\,\,\,\, \label{SAT15}
\ee
\bea
S_{R T,16} &=&2 m_H \, \Bigg\{\frac{4m_b}{\Delta_0}\Big[1+\frac{5\mu_\pi^2-9 \mu_G^2}{12 m_b^2}-\frac{10 \rhoD+9\rhoLS}{12 m_b^3} \Big] \nn \\
&+& \frac{2}{3 \Delta_0^2} \Big[14 v \cdot q \, \mu_\pi^2+6(m_b-2 v \cdot q)\mu_G^2+\frac{12(m_b- v \cdot q)}{m_b}\rhoD+\frac{3(3m_b-5 v \cdot q)}{m_b}\rhoLS \Big] \nn \\ 
&-& \frac{8}{3 \Delta_0^3} \Big[2m_b\,[q^2 - (q \cdot v)^2] \,  \mu_\pi^2-4v \cdot q (m_b - v \cdot q)\rhoD \label{SAT13} \\
&+& \frac{3[q^2-2 v \cdot q(m_b - v \cdot q)]}{2}\rhoLS \Big] \nn \\
&-&\frac{32}{3 \Delta_0^4}m_b\,(m_b- v \cdot q)\,[q^2 - (q \cdot v)^2] \,  \rhoD \Bigg\} \nn
\eea
\bea
S_{R T,17} &=&2 m_H \, \Bigg\{-\frac{2}{\Delta_0}\Big[1-\frac{5}{12 m_b^2}\mu_\pi^2+\frac{1}{4 m_b^2}\mu_G^2-\frac{1}{6m_b^3}\rhoD \Big] \nn \\
&-&\frac{1}{\Delta_0^2}\Big[\frac{2(4m_b+5 v \cdot q)}{3m_b}\mu_\pi^2-\frac{4 v \cdot q}{m_b}\mu_G^2 -\frac{4(m_b+2 v \cdot q)}{3 m_b^2}\rhoD-\frac{3v \cdot q}{m_b^2}\rhoLS \Big] \nn \\
&-& \frac{8}{3 \Delta_0^3} \Big[-[q^2 - (q \cdot v)^2] \, \mu_\pi^2+\frac{2m_b^2-m_b v \cdot q -(v \cdot q)^2}{m_b}\rhoD \nn \\
&+&\frac{3(m_b^2-m_b v \cdot q-(v \cdot q)^2)}{2m_b}\rhoLS \Big] \label{SAT17} \\
&+& \frac{16}{3 \Delta_0^4}\,(m_b- v \cdot q)\,[q^2 - (q \cdot v)^2] \,  \rhoD \Bigg\} \nn
\eea
\bea
S_{R T,18} &=&2 m_H \, \Bigg\{-\frac{2}{\Delta_0}\Big[1-\frac{5}{12 m_b^2}\mu_\pi^2+\frac{1}{4 m_b^2}\mu_G^2-\frac{1}{6m_b^3}\rhoD \Big] \nn \\
&-&\frac{1}{\Delta_0^2}\Big[\frac{2(4m_b+5 v \cdot q)}{3m_b}\mu_\pi^2-\frac{4 v \cdot q}{m_b}\mu_G^2 -\frac{8 v \cdot q}{3 m_b^2}\rhoD-\frac{3v \cdot q}{m_b^2}\rhoLS \Big] \nn \\
&-& \frac{8}{3 \Delta_0^3} \Big[-[q^2 - (q \cdot v)^2] \, \mu_\pi^2+\frac{2m_b^2-m_b v \cdot q -(v \cdot q)^2}{m_b}\rhoD  \nn \\
&+&\frac{3(m_b^2-m_b v \cdot q-(v \cdot q)^2)}{2m_b}\rhoLS \Big] \label{SAT18} \\
&+& \frac{16}{3 \Delta_0^4}\,(m_b- v \cdot q)\,[q^2 - (q \cdot v)^2] \,  \rhoD \Bigg\} \nn
\eea
  \bea
S_{R T,19} &=&2 m_H \,  \Bigg\{ \frac{2 }{\Delta_0^2} \, \Big[\frac{(4 \mu_\pi^2-6 \mu_G^2)}{3m_b}-\frac{4\rhoD+3\rhoLS}{3m_b^2} \Big]\nn \\ 
&+&\frac{1}{\Delta_0^3}\Big[\frac{16(m_b-v \cdot q)}{3m_b}\rhoD+\frac{4(m_b-2 v \cdot q)}{m_b}\rhoLS\Big]\Bigg\} \label{SAT19}
\eea
\bea
S_{R T,25} &=&2 m_H \, \Bigg\{-\frac{2}{\Delta_0}\Big[1-\frac{5}{12 m_b^2}\mu_\pi^2+\frac{1}{4 m_b^2}\mu_G^2-\frac{1}{6m_b^3}\rhoD \Big] \nn \\
&-&\frac{1}{\Delta_0^2}\Big[\frac{10 v \cdot q}{3m_b}\mu_\pi^2+\frac{4(m_b - v \cdot q)}{m_b}\mu_G^2 +\frac{8(2m_b - v \cdot q)}{3m_b^2}\rhoD\nn \\
&+&\frac{4m_b-3 v \cdot q}{m_b^2} \rhoLS \Big]  \label{SAT25} \\
&+& \frac{4}{3 \Delta_0^3} \Big[ 2\,[q^2 - (q \cdot v)^2] \, \mu_\pi^2-\frac{2v \cdot q (m_b - v \cdot q)}{m_b}\rhoD +\frac{3(m_b- v \cdot q)^2}{m_b}\rhoLS \Big] \nn \\
&+& \frac{16}{3 \Delta_0^4}
\,(m_b- v \cdot q)\,[q^2 - (q \cdot v)^2] \,  \rhoD  
\Bigg\} \nn
\eea
 and $S_{RT,(5,6,8,20,21,22,23,24)}=0$.
 Moreover, we have:
 \bea
T_{RT,i}&=&-T_{TR,i}   \,\,\,\,\,\,\,\,  (i=1,2,5,6) \nn \\
S_{RT,i}&=&-S_{TR,i} \,\,\,\,\,\,\,\, (i=1,2,9,11,12,13,14,15) \nn\\
T_{RT,i}&=&T_{TR,i}  \,\,\,\,\,\,\,\,\,\,\,\,\,  (i=3,4,7,8) \label{RTall} \\
S_{RT,i}&=&S_{TR,i} \,\,\,\,\,\,\,\,\,\,\,\,\, (i=3,4,7,10,16,17,18,19,25,26) . \hspace*{2cm}\nn
\eea
 
 \end{itemize}

%
\section{Coefficients in the $1/m_b$ expansion of the inclusive semileptonic decay width}\label{appC}
To provide the  coefficients  in Eq.~\eqref{fullwidth} we define the  variables
\be
\rho=\frac{m_U^2}{m_b^2} \,\,\, , \hskip 1.cm \rho_\ell=\frac{m_\ell^2}{m_b^2}\, .
\ee
In the formulae $\sqrt{\lambda}$ stays for $\sqrt{\lambda (1,\rho,\rho_\ell)}$. 
Factorizing the effective couplings in the Hamiltonian Eq.~(\ref{hnew}), we define for   $A=0,\,\mu_\pi^2,\,\mu_G^2,\rho_{D}^3$:
 $C^{(SM)}_A=|1+\epsilon_V|^2 \, {\cal C}^{(SM)}_A$, $C^{(i)}_A=|\epsilon_i|^2 \, {\cal C}^{(i)}_A$ for i=S,  P,  T,  R,   and
  $C^{(ij)}_A=2{\rm Re} (\epsilon_i\, \epsilon_j^*) \, {\cal C}^{(ij)}_A$ for (i,j)=(S,P),\ (SM,S), (SM,P), (SM,T),  (SMR), (S,T), (P,T),  (RS), (RP), (RT).
We also define: 
\be
{\cal L}_1=\log \big[\frac{(1+\sqrt{\lambda}-\rho+\rho_\ell)^2}{4 \rho_\ell} \big]  \,\,\, , \hskip 1.cm
{\cal L}_2=\log \big[\frac{(1+\sqrt{\lambda}+\rho-\rho_\ell)^2}{4 \rho} \big] \,.
\ee
Our results  for  SM  agree with 
\cite{Mannel:2017jfk} in the case of massless leptons and differ for the coefficient ${\cal C}_{\rho_D^3}^{(SM)}$ in the case $m_\ell \neq 0$ \footnote{
The result for  ${\cal C}_{\rho_D^3}^{(SM)}$ coincides with that recently reported in 
\cite{Moreno:2022goo,Rahimi:2022vlv}. }.
\begin{itemize}
\item {Standard Model:}
\bea
{\cal C}_0^{(SM)}  &=& -2 \, {\cal C}_{\mu_\pi^2}^{(SM)}=
\sqrt{\lambda} \Big[1 - 7 \rho - 7 \rho^2 + \rho^3 - (7 - 12 \rho + 7 \rho^2) \rho_\ell \nn \\
&-& 7 (1 + \rho) \rho_\ell^2 + \rho_\ell^3 \Big] \\
&+&
12 \Big\{(1-\rho^2)\rho_\ell^2{\cal L}_1+(1-\rho_\ell^2)\rho^2
 {\cal L}_2 \Big\} \nn
\eea
\bea
{\cal C}_{\mu_G^2}^{(SM)}& =& 
\frac{\sqrt{\lambda}}{2} \Big[ - 3 + 5 \rho - 19 \rho^2 + 5 \rho^3 
+ (5 + 28 \rho - 35 \rho^2) \rho_\ell \nn \\
& -& (19 + 35 \rho) \rho_\ell^2 + 5\rho_\ell^3 \Big]  \\
&+&  
6 \Big\{(1-5\rho^2)\rho_\ell^2{\cal L}_1+(1-5\rho_\ell^2)\rho^2
 {\cal L}_2 \Big\} \nn
\eea
\bea
{\cal C}_{\rho_D^3}^{(SM)}&=&\frac{2}{3}\sqrt{\lambda}\Big[17 + \rho - 11 \rho^2 + 5 \rho^3 +\rho_\ell( 4  + 18 \rho  - 
 32 \rho^2)  \nn \\
&+&\rho_\ell^2(- 23  - 35 \rho ) + 2\rho_\ell^3+3\rho_\ell[(1-\rho_\ell)^2-\rho^2]\Big] \\
&-&8  \Big\{\rho_\ell^2(-1+5\rho^2+\rho_\ell){\cal L}_1+[1-\rho_\ell+\rho_\ell^2(-1+5\rho^2+\rho_\ell)]{\cal L}_2 \Big\} \nn
\eea
\item{ S and P:}
\bea
{\cal C}_0^{(S)}&=&-2{\cal C}_{\mu_\pi^2}^{(S)}=\frac{\sqrt{\lambda}}{8} \Big[1 + 4 \sqrt{\rho} - 7 \rho + 40 \rho^{3/2} - 7 \rho^2 + 4 \rho^{5/2} + \rho^3 
\nn \\
&+& \rho_\ell (-7 - 20 \sqrt{\rho} + 12 \rho - 20  \rho^{3/2} - 7 \rho^2) 
+\rho_\ell^2(-7 - 8 \sqrt{\rho} - 7 \rho)+ \rho_\ell^3 \Big] \hskip 1cm \nn \\
&-&\frac{3}{2} \Big\{(-1 + \sqrt{\rho}) (1 + \sqrt{\rho})^3  \rho_\ell^2 {\cal L}_1  \\
&+&\rho^{3/2}\left[2 \rho + 2 (-1 + \rho_\ell)^2 +  \sqrt{\rho} (-1 +\rho_\ell^2) \right]{\cal L}_2 \Big\} \nn 
\eea
\bea
{\cal C}_{\mu_G^2}^{(S)}&=&\frac{\sqrt{\lambda}}{16} \Big[13 -132 \sqrt{\rho} +45 \rho -24 \rho^{3/2} - 27 \rho^2 + 12 \rho^{5/2} +5 \rho^3 
\nn \\
&+& \rho_\ell (-27 +84 \sqrt{\rho} + 68\rho - 60  \rho^{3/2} - 35 \rho^2) 
+\rho_\ell^2(-3 - 24 \sqrt{\rho} - 35\rho)+ 5\rho_\ell^3 \Big] \nn \\
&+&\frac{3}{4} \Big\{(1 + \sqrt{\rho})^2 (1 +4 \sqrt{\rho}-5\rho) \rho_\ell^2  \,{\cal L}_1  \\
&+& \rho^{1/2}\big[-2 \rho^2 + 4 (-1 + \rho_\ell)^2 + \rho (10 + 4 \rho_\ell - 6 \rho_\ell^2) \nn \\
&+&  \rho^{3/2} (1 - 5 \rho_\ell^2) + 4 \sqrt{\rho} (-1 + \rho_\ell^2) \big]{\cal L}_2 \Big\}\nn 
\eea

\bea
{\cal C}_{\rho_D^3}^{(S)}&=&\frac{\sqrt{\lambda}}{12} \Big[59 + 44 \sqrt{\rho} + 37 \rho - 28 \rho^{3/2} - 17 \rho^2 + 8 \rho^{5/2} + 
 5 \rho^3\nn \\
&+&\rho_\ell(-53 + 44 \sqrt{\rho} + 54 \rho - 40 \rho^{3/2} - 35 \rho^2)+\rho_\ell^2(13 - 16 \sqrt{\rho} - 35 \rho)+5\rho_\ell^3 \Big] \nn \\
&+&\Big\{- (1 + \sqrt{\rho})^2 (2 - 6 \sqrt{\rho}+ 5 \rho) \rho_\ell^2 {\cal L}_1  \\
&+&\left[- (2 + 2 \sqrt{\rho} + 5 \rho) + 4 \rho_\ell -   (1 + \sqrt{\rho})^2 (2 - 6 \sqrt{\rho} + 5 \rho) \rho_\ell^2 \right]
 {\cal L}_2 \Big\} \nn
\eea
In the pseudoscalar case the  coefficients  are obtained from the corresponding ones in the scalar case changing the sign of $m_U$ and of the odd powers of $\sqrt{\rho}$.

\item {T:}
\bea
{\cal C}_0^{(T)}  
&=& -2{\cal C}_{\mu_\pi^2}^{(T)}
=12
\sqrt{\lambda} \Big[1 - 7 \rho - 7 \rho^2 + \rho^3 - (7 - 12 \rho + 7 \rho^2) \rho_\ell \nn \\
&-& 7 (1 + \rho) \rho_\ell^2 + \rho_\ell^3 \Big]   \\
&+&144 \Big\{(1-\rho^2)\rho_\ell^2{\cal L}_1+(1-\rho_\ell^2)\rho^2
 {\cal L}_2 \Big\} \nn
\eea
\bea
{\cal C}_{\mu_G^2}^{(T)}&=&2\sqrt{\lambda} \Big[-25 - 25 \rho - 49 \rho^2 + 15 \rho^3 +\rho_\ell ( 47 + 44 \rho  - 
 105 \rho^2) \nn \\
&-&\rho_\ell^2( 73  + 105\rho) + 15 \rho_\ell^3 \Big]   \\
&+&24 \Big\{\rho_\ell^2(1 - 3 \rho) (3 + 5 \rho){\cal L}_1
+\rho (4 + 3 \rho - (4 + 15 \rho) \rho_\ell^2) {\cal L}_2 \Big\} \nn
\eea
\bea
{\cal C}_{\rho_D^3}^{(T)}&=&8 \sqrt{\lambda} \Big[3 - 11 \rho - 9 \rho^2 + 5 \rho^3 +\rho_\ell ( 23 + 6 \rho  - 
 31 \rho^2)  - 39 \rho_\ell ^2 (1+ \rho) + 5 \rho_\ell ^3 \nn \\
 &+&4(1-\rho)(1 + \rho -\rho_\ell) \rho_\ell  \Big] \nn \\
&-&32
   \Big\{ \rho_\ell^2 [-6 + 5 \rho (1 + 3 \rho) + 4  \rho_\ell] {\cal L}_1  \\
&+&   \big[2 - 5 \rho + [-6 + 5 \rho (1 + 3 \rho)] \rho_\ell^2 + 4 \rho_\ell^3\big]
{\cal L}_2 \Big\} \,\,\,\, \nn 
\eea

\item { S - P interference:}
The coefficients vanish.

\item { SM - S and SM - P interference: }
\bea
{\cal C}_0^{(SMS)}&=&-2{\cal C}_{\mu_\pi^2}^{(SMS)}\nn \\&=&\frac{\sqrt{\lambda}}{2}(1-\sqrt{\rho})\sqrt{\rho_\ell}\Big[1 + 3 \sqrt{\rho} - 2 \rho + 3 \rho^{3/2} + \rho^2 \nn \\
&+&\rho_\ell (10 +  15 \sqrt{\rho}  + 10 \rho) + \rho_\ell^2\Big] \,\,\,\,\,\,\, \nn \\
 &-&3\sqrt{\rho_\ell}\Big\{\rho_\ell(1 + \sqrt{\rho}) \left[(1 -\rho)^2 + (1 - \sqrt{\rho} + \rho) \rho_\ell\right]
 {\cal L}_1  \\
&+&\rho^{3/2}\left[\sqrt{\rho} (-1 + \rho_\ell) + (1 - \rho_\ell)^2 + \rho \rho_\ell \right]{\cal L}_2 \Big\} \nn
\eea
\bea
{\cal C}_{\mu_G^2}^{(SMS)}&=&\frac{\sqrt{\lambda}}{4} (1 - \sqrt{\rho}) \sqrt{ \rho_\ell}\Big[5 - 15 \sqrt{\rho} - 10 \rho + 9 \rho^{3/2} \nn \\
&+& 5 \rho^2 + \rho_\ell(2 + 
 45 \sqrt{\rho} + 50 \rho ) + 5\rho_\ell^2 \Big]\nn \\
 &-&\frac{3}{2}\sqrt{\rho_\ell}\Big\{\rho_\ell \left[(1 + 5  \sqrt{\rho}) (1 - \rho)^2 + (1 - 2  \sqrt{\rho} - 2 \rho + 
    5 \rho^{3/2}) \rho_\ell \right] {\cal L}_1  \nn \\
&+&\sqrt{\rho}\big[-2 + 2  \sqrt{\rho} + \rho - \rho^{3/2} +\rho_\ell (4 - 10 \rho  + \rho^{3/2} + 5 \rho^2 ) \hskip 2cm \\
&+&\rho_\ell^2(-2  - 2 \sqrt{\rho}  + 5 \rho )\big]{\cal L}_2 \Big\} \nn
\eea
\bea
{\cal C}_{\rho_D^3}^{(SMS)}&=&-\frac{\sqrt{\rho_\ell}}{6}\sqrt{\lambda}
\Big[-16 + 28  \sqrt{\rho}+ 2 \rho - 26 \rho^{3/2}  + 2 \rho^2 + 10 \rho^{5/2}  \nn \\
&+& \rho_\ell (41 - 95  \sqrt{\rho}  - 37 \rho + 103 \rho^{3/2} ) - \rho_\ell^2( 13  - 7  \sqrt{\rho})  \nn \\
&-&3 (1 + \sqrt{\rho}) (1 + \rho - \rho_\ell) \rho_\ell  \Big]  \\
&+&2\sqrt{\rho_\ell}(1-\sqrt{\rho})
\Big[\rho_\ell \big[2 - 2 \sqrt{\rho} - 5 \rho + 4 \rho^{3/2} + 5 \rho^2 +\rho_\ell(- 1 + 
 2 \sqrt{\rho}  + 5 \rho) \big]
{\cal L}_1\nn \\
&+&\big[-1 +\rho_\ell(1 +  \sqrt{\rho})^2 (2 - 6  \sqrt{\rho} + 5 \rho) + \rho_\ell^2 (-1 + 2 \sqrt{\rho}  + 5 \rho ) \big]
{\cal L}_2 \Big\} \nn
\eea

The  coefficients in the SM-P case are obtained from the corresponding ones in the SM-S case changing the sign of $m_U$ and  of the odd powers of $\sqrt{\rho}$.
\item { SM - T interference: }
\bea
{\cal C}_0^{(SMT)}&=&-2{\cal C}_{\mu_\pi^2}^{(SMT)}=12 \sqrt{\lambda} \sqrt{\rho \rho_\ell}\Big[-2 - 5 \rho + \rho^2 - 5 \rho_\ell (1-2 \rho) + \rho_\ell^2\Big]\nn \\
&+&72\sqrt{\rho  \rho_\ell}\, 
\Big\{ \rho_\ell [(1-\rho)^2+\rho \rho_\ell]
 {\cal L}_1+\rho [(1-\rho_\ell)^2+\rho \rho_\ell]
{\cal L}_2 \Big\} 
\eea
\bea
{\cal C}_{\mu_G^2}^{(SMT)}&=&6\sqrt{\lambda}\sqrt{\rho  \rho_\ell} \Big[-4 - 11 \rho+5 \rho^2 + \rho_\ell ( -3 + 50 \rho)  + 5 \rho_\ell^2 \Big]\nn \\
&+&12  \sqrt{\rho  \rho_\ell}
 \Big\{\rho_\ell [-1 - 14 \rho + 15 \rho^2 + (2  + 15 \rho) \rho_\ell]
 {\cal L}_1 \\
&+& [2 + 3 \rho +\rho_\ell(- 4  - 14 \rho  + 15 \rho^2) +  \rho_\ell^2(2 + 15 \rho)] {\cal L}_2 \Big\} \nn
\eea
\bea
{\cal C}_{\rho_D^3}^{(SMT)}&=&4\sqrt{\lambda}\sqrt{\rho \rho_\ell}\Big[-8 - 14 \rho + 10 \rho^2 +\rho_\ell(- 35 + 103 \rho ) \nn \\
&+& 7 \rho_\ell^2 -3\rho_\ell(1 + \rho - \rho_\ell) \Big]  \nn \\
&+& 48 \sqrt{\rho  \rho_\ell}\Big\{\rho_\ell \, \big[-5 \rho(1-\rho) -\rho_\ell (1-5\rho) \big]{\cal L}_1   \\
&+& \big[1 - 5 \rho \rho_\ell (1-\rho) -   \rho_\ell ^2 (1- 5 \rho) \big]
{\cal L}_2 \Big\} \nn
\eea

\item {T - S and T - P interference:}
The coefficients vanish. 

\item {R:}
\bea
{\cal C}_0^{(R)}  &=& -2 \, {\cal C}_{\mu_\pi^2}^{(R)}={\cal C}_0^{(SM)} \nn \\
{\cal C}_{\mu_G^2}^{(R)}& =& {\cal C}_{\mu_G^2}^{(SM)}  \\
{\cal C}_{\rhoD}^{(R)}& =& {\cal C}_{\rhoD}^{(SM)}  \nn
\eea

\item { SM - R interference: }
\bea
{\cal C}_0^{(SMR)}&=&-2{\cal C}_{\mu_\pi^2}^{(SMR)}=-2 \sqrt{\lambda}\sqrt{\rho} \left( 1 + 10 \rho + \rho^2 - 5 \rho_\ell- 5 \rho \rho_\ell - 2 \rho_\ell^2\right)  \nn \\
&+&12 \sqrt{\rho }\Big\{-\rho_\ell^2 (1-\rho){\cal L}_1 +\rho \big[\rho+(1-\rho_\ell)^2 \big]
{\cal L}_2 \Big\}
\eea
\bea
{\cal C}_{\mu_G^2}^{(SMR)}&=&-\frac{1}{3} \sqrt{\lambda} \sqrt{\rho}\left(13 - 14 \rho + 13 \rho^2 + 43 \rho_\ell - 77 \rho \rho_\ell - 86\rho_\ell^2 \right) \nn\\
&+&2 \sqrt{\rho }\Big\{\rho_\ell^2(-9 + 21 \rho + 4 \rho_\ell ){\cal L}_1  \\
&+&(2 - 3 \rho + 3 \rho^2 - 6 \rho \rho_\ell- 6 \rho_\ell^2 + 21 \rho\rho_\ell^2 + 4 \rho_\ell^3){\cal L}_2 \Big\} \nn
\eea
\bea
{\cal C}_{\rhoD}^{(SMR)}& =&-\frac{8}{3}\sqrt{\lambda} \sqrt{\rho} \left(11 - 7 \rho + 2 \rho^2 + 14 \rho_\ell- 13 \rho \rho_\ell- 19 \rho_\ell^2 \right)\nn \\
&+&16 \sqrt{\rho }\Big\{\rho_\ell^2 (-2 + 4 \rho +\rho_\ell){\cal L}_1 +\big[1 + \rho_\ell^2 (-2 + 4 \rho + \rho_\ell)\big]{\cal L}_2 \Big\} \,\,\,\,\,\,\,
\eea
\item {R - S  interference:}
\bea
{\cal C}_0^{(RS)}  &=& -2 \, {\cal C}_{\mu_\pi^2}^{(RS)}={\cal C}_0^{(SMS)} \nn \\
{\cal C}_{\mu_G^2}^{(RS)}& =& {\cal C}_{\mu_G^2}^{(SMS)}  \\
{\cal C}_{\rhoD}^{(RS)}& =& {\cal C}_{\rhoD}^{(SMS)}  \nn
\eea
\item {R - P  interference:}
\bea
{\cal C}_0^{(RP)}  &=& -2 \, {\cal C}_{\mu_\pi^2}^{(RP)}=-{\cal C}_0^{(SMP)} \nn \\
{\cal C}_{\mu_G^2}^{(RP)}& =-& {\cal C}_{\mu_G^2}^{(SMP)}  \\
{\cal C}_{\rhoD}^{(RP)}& =& -{\cal C}_{\rhoD}^{(SMP)}  \nn
\eea
\item { R - T  interference: }
\bea
{\cal C}_0^{(RT)}&=&-2{\cal C}_{\mu_\pi^2}^{(RT)}=12 \sqrt{\lambda}\, \sqrt{\rho_\ell}\left(1 - 5 \rho - 2 \rho^2 + 10 \rho_\ell - 5 \rho \rho_\ell + \rho_\ell^2 \right)  \nn \\
&-&72 \sqrt{\rho_\ell }\Big\{\rho_\ell \big[ (1-\rho)^2+\rho_\ell \big]{\cal L}_1 -\rho^2 (1-\rho_\ell)
{\cal L}_2 \Big\}
\eea
\bea
{\cal C}_{\mu_G^2}^{(RT)}&=&-2\sqrt{\lambda}\, \sqrt{\rho_\ell}\left( 17 - 7 \rho + 20 \rho^2 + 2 \rho _\ell+ 65 \rho \rho_\ell - 7 \rho_\ell^2 \right) \nn \\
&+&12 \sqrt{\rho_\ell }\Big\{\rho_\ell (5 + 6 \rho - 11 \rho^2 - 3 \rho_\ell - 2 \rho \rho_\ell){\cal L}_1 \\
& +&\rho \left(2 + 3 \rho - 11 \rho \rho_\ell - 2 \rho_\ell^2 \right)
{\cal L}_2 \Big\} \nn 
\eea
\bea
{\cal C}_{\rhoD}^{(RT)}& =&4\sqrt{\lambda}\, \sqrt{\rho_\ell}\big[8 + 14 \rho - 10 \rho^2 - 25  \rho _\ell - 43 \rho  \rho _\ell+ 5  \rho _\ell^2 + 3 \rho _\ell (1+\rho-\rho _\ell) \big] \nn \\
&+&48 \sqrt{\rho_\ell }\Big\{\rho_\ell \left(  2 + \rho - 3 \rho^2 - \rho_\ell - \rho \rho_\ell   \right){\cal L}_1  \\
&-&\left(  1 - 2  \rho_\ell - \rho  \rho_\ell + 3 \rho^2  \rho_\ell +  \rho_\ell^2 + \rho \rho_\ell^2 \right)
{\cal L}_2 \Big\} . \nn 
\eea

\end{itemize}
\bibliographystyle{JHEP}
\bibliography{refsFFP2}
\end{document}